\documentclass[aps,amsmath,twocolumn,amssymb,floatfix,showpacs,superscriptaddress,longbibliography]{revtex4-1}
\usepackage{amsmath}
\usepackage{mathtools}
\usepackage{braket}
\usepackage{bm}
\usepackage{graphicx}
\usepackage{amssymb}
\usepackage[dvipsnames]{xcolor}
\usepackage{multirow}
\usepackage{xfrac}
\usepackage{textcomp}
\usepackage{float}
\usepackage{enumerate}
\usepackage{subfigure}
\usepackage[colorlinks=true,linktoc=page,linkcolor=Plum,citecolor=blue,urlcolor=purple]{hyperref}

\newcommand{\ie}{{\it i.e.,\,\,}}
\newcommand{\eg}{{\it e.g.,~}}

\newcommand\bea{\begin{eqnarray}}
\newcommand\eea{\end{eqnarray}}
\newcommand\beq{\begin{equation}}  
\newcommand\eeq{\end{equation}}

\usepackage[normalem]{ulem}
\definecolor{lime}{HTML}{A6CE39}
\usepackage{sidecap,tikz}
\DeclareRobustCommand{\orcidicon}{\hspace{-1.0mm}
	\begin{tikzpicture}
	\draw[lime, fill=lime] (0.0,0.0) 
	circle [radius=0.15] 
	node[white] {{\fontfamily{qag}\selectfont \tiny \,ID}};
	\draw[white, fill=white] (-0.0525,0.095) 
	circle [radius=0.007];
	\end{tikzpicture}
	\hspace{-3.0mm}
}
\foreach \x in {A, ..., Z}{\expandafter\xdef\csname orcid\x\endcsname{\noexpand\href{https://orcid.org/\csname orcidauthor\x\endcsname}
		{\noexpand\orcidicon}}
}

\begin{document}
	
	
	\title{Tailoring phase transition from topological superconductor to trivial superconductor induced by magnetic textures of a spin-chain on a $p$-wave superconductor}  
	
	\author{Pritam Chatterjee}
	\email{pritam.c@iopb.res.in}
	\affiliation{Institute of Physics, Sachivalaya Marg, Bhubaneswar-751005, India}
	\affiliation{Homi Bhabha National Institute, Training School Complex, Anushakti Nagar, Mumbai 400094, India}
	\author{Saurabh Pradhan}
	\email{saurabh.pradhan@tu-dortmund.de}
	\affiliation{Lehrstuhl f\"ur Theoretische Physik II, Technische Universit\"at Dortmund Otto-Hahn-Str. 4, 44221 Dortmund, Germany}
	\author{Ashis K. Nandy}
	\email{aknandy@niser.ac.in}
	\affiliation{School of Physical Sciences, National Institute of Science Education and Research, An OCC of Homi Bhabha National Institute, Jatni 752050, India}
	\author{Arijit Saha\orcidC{}}
	\email{arijit@iopb.res.in}
	\affiliation{Institute of Physics, Sachivalaya Marg, Bhubaneswar-751005, India}
	\affiliation{Homi Bhabha National Institute, Training School Complex, Anushakti Nagar, Mumbai 400094, India}
	
	\begin{abstract} \noindent
		We theoretically investigate the phase transition from a non-trivial topological $p$-wave superconductor to a trivial $s$-wave like superconducting phase through a gapless phase, 
		driven by different magnetic textures as an one-dimensional spin-chain impurity, \eg  Bloch-type, in-plane and out-of-plane N\'eel-type spin-chains etc. In our proposal, the chain of 
		magnetic impurities is placed on a spin-triplet $p$-wave superconductor where we obtain numerically as well as analytically an effective $s$-wave like pairing due to spin rotation, resulting in 	
		gradual destruction of the Majorana zero modes present in the topological superconducting phase. In particular, when the impurity spins are antiferromagnetically aligned \ie spiral wave vector 
		$G_{s}=\pi$, the system becomes an effective $s$-wave superconductor without Majorana zero modes in the local density of states. The Shiba bands, on the other hand, formed due to the 
		overlapping of Yu-Shiba-Rusinov states play a crucial role in this topological to trivial superconductor phase transition, confirmed by the sign change in the minigap within the Shiba bands. 
		We also characterize this topological phase transition via gap closing and winding number analysis. Moreover, interference of the Shiba bands exhibiting oscillatory behavior within the 
		superconducting gap, $-\Delta_{p}$ to $\Delta_{p}$, as a function of $G_{s}$, also reflects an important evidence for the formation of an effective $s$-wave pairing. Such oscillation is absent in 
		the $p$-wave regime. We also analyse the case of two-dimensional $p$-wave superconductor hosting Majorana edge modes (in absence of any magnetic chain) and show that 
		initially Majorana zero modes (in presence of one-dimensional magnetic chain) can hybridize with such Majorana edge modes. Interestingly, in the topological regime with a fixed $G_{s}$ value, 
		the Mazorana zero modes survive at the ends of the magnetic chain even when the Majorana edge states disappear at some critical value of the chemical potential and exchange coupling 
		strength.
	\end{abstract}
	
	\maketitle

	\section{Introduction}\label{introduction}
	Topological phases of matter have been at the forefront of research in modern condensed matter physics for the past few decades.
	Due to nonlocal properties of these phases it is not possible to characterize the state of matter with a single physical
	order parameter. Instead, we need to define a topological invariant which can only take some discrete integer values.
	Quantization of the Hall conductance was the first signature of non-trivial topological states of matter~\cite{Thouless}. 
	However, the theoretical and experimental discovery of the quantum spin Hall effect in two-dimensional (2D) systems~\cite{kane2005quantum,kane2005quantumZ2,bernevig2006quantum,qi2011topological,
		doi:10.1126/science.1148047} characterized by $\mathcal{Z}_2$ topological invariant~\cite{KOHMOTO1985343,KANE20133} 
	have attracted a great deal of attention for the last two decades. These concepts have also been extended to 
	three-dimensional (3D) materials with non-trivial topological features in the band structures~\cite{moore2007topological,fu2007topological,roy2009topological,ando2013topological,fukaneTI2007,teofukaneTI2008,qi2008topological,Hosen2018,PhysRevB.106.125124} 
	The above ideas of topology was restricted to the system with a band gap and as a result these systems were insulators. 
	However, later it was realized that the topological classification are in principle valid for a manybody state with a band gap in the spectrum~\cite{PhysRevB.55.1142}.
	In this regard, Kitaev's elegant model was developed to characterize the topological superconductor (TSC) for an one-dimensional (1D) system~\cite{kitaev2001unpaired,kitaev2009periodic}. 
	Earlier, Read and Green proposed the idea of TSC in fractional quantum Hall state based on a 2D system~\cite{read2000paired}. Later on, Fu and Kane emphasized the realization 
	of TSC on the 2D surface of a 3D topological insulator, in close proximity to a $s$-wave superconductor and magnetic insulator~\cite{fu2008superconducting}. 
	
	Intriguingly, these TSCs host Majorana zero modes (MZMs) which are of their own anti-particle and satisfy non-abelian statistics~\cite{sato2003non,simon1983holonomy,avron1983homotopy,nayak2008non}.
	These MZMs are also suggested to be beneficial for topological quantum computation
	as they are free from decoherence by the environment~\cite{nayak2008non}. 
	In this context, theoretical proposals~\cite{PhysRevLett.105.177002,PhysRevLett.105.077001} to engineer such a TSC 
	from a semiconducting 1D nanowire with Rashba spin-orbit coupling have stimulated a lot of recent exciting experiments 
	towards realizing this exotic phase hosting MZMs. The zero bias peak, observed in several transport experiments based on hybrid 
	superconductor-semiconductor systems, has been interpreted as the indirect signatures of Majorana fermions~\cite{mourik2012signatures,das2012zero,rokhinson2012fractional,finck2013anomalous,Albrecht2016,doi:10.1126/science.aaf3961}.
	However, this conclusion still remains openly debated. 
	
	In recent times, much attention has been directed at an alternative proposal, where a TSC phase can be realized in a chain of magnetic impurities placed on the surface of a $s$-wave superconductor~\cite{Ali,nanoplates,Majoranacontrol,Felix,Felix2,natcomms,RKKY,pascal,Daniel_Loss,PhysRevB.100.075420,PhysRevB.88.205402,PhysRevB.99.024505,PhysRevB.102.104501,PhysRevB.102.165312,PhysRevB.100.214504,Hui2015,PhysRevB.104.L121406,PhysRevB.94.144509,PhysRevB.94.094515,PhysRevB.102.165312,refId0}. 
	Physically, magnetic impurities placed on a conventional superconductor give rise to Yu-Shiba-Rusinov (YSR) states~\cite{10.1143/PTP.40.435,PhysRevB.103.165403,PhysRevB.103.205424,PhysRevB.105.245403}. The individual localized YSR states can hybridize and form a band called Shiba band. The helical spin texture plays the combined role of the spin-orbit coupling and external magnetic field in the 1D nanowire proposal~\cite{PhysRevLett.105.177002,PhysRevLett.105.077001}. As a result, the topological superconducting phase can be effectively realized in the Shiba bands, akin to spinless $p$-wave superconductors (Kitaev model) hosting MZMs at the two ends of the 1D spin-chain~\cite{kitaev2001unpaired,Beenakker,Alicea_2012,Leijnse_2012}. 
	A particular advantage of the YSR chain proposal is that the signature of YSR states or topological MZMs can be experimentally detected by scanning tunneling microscopy (STM) measurements~\cite{PhysRevLett.126.076802,PhysRevLett.83.176,doi:10.1126/science.275.5307.1767,Yazdani_2015,science}.
	
	In our work, we take the opposite route in which a chain of magnetic impurities is placed on a spin-triplet $p$-wave superconducting substrate.
	Although (pseudo) spin triplet or odd-parity superconductors are rarely observed in nature, various successful attempts are made to host unconventional superconductivity in materials \eg 
	heterostructures by growing a magnetic layer on a Rashba superconductor~\cite{Menard2017}, 
	superconducting doped topological insulator~\cite{PhysRevLett.104.057001,PhysRevLett.105.097001,PhysRevB.83.224516} etc. 
	However, here we would not like to make any direct connection between our theoretical toy model and promising candidate material and relevant experiments.
	We rather investigate the effect of various magnetic textures in the form of in-plane and out-of-plane N\'eel-type, Bloch-type  etc  spin-spirals (SSs) 
	on the topological superconducting phase. 
	Depending on the spin configuration of the magnetic impurities, we observe a phase transition from TSC to trivial superconductor. As a result, the MZMs
	disappear via a gap closing within the Shiba bands. The system periodically returns back to the original TSC phase as the spins rotate back to their initial configuration. 
	Considering a lattice model, we also show that the Majorana peak height in the local density of states (LDOS) gradually becomes smaller in magnitude as we rotate 
	the spin configuration and eventually vanishes in the trivial phase. This also becomes evident from similar analysis in our 2D model. Initially, the MZMs can hybridize with the 
	Majorana edge modes (MEMs) present for a pure 2D $p$-wave superconducting substrate \ie in the absence of any magnetic spin chain. Such MEMs disappear at some critical value of the 
	chemical potential and exchange coupling strength and hence, only the MZMs survive within the Shiba bands. These MZMs are further destroyed as we tune the spin configuration of the 
	magnetic impurities. We further analytically emphasize that the spin configuration of the magnetic impurities induces an effective $s$-wave pairing in the non-trivial phase of the magnetic chain, 
	resulting in a phase transition between topologically non-trivial to trivial superconductors.
	
	The remainder of the paper is organized as follows. In Sec.~\ref{Model Hamiltonian}, we present our 1D lattice model 
	including various possible configurations of the impurity spin-chains. We discuss our numerical results for the band spectrum, LDOS, topological invariant
	and analytical results for the effective pairing gap in Sec.~\ref{results}. In Sec.~\ref{2D results}, we discuss our 2D results based on a lattice model. 
	Finally, we summarize and conclude our paper in Sec.~\ref{Summary and Discussion}.
	

	\section{Model Hamiltonian and Impurity spin texture}\label{Model Hamiltonian}
	In this section we construct a 1D tight-binding lattice model considering different orientations
	of the impurity spin-chains and our supporting observable. 
	\subsection{Tight-binding Model and different impurity spin configurations} 
\begin{figure}[H]
		\begin{center}
			\includegraphics[width=0.45\textwidth]{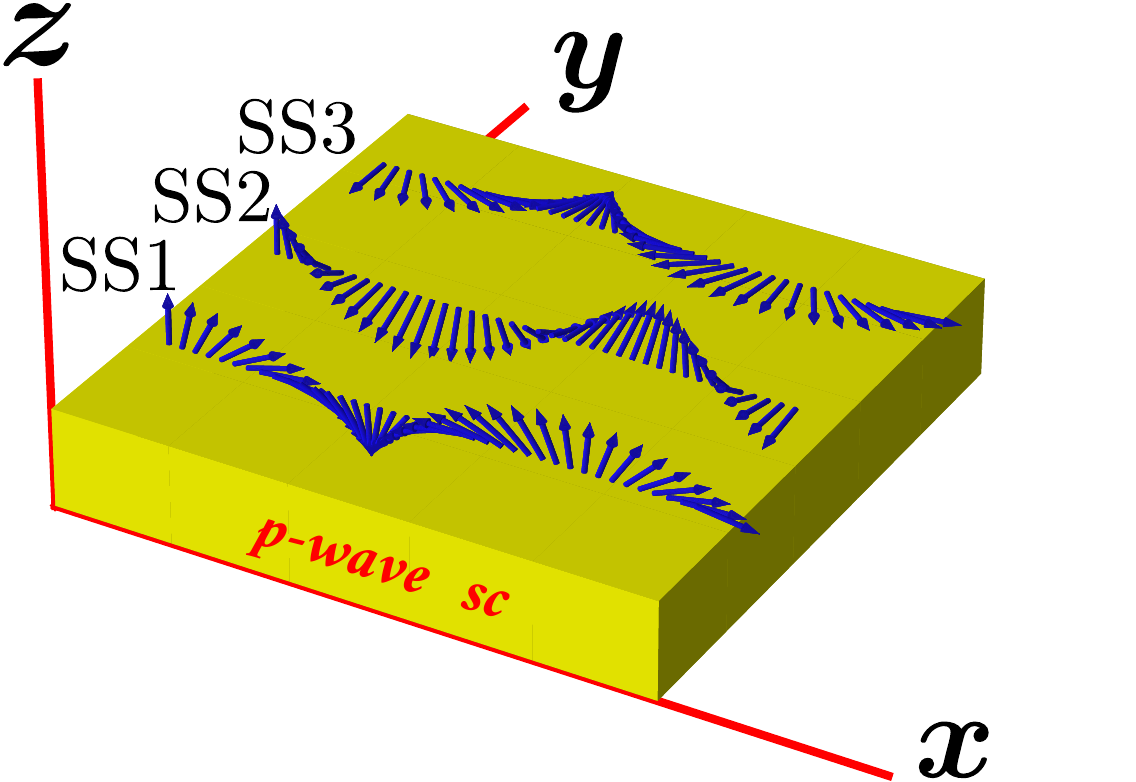}
		\end{center}
		\caption{\hspace*{-0.1cm} Schematic setup of our model where different 1D SSs, representing chain of magnetic impurities, are placed on a $p$-wave superconductor. Different spiral 
			textures are schematically depicted by the SSs with propagation vector along the $x$-axis with different rotational axes as (SS1) out-of-plane N\'eel-type, (SS2) Bloch-type and (SS3)  
			in-plane N\'eel-type impurity spin-chains.
		}
		\label{spintexture}
	\end{figure}

	We consider a 1D chain of magnetic impurities placed on a superconducting substrate essentially with $p$-wave (spin-triplet) pairing. 
	Our setup is schematically shown in Fig.~\ref{spintexture}. 
	We can effectively describe this system using a 1D lattice model with the following Hamiltonian,
	\begin{eqnarray}
	H&=&\sum_{l,\alpha}(t_{l}c_{l,\alpha}^{\dagger}c_{l+1,\alpha}^{}\!+\! h.c.)\!-\! \mu
	\sum_{l,\alpha}c_{l,\alpha}^{\dagger}c_{l,\alpha}^{}\!+\! \nonumber\\
	&&\sum_{l,\alpha,\beta}\!\!{(\vec{B_{l}}.\vec{\sigma})}_{\alpha,\beta}
	c_{l,\alpha}^{\dagger}c_{l,\beta}^{}\!+\!\Delta_{p}\! \sum_{l,\alpha}
	(c_{l,\alpha}^{\dagger}c_{l+1,\alpha}^{\dagger}\!+\! h.c.)
	\label{NuModel}\ 
	\end{eqnarray}
	where, $c^\dagger$ and $c$ correspond to electron creation and annihilation operators, respectively for the superconductors, $t$ is the electron hopping amplitude 
	between adjacent spins in the 1D chain, $\mu$ is the chemical potential and $\Delta_{p}$ is the spin symmetric triplet order parameter of the superconductor~\cite{nontriviality,Leijnse_2012}. 
	The doping of electron or hole depends on the sign of $\mu$. Here, we assume $t_{l}=t_{l}^{*}=t$ and set $t=1$ for the overall energy scale of our system.
	Third term represents the exchange coupling between the magnetic impurity spin and the electron spin of the superconductor.
	We assume that all magnetic impurity spins in the chain are classical spins and each spin is represented by a three-dimensional vector with a magnitude $S$.
	This assumption only remains valid for large $S$ were the quantum fluctuations of impurity spins can be neglected. 
	Hence, the impurity spins interacting with the electronic spin $\vec{\sigma}$ can be replaced by an effective local magnetic field as $\vec{B_{l}}=B_{0}\hat{S_{l}}$ in our model, 
	where $B_{0}=JS$, $J$ is the magnetic exchange coupling strength, $\hat{S_{l}}$ is the site dependent impurity spin direction.
	Also, $l$ and $\alpha, \beta$ correspond to the lattice sites and spin indices, respectively. Then one can write the unit vector along impurity spin 
	direction in spherical polar coordinate system~\cite{Ali,nanoplates} as,
	\begin{eqnarray}
	\hat{S_{l}}= \sin\theta_{l}\cos\phi_{l}\hat i+
	\sin\theta_{l}\sin\phi_{l}\hat{j} + \cos\theta_{l}\hat{k} \ ,
	\label{impspin}
	\end{eqnarray}
	where $\theta_{l}=G_{s}x_{l}$~\cite{theta} and $\phi_{l}=G_{h}x_{l}$~\cite{phi_Franz}, $x_{l}=la$, $a$ is the lattice constant which has been considered to be unity.
	Note that, $\theta$ (cone angle) and $\phi$ are two important quantities to determine the  noncollinear spiral configuration. For a fixed value of $\phi_l$ ($\theta_l), G_{s}$ ($G_{h}$) 
	determines the period of SS. In our model, the impurity spin-chain propagates along the $x$ direction as shown in Fig.~\ref{spintexture}. 
	There is a number of cases arise depending on different configurations of the inpurity spins~\cite{Majoranacontrol} and few cases are mentioned below.

	\begin{enumerate}
		\item When $\phi_{l}=0$ and $\theta_{l}=G_{s}x_{l}$, then the impurity spin rotates in the $xz$-plane as shown in Fig.~\ref{spintexture}, named SS1. 
		Therefore, spin rotation axis lies perpendicular to the direction of propagation of the SS. 
		This is called out-of-plane N\'eel-type SS configuration~\cite{PhysRevLett.116.177202}.
		\item On the other hand, spiral configuration with $\phi_{l}=\pi/2$ and $\theta_{l}=G_{s}x_{l}$ correspond to rotation axis of impurity spins parallel to the direction of propagation 
		\ie in the $yz$-plane ~\cite{Felix,Felix2,PhysRevResearch.2.043366}.
		Such configuration with helical modulation is called Bloch-type SS configuration, named SS2 in Fig.~\ref{spintexture}.
		\item Now, if $\theta_{l}$, the cone angle is fixed to $\pi/2$ and $\phi_{l}=G_{h}x_{l}$, inpurity spins rotate in the $xy$ plane with rotational axis perpendicular the $x$-axis. 
		Such configuration is known as in-plane N\'eel-type SS, named SS3 in Fig.~\ref{spintexture}. This type of SS configuration is also commonly known as flat spiral.
		Additionally, for $0 < \theta_l < \pi/2$ and $\phi_{l}=G_{h}x_{l}$, the spiral structure takes the conical SS form.
		\item One can in general consider the situation where both $\theta_{l}$ and $\phi_{l}$ change with respect to the lattice site index $l$, resulting in a complex spin texture 
		(see Fig.~\ref{ldosgsgh}(b) as a schematic example).
	\end{enumerate}
	
	In our analysis, we mainly focus on case 1 and 4 throughout the paper. 
	After transforming to the Bogoliubov basis 
	$\Phi_{l}=(c_{l,\uparrow},c_{l,\downarrow},c_{l,\downarrow}^{\dagger},-c_{l,\uparrow}^{\dagger})$,
	the Hamiltonian in Eq.~(\ref{NuModel}) takes the form
	
	\begin{eqnarray}
	H = \sum_{l}(\Phi_{l}^{\dagger}\tilde{\mathcal {T}_{l}}\Phi_{l+1}+h.c)+
	\Phi_{l}^{\dagger}
	\tilde{\mathcal{E}_{l}}\Phi_{l}\ ,
	\label{transModel}
	\end{eqnarray}
	where the matrices $\tilde{\mathcal {T}_{l}}$ and $\tilde{\mathcal{E}_{l}}$ can be written as,
	\begin{eqnarray}
	\tilde{\mathcal {T}_{l}} &=
	\begin{pmatrix} t & 0 & 0 & -\Delta_{p}\\ 0 & t & \Delta_{p} &
	0\\ 0 & -\Delta_{p} & -t & 0\\ \Delta_{p} & 0 & 0 & -t 
	\end{pmatrix} \ ,
	\label{hoppmat} 
	\end{eqnarray} 
	\begin{eqnarray}
	\tilde{\mathcal{E}_{l}} &= \begin{pmatrix}
	-\mu\sigma_{0}+\vec{B_{l}}.\vec{\sigma} & O_{2\times 2}
	\\
	O_{2\times 2}                           &
	\mu\sigma_{0}+\vec{B_{l}}.\vec{\sigma}
	\end{pmatrix} \ .
	\label{onsitemat}
	\end{eqnarray}
	
	Here, $\sigma_{0}$ and $\vec{\sigma}=(\sigma_{x},\sigma_{y},\sigma_{z})$ 
	are $2\times 2$ identity marix and Pauli matrices in spin space, 
	respectively and $O_{2\times 2}$ is a null matrix. Index $l$ runs from $0$ to $N$ 
	(any finite number of lattice sites). We consider $N=48$ throughout our numerical computation
	employing open boundary condition (OBC).
	
	\subsection{Observable} 
	To understand the physical presence of MZMs in a real system, one needs to calculate the LDOS~\cite{Ali,science}. 
	This identifies the topological superconducting phase by manifesting a zero-bias peak (ZBP) associated with the MZMs at the end of the impurity spin-chain, 
	whereas, the middle of the system exhibits the presence of Shiba bands within the superconducting gap 
	~\cite{science,Ali}. 
	Energy difference between the two Shiba bands main peaks (on either side of the Majorana ZBP) 
	can be defined as a bandgap, called ``minigap" $\Delta_{m}$~\cite{Ali}. Sign change of this minigap corresponds to non-topological 
	to topological phase transition~\cite{Ali,Felix}.
	The quasiparticle LDOS for $i^{\rm th}$ site and at a given energy $E$ can be defined as~\cite{natcomms,Ali}, 
	\begin{eqnarray}
	D_{i}(E)=\sum_{n\sigma}(\lvert
	u^{n}_{i\sigma}\rvert^{2}+\lvert
	v^{n}_{i\sigma}\rvert^{2})\delta(E-E_{n}) \label{ldos}\ ,
	\end{eqnarray}
	where, $u^{n}_{i\sigma}$ and $v^{n}_{i\sigma}$ are the electron like and hole like 
	quasiparticle amplitudes,  respectively.
	Here, $n$ and $\sigma$ being the eigenvalue and spin indices, respectively. 
	
	\section{Results}\label{results}
	In this section, we discuss our numerical results based on the lattice model as well as the analytical treatment by considering a continuum model to support the former. 
	We mainly focus on the discussions of how the topological properties of the system, in particular, the MZMs in the TSC phase are gradually destroyed due to the effect of different 
	SS configurations, captured in their LDOS behavior. 
	This essentially establishes that the $p$-wave superconducting gap contribution gradually decreases
	and the $s$-wave pairing gap amplitude grows with the controlled modulation of 
	$G_{h}$ and $G_{s}$ in spin-chains. 
	\subsection{\bf{LDOS at $E$=0} in $\bf{G_{s}}$-$\bf{G_{h}}$ plane}
\begin{figure}[H]
		\begin{center}
			\includegraphics[width=0.5\textwidth]{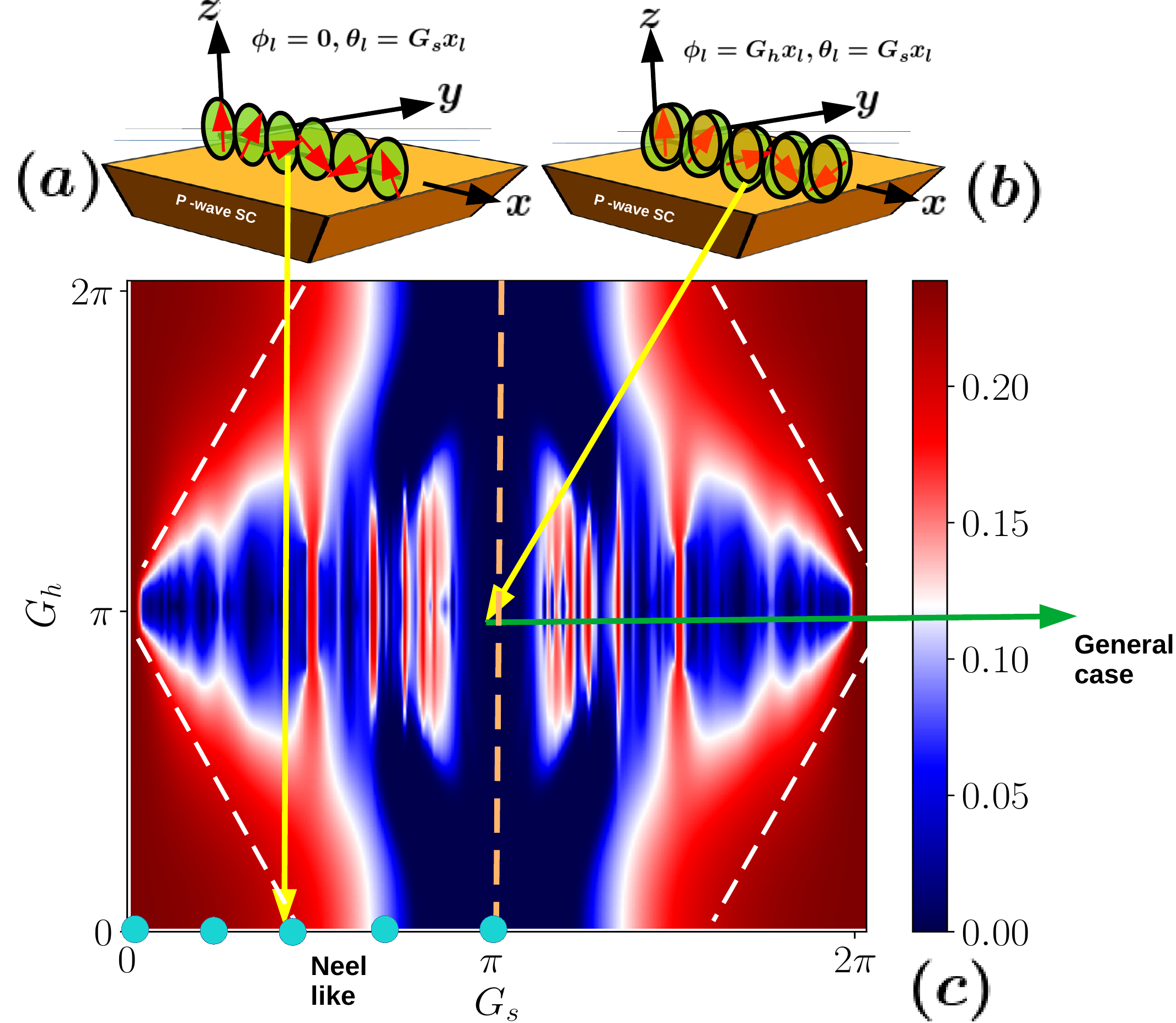}
		\end{center}
		\caption{\hspace*{-0.1cm} 
			(a) Spin-chain configuration for the N\'eel-type rotation in the $xz$-plane for $G_{h}=0$. 
			(b) Spin-chain configuration for the general type of rotation on a sphere, where both $G_{s}, G_{h}\neq0$. 
			(c) Numerical results for LDOS peak height at $E$=0 is shown in the  $G_{s}$-$G_{h}$ plane. The color bar at the right most indicates the amplitude of LDOS peak in an arbitrary unit.
			Here, we choose the other model parameters as $\mu=4t$, $B_{0}=5t$, $\Delta_{p}=t$.} 
		\label{ldosgsgh}
	\end{figure}

	In this subsection, we discuss the results of our model mainly for a general case where a chain of magnetic impurities along $x$-direction is engineered via controlled variations of  
	$G_s$ and $G_h$ on top of a $p$-wave superconducting substrate. Due to the freedom of choosing different rotational configurations of impurity spins on the surface of a sphere, governed by  
	$G_{s}, G_{h}$ values, interesting phenomena can occur on the Majorana ZBP and the corresponding topological phase. 
	In Fig.~\ref{ldosgsgh}, we summarize our results in the $G_s$-$G_h$ plane. Fig.~\ref{ldosgsgh}(a) illustrates the schematic of a special case of SS with the rotation axis perpendicular 
	to the $xz$-plane. 
	Hence, along the $G_{h}=0$ line ($\phi_{l}=0$), the linearly varying $\theta_{l}$ with lattice sites keeps N\'eel-type spin rotation with a period $\frac{2\pi}{G_s}$ in the unit of lattice constant. 
	Furthermore, Fig.~\ref{ldosgsgh}(b) schematically represents the more general case of a SS configuration with an arbitrary rotational axis direction due to both $G_{s}, G_{h} \neq 0$ \ie both 
	$\theta_{l}$ and $\phi_{l}$ vary linearly with the lattice sites.
	
	In Fig.~\ref{ldosgsgh}(c), we show the variation of zero-energy LDOS 
	as a function of $G_{s}$ and $G_{h}$ such that both $\theta_{l}$ and $\phi_{l}$ 
	vary with the 1D lattice sites. Here, LDOS is computed employing the formula mentioned in Eq.~(\ref{ldos}). 
	Also this LDOS is presented corresponding to the end of the finite chain where MZMs are present in the topological regime. 
	At the four corner points in Fig.~\ref{ldosgsgh}(c) corresponding to 
	$G_{s}, G_{h}=(0,0), (0,2\pi), (2\pi,0)$ and ($2\pi,2\pi$), the alignment of all spins is ferromagnetic 
	in the chain. Those corner points and their nearby regime indicate Kitaev chains~\cite{kitaev2001unpaired,Alicea_2012} with pure spin symmetric triplet $p$-wave pairing gap $\Delta_{p}$, 
	highlighted by maroon color regime within the dashed white line in Fig.~\ref{ldosgsgh}(c).
	The LDOS at zero-energy becomes maximum at those points, indicating topological superconducting phase hosting MZMs at the ends of the chain.
	Additionally, the red region beyond the white dashed line in Fig.~\ref{ldosgsgh}(c) where the noncollinear spin configurations are determined by $G_{s}, G_{h} \neq 0$, 
	the $p$-wave pairing is significantly strong enough to host MZMs which, however, gradually disappears in the blue region.   
	This is due to the fact that an effective $s$-wave pairing is generated, giving rise to a trivial phase in the deep blue region.
	As a result, the LDOS peak height at $E=0$ decreases significantly to a small value from its maximum value of about 0.25 (in arbitrary unit~\cite{Ali}). 
	Therefore, the thin white regime separating red and blue regions in Fig.~\ref{ldosgsgh}(c), represents roughly the phase boundary between topologically non-trivial and trivial superconducting 
	phases via a non-trivial gap closing. 
	Note that, when it enters into the topologically trivial phase, the zero energy LDOS peak height becomes vanishingly small and a trivial $s$-wave gap opens along $G_{s} = \pi$ line, as indicated 
	by a vertical dashed line at the middle of Fig.~\ref{ldosgsgh}(c).  In this regime, no MZMs are present as the $p$-wave pairing contribution is destroyed and the system hosts an effective 
	$s$-wave pairing (see further discussion on this effective $s$-wave contribution in the subsequent sub-sections).
     
\subsection{Topological characterization}
\begin{figure}[H] \begin{center}
		\includegraphics[width=0.5\textwidth]{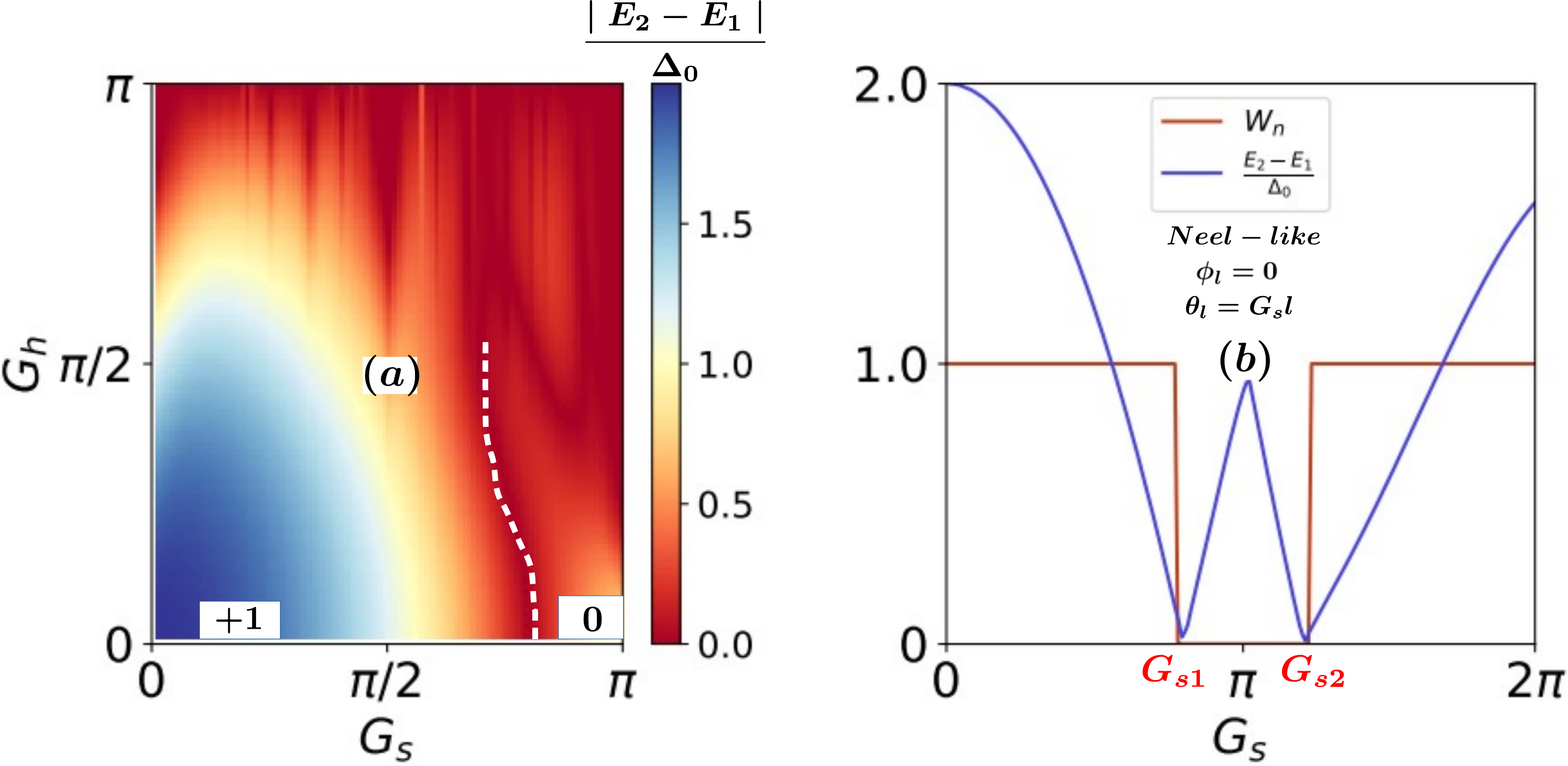} \end{center}
	\caption{\hspace*{-0.1cm} (a) Bandgap ($E_{2}-E_{1}$) (normalized by $\Delta_{0}$)  is shown as a function of two spiral wave vectors $G_{s}$ and $G_{h}$ considering purely $p$-wave pairing.
		Here, we choose $\mu=4t$, $B_{0}=5t$.
	(b) Winding number $W_{n}$ is depicted as a funtion of $G_{s}$ for the N\'eel-like ($\phi_{l}=0$) rotation of the magnetic impurities. The gap closing points and the concomminant jumps 
	of $W_{n}$ indicate the topological phase transition. 
	}
	\label{t_c}
\end{figure}
	
               From the features of LDOS, one can only observe how the Majorana peak height gradually 
	      decreases as we tune the spiral wave vector configuration. In this regard, we illustrate the variation of gap structure $\lvert E_2-E_1\rvert/\Delta_{0}$ 
               in $G_s$-$G_h$ plane where the maroon lines (one such marked by the dashed white line for guide to the eye) indicate the boundary region between topologically non-trivial ($W_n=+1$) and 
               trivial ($W_n=0$) phase (see Fig.~\ref{t_c}(a)). It has been clearly shown that when $G_h=0$, gap closing takes place around $G_s\sim 2.5$. In order to characterize the topological phase 
               of our model, we also calculate the topological invariant considering the N\'eel type SS configuration. As our lattice model doesn't satisfy translation symmetry, but continues to preserve 
               the chiral symmetry in the presence of the magnetic impurities, we follow Refs.~\cite{Demler2010, PhysRevLett.109.150408} to compute the winding number $W_{n}$ in real space. 
               The corresponding winding number $W_{n}$ is shown for the case when $\phi_{l}=0$ and $\theta_{l}=G_{s}l a$ (N\'eel type as shown in Fig.~\ref{spintexture}) in Fig.~\ref{t_c}(b). 
               It is evident that there are sharp topological phase transition around $G_{s1}= 2.5$ and $G_{s2}=3.7$ where $W_{n}$ jumps from 1$\rightarrow$0 in accordance with the gap 
               $\lvert E_2-E_1\rvert$ closing points. This also matches with the $G_{h}=0$ line of Fig.~\ref{t_c}(a). The topological superconducting phase with $W_{n}=+1$ hosts Majorana zero modes. 
               This gives a clear signature of the topological phase transition in our model. Furthermore, for the N\'eel type rotation, an analytical understanding of the gap structure is presented in 
               Appendix~\ref{gap_structure}.

	\subsection{Detailed study of the topologically non-trivial to trivial superconductor phase transition for N\'eel type rotation in the magnetic impurity-chain}
\begin{figure*}[]
		\begin{center}
			\includegraphics[width=1.0\textwidth]{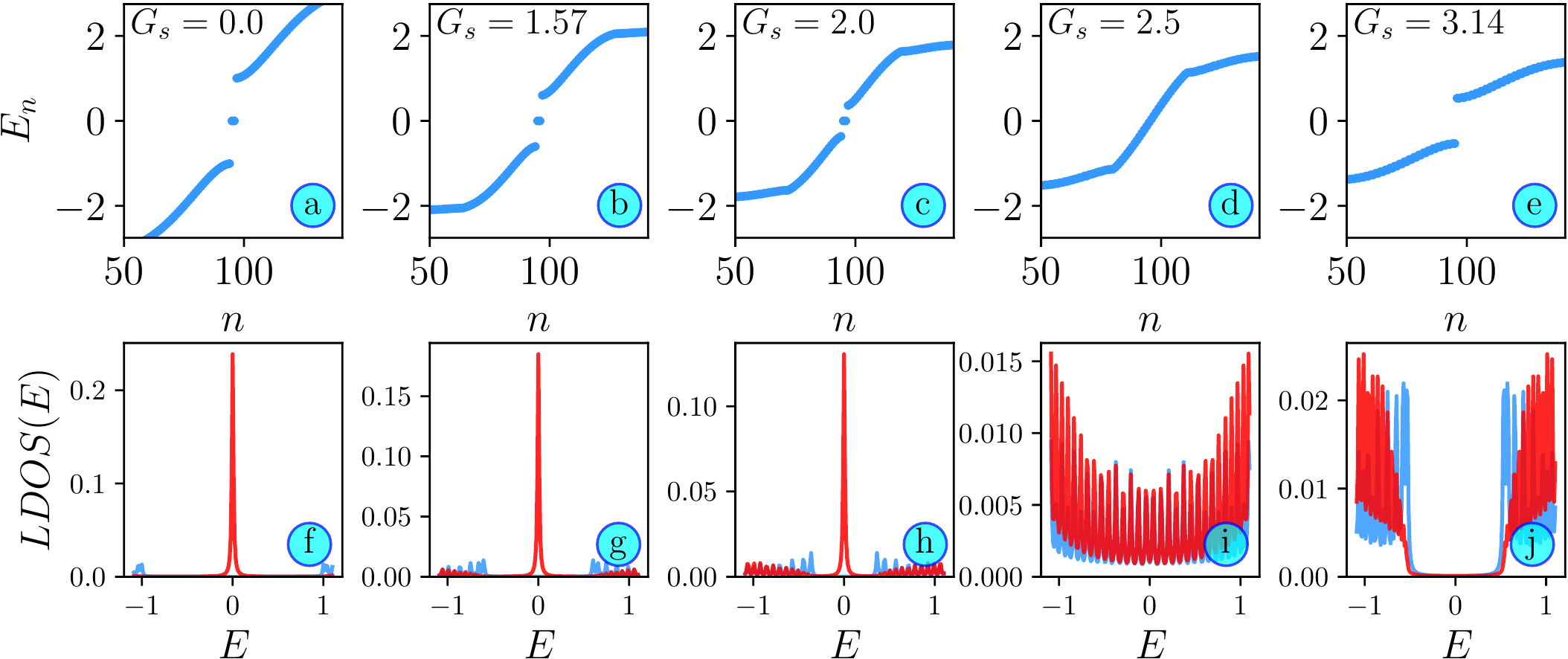}
		\end{center}
		\vspace{-0.5cm}
		\caption{\hspace*{-0.1cm} (i) The first row,
			Fig.~\ref{shiba}(a)-(e), represents the eigenvalue spectrum for a finite size 1D chain considering
			different values of spiral wave vector
			$G_{s}$=0.0, 1.57, 2.0, 2.5, 3.14. Here, we choose the other model parameters as
			$\mu=4t$, $B_{0}=5t$, $\Delta_{p}=t$. (ii) In the second row
			Fig.~\ref{shiba}(f) -(j), we show the
			LDOS as a function of energy for the same set of SS wave
			vector $G_{s}$ and choosing the same parameter values as mentioned above. 
			The Majorana ZBP is indicated by the red plots in
			panels (f)-(h) when LDOS is computed at the end of the finite chain. Whereas, blue plots denote the
			Shiba bands in LDOS when it is measured at the middle of the chain.
		}\label{shiba}
	\end{figure*}

	In this subsection, we further discuss the outcome for a N\'eel-type rotation of impurity spin-chain, SS1 depicted in Fig.~\ref{spintexture}.
	In this case, we obtain the variation of different physical quantities for a SS with varying period by changing $G_{s}$ while keeping $\phi_{l}=0$. 
	In Fig.~\ref{shiba}(a)-(e), we show the eigenvalue spectrum for different values of $G_{s}$, considering a finite size system with OBC. 
	Those $G_s$ values are approximately marked by the cyan dots on the $G_s$ axis of Fig.~\ref{ldosgsgh}(c).
	This enables one to understand the appearance and disappearance of MZMs under the influence of SS period.
	For $G_{s}=0$ and $2\pi$, a pair of MZMs appears at the two ends of a finite chain due to pure $p$-wave superconductivity \ie the Kitaev limit~ 
	\cite{kitaev2001unpaired,Beenakker,Alicea_2012,Leijnse_2012}. 
	Around these points, MZMs appear with the following condition $B_{0}\geq\sqrt{\mu^{2}+\Delta_{p}^{2}}$~\cite{PhysRevLett.105.177002,PhysRevLett.105.077001},
	akin to the 1D nanowire case. Due to the modulation of SS wave vector by changing $G_{s}$, the topological superconducting gap or the minigap, $\Delta_{m}$, within the Shiba bands 
	gradually decreases and the MZMs lying in the gap disappear from the system for $G_s=2.45$. 
	Interestingly, a nontrivial gap closing takes place within the Shiba bands for a particular period of the SS corresponding to $G_{s}=2.45$ as shown in Fig.~\ref{shiba}(d).
	However, further modulation of $G_{s}$ upto $\pi$ opens up a trivial superconducting gap without MZMs in the system. 
	The $G_{s}=\pi$ represents an antiferromagnetic spin-chain configuration and with further rotation of spins from $\pi$, the spin-chain will again become a ferromagnetic chain 
	at $G_{s}=2\pi$ point. Therefore, within $2\pi > G_{s} > \pi$, the phenomena repeats as we move away from $2\pi$ point towards $\pi$ point and the Kitaev phase reappears at $G_{s}=2\pi$.
	
	The main underlying physics of our model is governed by the modulated spin textures which can be characterized by $G_h$ and $G_s$. 
	Here, it is worth to mention that noncollinear magnetism~\cite{PhysRevResearch.2.043366,Charanpreet} 
	including spin-spiral order~\cite{PhysRevLett.116.177202} occurs due to the competition between underlying Heisenberg exchange interaction and the Dzyaloshinskii-Moriya (DM) interaction. 
	In most cases where magnetic layer/chain is grown on metallic or semiconducting substrate, the ubiquitous Ruderman-Kittel-Kasuya-Yosida (RKKY) exchange tensor between magnetic 
	atoms contain both Heisenberg exchange and DM interactions. The antisymmetric DM interaction is the result of spin-orbit coupling that occurs in magnetic heterostructures due to the 
	broken inversion symmetry. In addition, there may exist another spin-orbit coupling induced interaction parameter called axial magnetic anisotropy constant. In order to engineer $G$, one needs 
	to have control on these interaction parameters. Indeed, these parameters are highly tunable through intercalation~\cite{Zheng2021}, 
	electron doping~\cite{Charanpreet}, thickness~\cite{https://doi.org/10.1002/adma.202002043}, 
	strain~\cite{PhysRevLett.116.017201}, external electric filed~\cite{Yang2018}, 
	and adjusting the interatomic distances~\cite{Khajetoorians2016} etc. 
	Hence, the smooth variation in $G$ can be attributed to the high tunability of RKKY exchange tensor and magnetocrystalline anisotropy energy.
	
	In the presence of superconducting substrate, the exchange frustration arises from the 
	similar RKKY-type exchange interactions between the impurity spins~\cite{Felix, RKKY,PhysRevB.90.125443}.
	Such RKKY-type interaction between two magnetic impurities at a distance $x_{ij}$ apart can be mediated via the virtual exchange of electron-hole excitations whereas superconductivity 
	prohibits this pairs with energy less than the superconducting gap. As a result, system exhibits an effective pairing gap which depends on $x_{ij}$ and spin spiral wave vector. 
	Hence, for our case, the physical reason behind the disappearance of zero-energy MZMs can be attributed to the fact that
	interplay of N\'eel-type spin spiral order and $p$-wave superconductivity generates an effective pairing gap which is $s$-wave in nature. 
	This can be written in real space as
	\begin{eqnarray}
	{{\tilde{\Delta}}^{\rm eff}}_{ij} &=&
	-(\epsilon_{0}/2)\delta_{ij}-(1-\delta_{ij})
	\nonumber\\
	&&\left[\frac{\Delta_{t}}{\sqrt{1+\tilde{\Delta}_{p}^{2}}}\sin(k_{F}^{\prime}
	\lvert
	x_{ij}\rvert)
	+\gamma\Delta_{t}\cos(k_{F}^{\prime}x_{ij})\right]\nonumber\\ 
	&&\times~
	e^{-\frac{\mid x_{ij}\mid}{\xi_{0}}}
	\cos\left(\frac{G_{s}}{2}x_{ij}\right)\ ,
	\label{effdel}
	\end{eqnarray}
	where, $x_{ij}=x_{i}-x_{j}$, $k_{F}^{\prime}=\frac{k_{F}}{\sqrt{1+\tilde{\Delta}_{p}^{2}}}$,
	$\tilde{\Delta}_{p}=\Delta_{p}/V_{F}$,\\
	$\Delta_{t}=\frac{\Delta_{p}k_{F}}{\sqrt{1+\tilde{\Delta}_{p}^{2}}}$, and
	$\gamma=\frac{\tilde{\Delta}_{p}}{\sqrt{1+\tilde{\Delta}_{p}^{2}}}$.
	
	\begin{figure}[H]
		\begin{center}
			\includegraphics[width=0.48\textwidth]{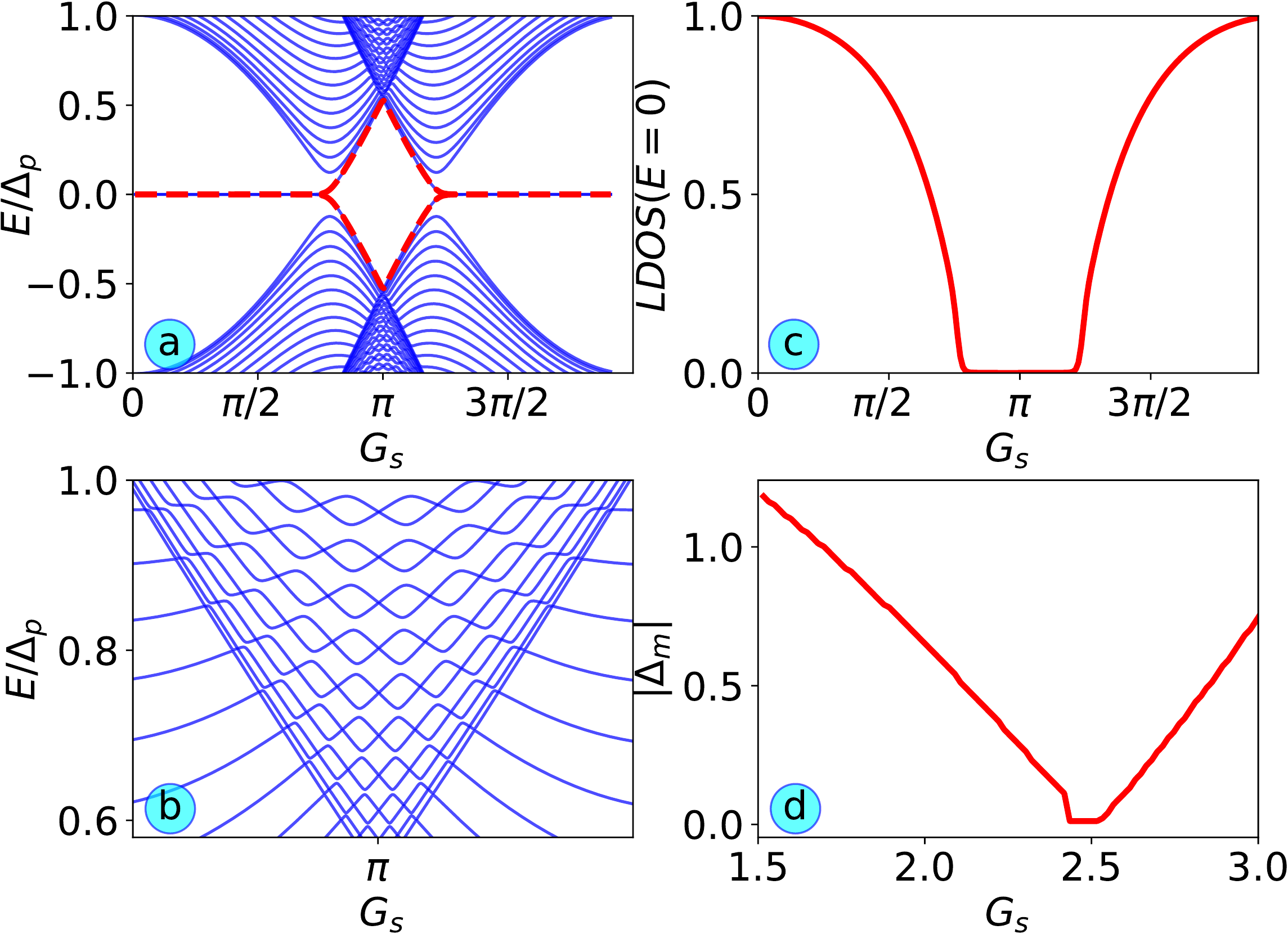}
		\end{center}
		\caption{\hspace*{-0.1cm} (a) Band spectrum where the energy, $E$, is plotted as a function of $G_{s}$ within the energy range $-\Delta_{p}$ to $\Delta_{p}$. 
			(b)  Enlarged view of the energy spectrum is shown as a function of $G_{s}$, around the trivial gapped region in the vicinity of $G_s=\pi$.
			(c) LDOS at $E=0$ is shown as a function of $G_{s}$ and the vanishing LDOS represents the trivial $s$-wave gapped phase.
			(d) Variation of the absolute value of the minigap ($\Delta_{m}$) is illustrated as a function of $G_{s}$. 
			Here, other model parameters are chosen as $\mu=4t$, $B_{0}=5t$, $\Delta_{p}=1.0t$. 
		}\label{bands}
	\end{figure}
	
	Here, $\xi_{0}=V_{F}/\omega(0)$ is the superconducting coherence length, $V_{F}$ and $k_{F}$ are the Fermi velocity and Fermi momentum respectively, 
	and $\omega(0)=\Delta_{t}/\sqrt{1+\tilde{\Delta}_{p}^{2}}$. $\epsilon_{0}$ is
	the YSR bound state energy in the deep Shiba limit when a single magnetic
	impurity is placed on a $p$-wave superconductor~\cite{Kaladzhyan_2016}.
	Detailed analytical derivation of Eq.~(\ref{effdel}) is presented in Appendix~\ref{eff_pairing}
	based on a low-energy continuum model. It is evident from Eq.~(\ref{effdel}) that the nature of the 
	effective gap is $s$-wave like as $\tilde{\Delta}^{\textrm{eff}}_{ij}=\tilde{\Delta}^{\textrm{eff}}_{ji}$. 
	Due to the emergence of effective $s$-wave pairing, the system approaches to a trivial superconducting phase with the modulation of $G_{s}$ within the range $\pi  \gtrsim G_{s} \gtrsim 2.5$.
	Particularly at $G_{s}=\pi$, the SS corresponds to the antiferromagnetic spin-chain and the effective gap is purely $s$-wave type, consistent with the earlier reports~\cite{phi_Franz}.
	
	To elaborate further, in Fig.~\ref{shiba}(f)-(j), we discuss about the variation of LDOSs as a function of energy with different values of $G_{s}$.  
	In the topological regime, if we compute the LDOS at the end of the finite chain, we obtain Majorana ZBP (red color) at exactly zero-energy ($E=0$). 
	On the other hand, feature of LDOS at the middle of the chain indicate non-topological Shiba band (blue color) within the superconducting gap $-\Delta_{p}$ to $\Delta_{p}$~\cite{science,Ali}. 
	Distance between the two closest Shiba peaks on either sides of the $E=0$ peak is called the minigap, $\Delta_{m}$~\cite{Ali}.
	From Fig.~\ref{shiba}(f), it is evident that the height of the Majorana ZBP for $G_{s}=0$ is maximum, $\approx 0.25$ in an arbitrary unit. 
	Concomitantly, no Shiba peak appears within $-\Delta_{p}$ to $\Delta_{p}$ as the impurities are ferromagnetically aligned. 
	Height of the Majorana peak at $E=0$ decreases with the enhancement of $G_{s}$. Eventually at $G_{s}=2.5$, 
	Majorana ZBP disappears and it refers to a gapless phase in LDOS. 
	Finally, we obtain some trivial gap in LDOS spectrum at $G_{s}=\pi$.
	Simultaneously, the signature of Shiba bands are reflected in the LDOS behavior
	within $-\Delta_{p}$ to $\Delta_{p}$ (see Fig.~\ref{shiba}(g)-(j)) when $G_{s}\neq 0$. 
	This can be understood as we increase $G_{s}$, the magnitude of minigap $\Delta_{m}$ decreases with $G_{s}$ and at $G_{s}=2.45$, it vanishes. Afterwards, further increase of 
	$G_{s}$ enables the minigap to change sign and $\lvert \Delta_{m}\rvert$ increases further in magnitude.  
	Sign change of minigap ($\Delta_{m}$) indicates the topological to non-topological phase transition.

	In Fig.~\ref{bands}, we explain the variations of important physical quantities as a function of $G_{s}$ for out-of-plane N\'eel-type SSs. 
	In Fig.~\ref{bands}(a), we depict the variations of the energy spectrum within the range $-\Delta_{p}$ to $\Delta_{p}$, as a function
	of $G_{s}$. The spectrum exhibits 2$\pi$ periodicity with respect to $G_{s}$.
	For $G_{s}=0$, the system is in the Kitaev limit manifesting topological superconducting phase and thus we always obtain the MZMs, indicated by the red line in Fig.~\ref{bands}(a).
	Majorana modes merge into the bulk states and disappear completely around $G_{s}=2.5$ as after that the system purely behaves like a trivial $s$-wave superconductor.
	Interestingly, we observe that Shiba bands within $-\Delta_{p}$ to $\Delta_{p}$ (precisely, above $|E/\Delta_p|\gtrsim 0.6$) in the trivial $s$-wave superconducting phase oscillate 
	as they interfere with each other in the presence of a magnetic impurity chain.
	This can be clearly visible in Fig.~\ref{bands}(b) where the interference of Shiba bands exhibits oscillations around $G_{s}=\pi$ SS configuration, within a range $3.8 \gtrsim G_{s} \gtrsim 2.5$. 
	Such type of oscillations can be another evidence of the formation of an effective pure $s$-wave pairing
	and is absent in the non-trivial topological $p$-wave superconducting phase. 
	In Appendix~\ref{bands_s-wave}, we discuss in detail that similar oscillations of  
	Shiba bands can be obtained when 1D spin spiral is placed on a purely $s$-wave superconductor.
	Further increase in $G_{s}$ beyond $G_{s}\approx 3.8$ value, the system starts to tunes itself and the initial 
	topological superconducting phase with a dominant $p$-wave pairing reappears as shown by the normalized LDOS  peak height  behavior with change of $G_s$ in Fig.~\ref{bands}(c). 
	The Majorana ZBP height is maximum for the ferromagnetic spin-chain corresponding to $G_s=0$ and $2\pi$.
	As we tune the $G_s$ towards $\pi$, a smooth variation of the Majorana peak height is observed as a function of $G_{s}$. 
	Interestingly, it monotonically decreases upto $G_{s}\approx 2.5$ ($G_{s}\approx 3.8$) starting from $G_{s}\approx 0$ ($G_{s}\approx 2\pi$)
	and after that the Majorana peak disappears. 
	The corresponding gap in the LDOS signifies emergence of topologically trivial phase with an effective $s$-wave pairing. Therefore, in a real system, one can control the 
	appearance and disappearance of MZMs using the variation of SS wave vector.
	In Fig.~\ref{bands}(d), we depict the variation of the absolute value of the minigap ($\Delta_{m}$) as a function of the $G_{s}$. 
	The absolute value of $\Delta_{m}$ gradually decreases from its maximum value at $G_{s}=0$ and vanishes around $G_{s}=2.5$, referring to the topological phase in the system. 
	A small plateau corresponding to $\Delta_{m}=0$ indicates a gapless phase near $G_s=2.5$ (see also Fig.~\ref{shiba}(d) and (i)) and afterwards it's magnitude rises again as one changes $G_{s}$.
	Hence, it is evident that there is a sign change of $\Delta_{m}$ within the Shiba bands, indicating a topological phase transition from a non-trivial $p$-wave to a trivial 
	$s$-wave superconductor through a gapless phase.
	
	\begin{figure*}[ht!]
		\begin{center}
			\includegraphics[width=1.0\textwidth]{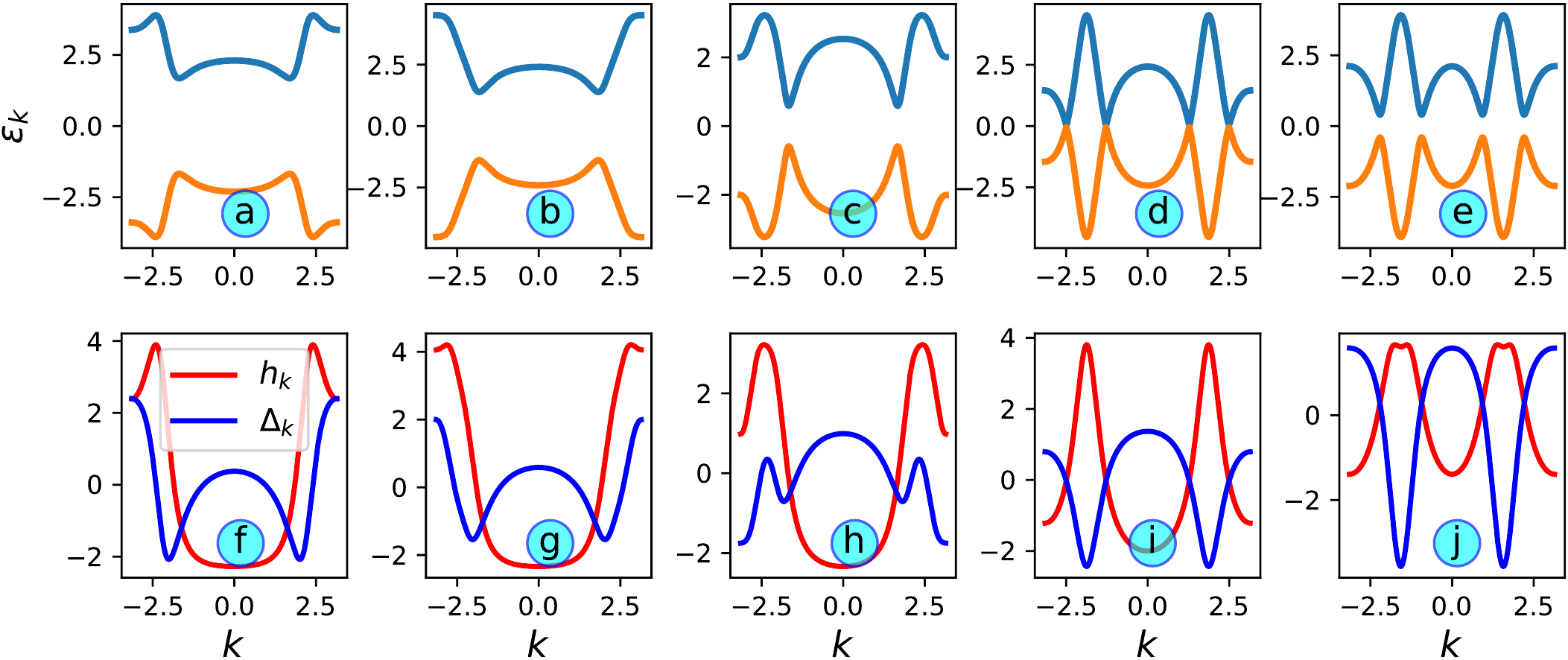}
		\end{center} \vspace{-0.5cm}
		\caption{\hspace*{-0.1cm} (i) In the first row, Fig.~\ref{analytical}(a)-(e),
			we show the behavior of the energy eigenvalue spectrum $\epsilon_{k}$ for five different values of $G_{s}=0.0,1.57,2.0,2.55,3.14$. (ii) Second row, Fig.~\ref{analytical}(f)-(j),
			represents the variation of the on-site term $h_{k}$ and the effective pairing gap $\Delta_{k}$ 
			as a function of the propagating vector $k$, considering the same set of $G_{s}$ values as mentioned above.  
			Here, we choose the other parameters as $\epsilon_{0}=0$, $\xi_{0}=2a$, $k_{F}a=(2\pi+G_{s}/2)$. 
		}\label{analytical}
	\end{figure*}
	
	\subsection{Analytical results based on continuum model}
\begin{figure}[]
		\begin{center}
			\includegraphics[width=0.53\textwidth]{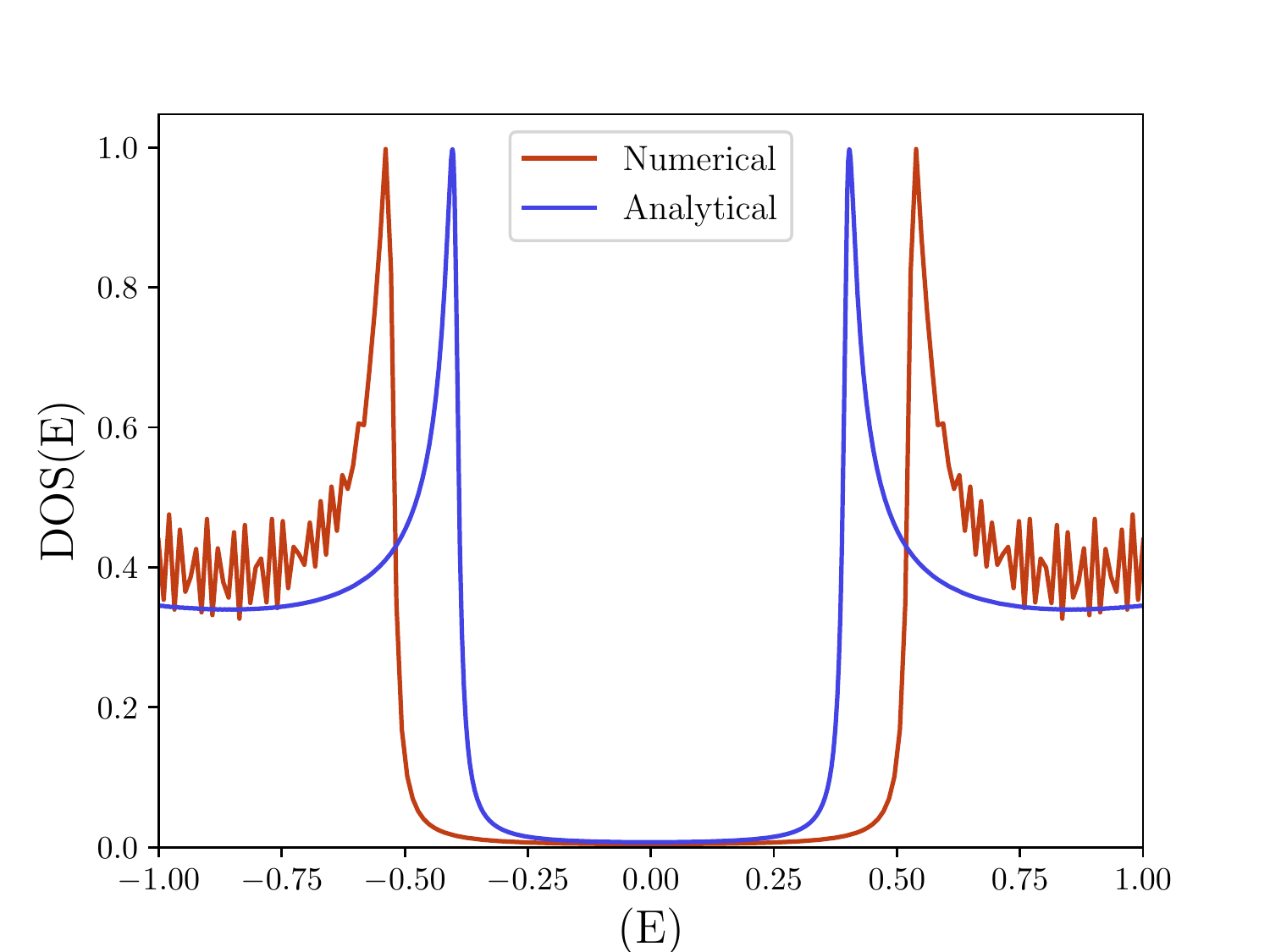}
		\end{center}
		\caption{\hspace*{-0.1cm}
			Normalized total DOS is shown as a function of energy $E$ for $G_{s}=\pi$ \ie when 
			impurity spins follow antiferromagnetic alignment. Here, numerical and analytical results
			are indicated by red and blue colors, respectively. 
			We choose the other model parameters as $\mu=4t$, $B_{0}=5t$, $\Delta_{p}=1.0t$ and 
			$\epsilon_{0}=0$, $\xi_{0}=2a$, $k_{F}a=(2\pi+G_{s}/2)$ for our numerical and analytical
			computation, respectively.
		}
		\label{dos}
	\end{figure}

	In this subsection, we discuss our analytical approach for the N\'eel-type rotation in order to support our numerical results presented in the previous subsection.
	Our analytical calculation is based on an effective continuum theory of our lattice model in Eq.~(\ref{NuModel}). 
	Similar approach was reported earlier in case of a chain of magnetic impurities placed on a $s$-wave superconductor~\cite{Felix,pascal,Daniel_Loss}.
	From the features of the energy spectrum, the effective gap structure and total density of states (DOS) presented in Fig.~\ref{analytical} and  Fig.~\ref{dos}, respectively, we confirm 
	the topological phase transition along with the nature of an effective $s$-wave pairing gap.
	
	We begin from the low energy continuum model (Bogoliubov-de Gennes Hamiltonian (BdG)) in which a 
	1D chain of localized magnetic impurities is placed on a $p$-wave superconductor,
	\begin{eqnarray}
	H=\xi_{k}\tau_{z}-J\sum_{j}(\vec{S_{j}}.\vec{\sigma})\delta(x-x_{j})
	+\Delta_{p}k\tau_{y}\ ,
	\label{lowHamil}
	\end{eqnarray}
	where, $\xi_{k}=\frac{k^{2}}{2m}-\mu$, $k$ is the wave-vector corresponding to a model with 1D spin-chain placed on a superconducting substrate
	with pairing gap $\Delta_{p}$. 
	Particle-hole and spin degrees of freedom can be expressed in terms of the Pauli matrices $\tau$ and $\sigma$ respectively.
	Starting from this model one can obtain an effective pairing gap in real space as mentioned in Eq.~(\ref{effdel}).
	The calculational details are presented in Appendix~\ref{eff_pairing}.
	Hence, taking Fouriar transform of this effective gap and diagonal components of the real-space Hamiltonian,
	we obtain diagonal elements ($h_{k}$) and effective pairing gap ($\Delta_{k}$) as non-diagonal elements 
	in momentum space. 
	The analytical expressions of $h_{k}$ and $\Delta_{k}$ for our 1D chain can be written as
	\begin{eqnarray} h_{k}
	&=&
	\frac{\epsilon_{0}}{2}-\frac{1}{2}\Bigg\{k_{F}^{\prime}\Delta_{p}\Bigg[F_{1}\Bigg(k+\frac{G_{s}
	}{2}\Bigg)+F_{1}\Bigg(k-\frac{G_{s}}{2}\Bigg)\Bigg]\nonumber\\
	&&+\gamma\Delta_{t}\Bigg[F_{2}\Bigg(k+\frac{G_{s}}{2}\Bigg)+F_{2}\Bigg(k-\frac{G_{s}}{2}\Bigg)\Bigg]\Bigg\}\ ,
	\label{Feffonsite}
	\end{eqnarray}
	\begin{eqnarray} \Delta_{k} &\!\!=\!\!&
	-\frac{\epsilon_{0}}{2}\!+\!\frac{1}{2}\Bigg\{\frac{\Delta_{t}}{\sqrt{1+\tilde{\Delta}_{p
			}^{2}}}\Bigg[F_{4}\Bigg(k \!+\!\frac{G_{s}}{2}\Bigg)\!+\!F_{4}\Bigg(k \!-\!\frac{G_{s}}{2}\Bigg)\Bigg]\nonumber\\
	&&+\gamma\Delta_{t}\Bigg[F_{3}\Bigg(k+\frac{G_{s}}{2}\Bigg)+F_{3}\Bigg(k-\frac{G_{s}}{2}\Bigg)\Bigg]\Bigg\}\ .
	\label{FeffDelta}
	\end{eqnarray}
	
	Therefore, from the BdG Hamiltonian (see Appendix~\ref{eff_pairing} for details) we can obtain the effective energy spectrum of our system as 
	$\epsilon_{k}=\pm\sqrt{h_{k}^{2}+\Delta_{k}^{2}}$. 
	In Eq.~(\ref{Feffonsite}) and (\ref{FeffDelta}), $\epsilon_{0}=2\Delta_{t}/\alpha$ is the Shiba bound state energy for a single magnetic impurity placed on a $p$-wave superconductor~\cite{Kaladzhyan_2016}, $\alpha=JS/(V_{F}\sqrt{1+\tilde{\Delta}_{p}^{2}})$ is the impurity strength and the renormalized Fermi momentum, $k_{F}^{\prime}=k_{F}/\sqrt{1+\tilde{\Delta}_{p}^{2}}$. 
	The corresponding functions which we have defined in Eq.~(\ref{Feffonsite}) and (\ref{FeffDelta}) read,
	\begin{eqnarray}
	F_{1}(k)
	&=&\Im\Bigg[\!\frac{i}{1-e^{\frac{a}{\xi_{0}}-i(k_{F}^{\prime}+k)}} 
	\!+\! \frac{i}{1-e^{\frac{a}{\xi_{0}}-i(k_{F}^{\prime}-k)}}\Bigg]\!\ \!, \\
	F_{2}(k) &=&
	\Re\Bigg[\!\frac{i}{1-e^{\frac{a}{\xi_{0}}-i(k_{F}^{\prime}+k)}}
	\!+\!\frac{i}{1-e^{\frac{a}{\xi_{0}}-i(k_{F}^{\prime}-k)}}\Bigg]\!\ \! ,
	\end{eqnarray}
	\begin{eqnarray}
	F_{3}(k) &=&
	\Re\Bigg[\!\frac{1}{1-e^{\frac{a}{\xi_{0}}-i(k_{F}^{\prime}+k)}}
	\!+\!\frac{1}{1-e^{\frac{a}{\xi_{0}}-i(k_{F}^{\prime}-k)}}\Bigg]\!\ \!,\\
	F_{4}(k) &=&
	\Im\Bigg[\!\frac{1}{1-e^{\frac{a}{\xi_{0}}-i(k_{F}^{\prime}+k)}}
	\!+\!\frac{1}{1-e^{\frac{a}{\xi_{0}}-i(k_{F}^{\prime}-k)}}\Bigg]\!\ \! ,
	\end{eqnarray}
	where, $\xi_{0}$ is the superconducting coherence length. 
	In principle, in our model Hamiltonian (Eq.~(\ref{NuModel})), we always consider $\xi_{0}\to\infty$ limit so that mean field theory of superconductivity remains valid~\cite{CastroNeto}.
	Therefore, in the theoretical model (Eq.~(\ref{lowHamil})) we always consider $\xi_{0}> a$ throughout our analysis.
	
	In Fig.~\ref{analytical}(a)-(e), we depict the variation of the energy spectrum $\epsilon_{k}$ for N\'eel-type rotation of the magnetic impurities, choosing different values of $G_{s}$. 
	Moreover,  Fig.~\ref{analytical}(f)-(j) represent the variation of both the effective gap $\Delta_{k}$ and $h_{k}$ as a function of $k$, employing Eq.~(\ref{Feffonsite}) and (\ref{FeffDelta}), respectively. 
	For $G_{s}=0$ case, the superconducting gap is $p$-wave in nature. 
	Therefore, we obtain a topologically non-trivial gap in the energy spectrum, see Fig.~\ref{analytical}(a). 
	This phase hosts MZMs. 
	Afterwards, this gap became smaller in magnitude with increasing $G_{s}$,  as one compares the gap around $\epsilon_k=0$ in Fig.~\ref{analytical}(b) and (c).
	Eventually, this gap closes at $G_{s}=2.55$ and after that it again open up as evident from Fig.~\ref{analytical}(d) and (e). 
	Therefore, a topological phase transition takes place at $G_{s}=2.55$. 
	This phenomena can also be understood from the individual behavior of $\Delta_{k}$ and $h_{k}$ for the same set of $G_{s}$ values. 
	Most importantly, this gap closing and reopening phenomena within our analytical treatment matches with our numerical results where it has been shown to take place at slight 
	lower value of $G_{s}$, which is $2.5$.
	
	For $G_{s}=\pi$, when impurity spins form an antiferromagnetic spin-chain, the gap in the energy spectrum is trivial, see Fig.~\ref{analytical}(e). 
	However, this gap is not insulating as evident from the variation of $\Delta_{k}$ at $G_{s}=\pi$ in Fig.~\ref{analytical}(j). 
	Hence, one can conclude that it should be a trivial superconducting phase rather than a trivial insulator. 
	To confirm that this superconducting phase is effective $s$-wave in nature, we compute the total DOS, both numerically as well as analytically, and compare their behavior 
	at $G_{s}=\pi$ as a function of energy as shown in Fig.~\ref{dos}.
	We employ the following formula to numerically compute the total DOS as 
	\begin{eqnarray}
	D(E)=\frac{1}{\pi}\sum_{n}\delta(E-E_{n})\ ,
	\label{NTDOS}
	\end{eqnarray}
	where, the sum is taken over the energy eigenvalues ($E_{n}$) of Eq.~(\ref{NuModel}) and $\delta(E-E_{n})$ is modeled using a Lorentzian with broadening $0.01t$.
	Analytically, normalized total DOS can be calculated from the formula,

	\begin{eqnarray}
	D(E)=\int_{-\pi}^{\pi}\frac{dk}{2\pi}\delta(E-\epsilon_{k})\ ,
	\label{ATDOS}
	\end{eqnarray}
	where, $\epsilon_{k}=\pm\sqrt{h_{k}^{2}+\Delta_{k}^{2}}$ and we use Eq.~(\ref{Feffonsite}) and (\ref{FeffDelta}). 
	Note, we obtain qualitatively similar features of total DOS for both the cases at $G_{s}=\pi$. 
	In particular, quasi-particle peaks appear at $\Delta_{\rm {eff}}\simeq \Delta_{p}/2$ and after that 
	they decay as can be seen from Fig.~\ref{dos}. 
	Hence, we conclude that this superconducting phase is conventional $s$-wave like with no MZMs present.
	
\section{Discussion considering 2D $p$-wave superconductor} \label{2D results}
\begin{figure*}[]
\begin{center}
	\includegraphics[width=1.0\textwidth]{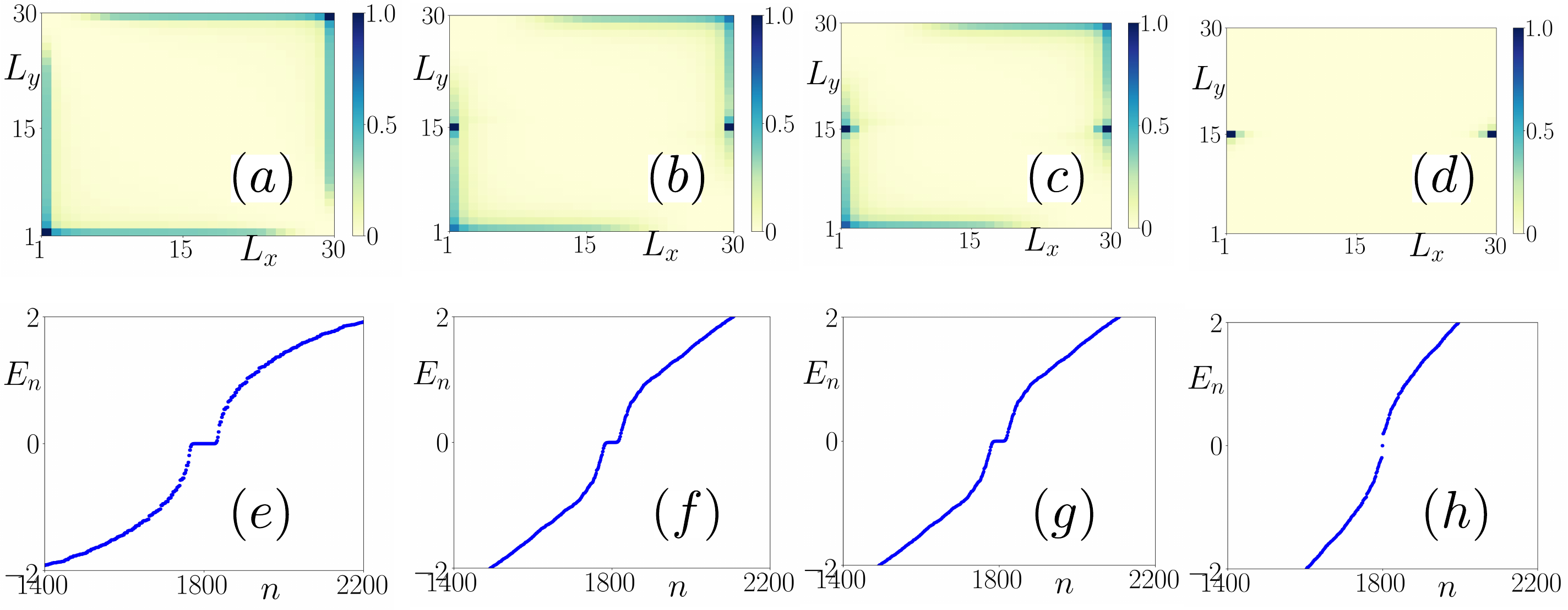}
\end{center} \vspace{-0.5cm}
\caption{\hspace*{-0.1cm} Panels (a), (b), (c) and (d) correspond to LDOS $({\rm {at}}~E=0)$ in $L_x-L_y$ plane for four different combinations of the parameters (i) $\mu=2.0t, B_0=0.0$ 
	(ii) $\mu=3.0t, B_0=3.0t$ (iii) $\mu=3.0t, B_0=4.0t$ and (iv) $\mu=4.0t, B_0=5.0t$ respectively. Panels (e), (f), (g) and (h) correspond to the eigenvalue spectrum for the same parameter regime  
	respectively. The other remaining parameters take the value: $G_s=0.5$ and $\Delta_p=t$. 
	}
\label{2D_1}
\end{figure*}

In this section, we consider a 1D magnetic spin chain (locally varying) on top of a 2D $p$-wave superconductor. We can describe this 2D system using the following Hamiltonian as,
\begin{eqnarray}
H_{\rm{2D}}&=&\sum_{l,\alpha}(t_{l}c_{l,\alpha}^{\dagger}c_{l+1,\alpha}^{}\!+\! h.c.)\!-\! \mu
\sum_{l,\alpha}c_{l,\alpha}^{\dagger}c_{l,\alpha}^{}\!+\! \nonumber\\
&&\sum_{l,\alpha,\beta}\!\!{(\vec{B_{l}}.\vec{\sigma})}_{\alpha,\beta}
c_{l,\alpha}^{\dagger}c_{l,\beta}^{}\!+\!\Delta_{p}\! \sum_{l,\alpha}
(c_{l,\alpha}^{\dagger}c_{l+1,\alpha}^{\dagger}\!+\! h.c.)\ ,~~~~
\label{2DNuModel}\ 
\end{eqnarray}
Here, the index $l$ accounts for the $x$, and $y$ coordinates of the corresponding 2D lattice sites. In particular, we choose $l = (n, m)$ in 2D, and the number of sites along $x (y)$ direction is
$N_x (N_y )$. Other symbols in Eq.~(\ref{2DNuModel}) indicate the same meaning as mentioned after Eq.~(\ref{NuModel}). A pure 2D $p$-wave superconductor (in absence of any magnetic spin chain) exhibits a gapless topological superconducting phase at the bulk of the system hosting MEMs of flat type~\cite{Zhang2019,Wang2017}. Afterwards we incorporate a 1D magnetic spin chain on top of this 2D $p$-wave superconductor. Here, the ends of the chain lie exactly on the edge of the superconducting substrate. Initially, for some specific values of the exchange coupling strength ($B_0=JS$) and chemical potential ($\mu$) of the system (when the chain is in the topological regime), MZMs (localized at the two ends of the magnetic chain) can hybridize with the MEMs. Hence, one cannot 
distinguish between the MEMs and MZMs in this regime. However, MEMs disappear at some specific values of the $\mu$ and $B_0$ even when the chain is in the topological regime. Only MZMs can survive and they are localized at the two ends of the chain. In this case, the Shiba states are energetically separated from the rest of the system. This is the prime reason for the suppression of the hybridization of the MZMs with the MEMs. Hence, this features hint towards an interesting interplay between $\mu$ and $B_0$, which makes the distinction between MEMs and MZMs possible when 
the chain is in the topological regime. However, it is difficult to find an analytical condition between $\mu$ and $B_0$ in the case of a 2D $p$-wave superconductor in presence of the magnetic spin chain. Although, the above mentioned features are evident from our numerical results based on the 2D $p$-wave superconductor. In Fig.~\ref{2D_1}(a), we depict the LDOS at $E=0$ in $L_x-L_y$ plane 
choosing $B_0=0.0$ and $\mu=2.0t$. In absence of any spin chain, the system is a purely 2D $p$-wave superconductor where MEMs appear and it is also reflected in the corresponding eigenvalue spectrum [see Fig.~\ref{2D_1}(e)]. On the other hand, at any intermediate values of $\mu$ and $B_0$, the MZMs hybridize with the MEMs as shown via LDOS [Figs.~\ref{2D_1}(b)-(c)] and the corresponding eigenvalue spectrum [Figs.~\ref{2D_1}(f)-(g)]. Further, MEMs disappear when $\mu=4.0t$ and $B_0=5.0t$, only MZMs present and localized at the two ends of the chain 
[see Fig.~\ref{2D_1}(d)]. In this case, we obtain gapped eigenvalue spectrum and only two localized MZMs appear at exactly $E=0$ [see Fig.~\ref{2D_1}(h)]. In the above cases, we fix the spiral wave 
vector at $G_s=0.5$ \ie the chain is in the topological regime.

\begin{figure*}[]
	\begin{center}
		\includegraphics[width=1.0\textwidth]{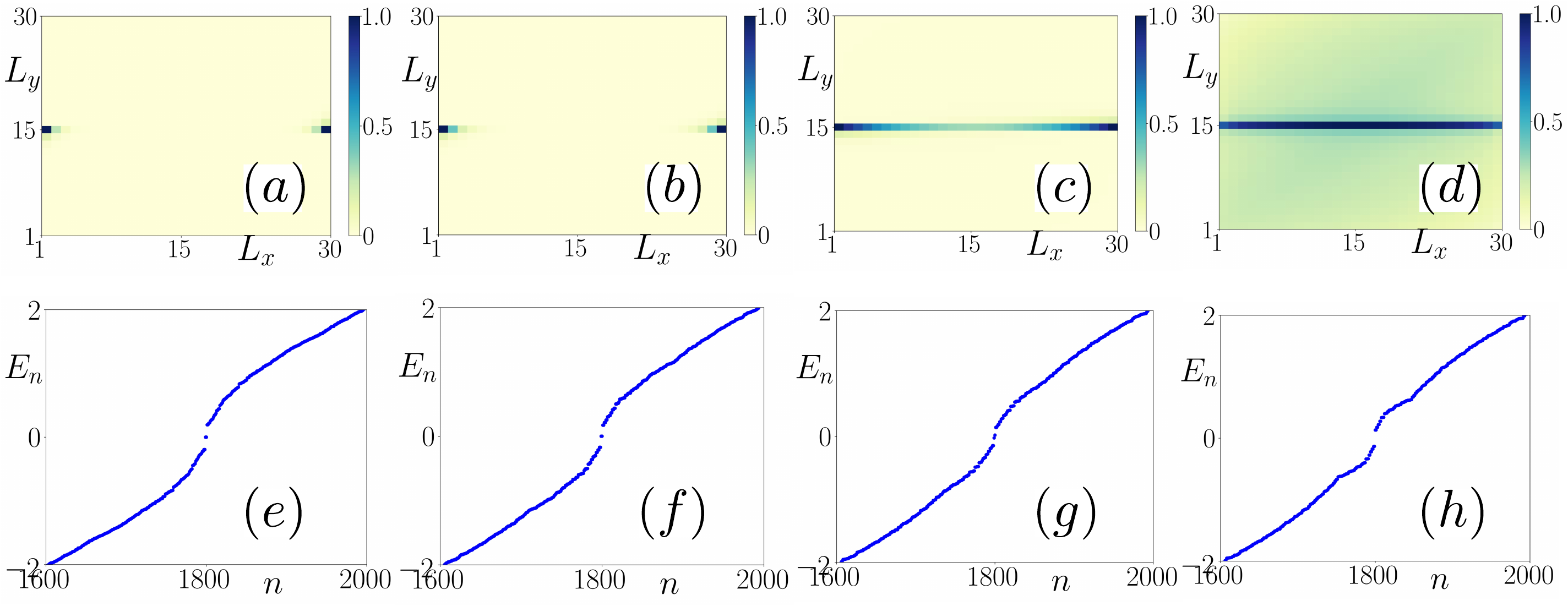}
	\end{center} \vspace{-0.5cm}
	\caption{\hspace*{-0.1cm} Panels (a), (b), (c) and (d) correspond to LDOS $(E=0)$ in $L_x-L_y$ plane for four different values of the spiral wave vactor $G_s=0.5, 1.57, 2.4, 3.14$ respectively. 
	Panels (e), (f), (g) and (h) correspond to the eigenvalue spectrum for the same values of $G_s$ respectively.  We choose the remaining parameter values as: $\mu=4.0t, B_0=5.0t, \Delta_p=t$.
	}\label{2D_2}
\end{figure*}

As we start increasing $G_s$, the MZMs are gradually destroyed by the modulation of the spiral wave vector.  
It is clear from Fig.~\ref{2D_2} that $G_s$ drives the system from topological superconductor to trivial superconductor ($s$-wave like) within the magnetic spin chain (Shiba band). Note that, the magnitude of $E=0$ LDOS (unnormalized) is vanishingly small in Fig.~\ref{2D_2}(d) due to the absence of any zero-energy state in the trivial phase as shown in Fig.~\ref{2D_2}(h). Therefore, when the chain is in the trivial regime, neither MEMs nor MZMs appear in the system.

	
	\section{Summary and conclusions} \label{Summary and Discussion}
	To summarize, in this article, we consider a 1D chain of magnetic impurities placed on a $p$-wave unconventional superconducting substrate and discuss the effect of different 
	spin textures on the topological superconducting phase.
	First, we numerically analyse various 1D SS configurations by varying both $G_{s}$ and $G_{h}$ (controlling parameters for changing $\theta_l$ and $\phi_l$) as the general case.
	We find that the effect of $p$-wave pairing gradually becomes weaker while an effective $s$-wave pairing is simultaneously generated, giving rise to a trivial superconducting phase at 
	$G_{h}, G_{s} = (\pi, \pi)$. 
	This is also reflected in the zero energy LDOS behavior when it is calculated at the ends of the finite chain.
	If we continuously increase the pitch vectors of the SS by changing ($G_{h}, G_{s}$) till $(\pi, \pi)$, then the Majorana peak height in the LDOS slowly decreases and eventually vanishes 
	at ($\pi, \pi$). Hence, we find that the reason behind the complete destruction of MZMs is the formation of an effective $s$-wave pairing gap. Moreover, we analyse the 
	gap structure $\lvert E_2-E_1\rvert/\Delta_{0}$ in $G_s$-$G_h$ plane to find the gap closing points in the parameter phase. Interestingly, considering the N\'eel-type SS configuration 
	($G_{h}=0, G_{s} \neq 0$), we compute the topological winding number $W_{n}$ and show that sharp topological phase transition takes place
	around $G_{s1}= 2.5$ and $G_{s2}=3.7$ where the band gap $\lvert E_2-E_1\rvert$ closes. 
	In order to have a better understanding of this phenomena, we analyse this N\'eel-type rotation ($G_{h}=0, G_{s} \neq 0$) further in both numerical and analytical model studies.
	Interestingly, we find that the rotation of magnetic impurity spins in the chain and $p$-wave superconductivity generates an effective $s$-wave superconducting phase at $G_{s}=\pi$ \ie 
	when the impurity spins form an antiferromagnetic spin-chain.
	Thus, one can realize a phase transition from topological to trivial superconductor in which the signature of MZMs vanishes in LDOS. 
	In our analytical model, we begin from a low-energy model of a chain of magnetic impurities in the presence of $p$-wave superconducting pairing and obtain an analytical expression of the 
	effective pairing $\tilde{\Delta}_{ij}^{\rm eff}$. 
	We also calculate the total DOS from our analytically obtained energy spectrum $\epsilon_{k}$ at $G_{s}=\pi$.
	This exhibits a conventional $s$-wave like behavior which qualitatively matches with our numerical results.
	Moreover, we show that the non-topological Shiba bands (within $-\Delta_{p}$ to $\Delta_{p}$) oscillate with the spiral wave vector $G_{s}$, when the system reaches to the trivial 
	$s$-wave superconducting phase. Such kind of oscillation can be obtained when magnetic impurities are placed on a $s$-wave superconductor. 
	We also consider a 2D $p$-wave superconductor and show via the LDOS and eigenvalue spectrum that initially MZMs can hybridize with the MEMs. However, even when the 
	magnetic chain is in the topological regime, MEMs disappear at some critical value of the chemical potential $\mu$ and exchange coupling strength $B_0=JS$ and only MZMs survive. 
	These MZMs also disappear when the chain is in the trivial regime with further modulation of the SS wave vector $G_{s}$.

	Experimental realization of topological Shiba bands in atomic spin-chains has been reported in very recent experiments~ 
	\cite{science,PhysRevLett.126.076802,Kaesster2021,doi:10.1073/pnas.2210589119,Schneider2021,Beck2021,Schneider2022}.
	In those experiments, different transition metals like Fe~\cite{Schneider2021}, Mn~\cite{Schneider2022} etc. have been designed as a chain of magnetic impurity 
	atoms on the surface of conventional $s$-wave superconducting substrates like Pb (110), Nb (110)~\cite{science,Schneider2021} etc. 
	In such systems, the value of the $s$-wave pairing gap (\eg Nb) is approximately $\Delta_{s}\sim$~1.52~$\rm {meV}$~\cite{Schneider2021} and experiments have been performed at low 
	temperatures ($1.4-1.6$ K)~\cite{science}. 
	They have observed the Majorana ZBP in topological Shiba bands via $(\rm {dI/dV})$ measurement with different positions of the scanning tunnelling microscope tip~\cite{science,Schneider2021}. 
	Therefore, given the so far experimental progress in this research field, we believe that our theoretical model 
	proposal is timely and may be possible to realize in future experiments based on $p$-superconductors engineered in materials \eg heterostructures by growing a magnetic 
	layer on a Rashba superconductor~\cite{Menard2017}, doped topological insulator~\cite{PhysRevLett.104.057001,PhysRevLett.105.097001,PhysRevB.83.224516} etc. Furthermore, the smooth 
	variation of $G$ can be attributed to the high tunability of RKKY exchange tensor and magnetocrystalline anisotropy energy by various routes~\cite{Charanpreet,
	https://doi.org/10.1002/adma.202002043,PhysRevLett.116.017201,Yang2018,Khajetoorians2016}.

	\acknowledgments{} A.K.N and A.S acknowledge the support from the Department of Atomic Energy, Government of India. PC acknowledges Sudarshan Saha, Arnob Kumar Ghosh, 
	Debashish Mandal for useful discussions.

	\appendix
	\section{Oscillation of the Shiba bands in case of $s$-wave superconductor}\label{bands_s-wave}
\begin{figure}[H] \begin{center}
			\includegraphics[width=0.5\textwidth]{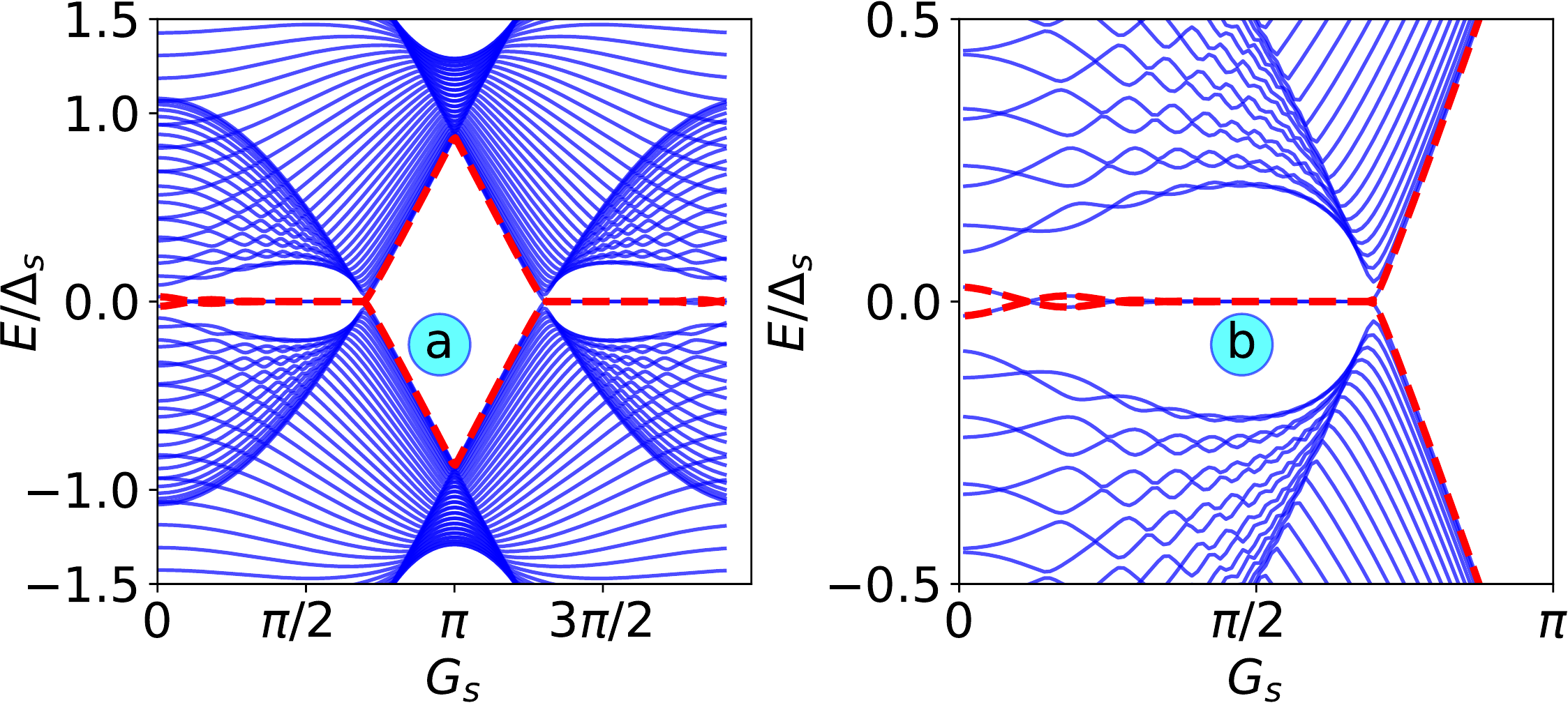} \end{center}
		\caption{\hspace*{-0.1cm} (a) Band spectrum ($E$) is shown as a function of spiral wave vector
			($G_{s}$) within the energy range -1.5$\Delta_{s}$ to 1.5$\Delta_{s}$ considering purely $s$-wave pairing.
			Here, we choose $\mu=4t$, $B_{0}=5t$, $\Delta_{s}=1.0t$.
			(b) Close view of the energy spectrum (within the range $-0.5\Delta_{s} \to 0.5 \Delta_{s}$) 
			is depicted when $G_{s}$ varies from $0$ $\to$ $\pi$.
		}
		\label{bands_s}
	\end{figure}

	Here, we discuss the nature of the Shiba bands when magnetic impurities are placed 
	on a conventional $s$-wave superconductor. It is evident from Fig.~\ref{bands_s}(a) and Fig.~\ref{bands_s}(b) 
	that Shiba bands are formed within $-\Delta_{s}$ to $\Delta_{s}$ and they are non-topological in nature when $G_{s}=0$. 
	We obtain the oscillation of the Shiba bands with respect to $G_{s}$ when the Shiba states
	corresponding to the individual impurity spin interferes with each other. 
	It is an important observation for the formation of $s$-wave superconductor 
	as mentioned in Fig.~\ref{bands}(a) of the main text. 
	It is evident from Fig.~\ref{bands_s}(a) that such oscillation of Shiba band persists until $G_{s}\le\pi/2$ 
	\ie within the trivial regime. When $G_{s}\ge\pi/2$, the system enters into the topological superconducting 
	regime due to the formation of an effective $p$-wave gap and such oscillation of the Shiba bands disappears
	(see Fig.~\ref{bands_s}(b) for more clarity).
	
	\section{Analytical calculation of the effective pairing gap based on low energy effective model}\label{eff_pairing} 
	In order to support our numerical results based on the lattice model (Eq.~(\ref{NuModel})),
	we also carry out analytical calculation considering low energy BdG Hamiltonian that describes
	magnetic impurities on a $p$-wave superconductor. 
	Our continuum Hamiltonian can be written as,
	\begin{eqnarray}
	H=\xi_{k}\tau_{z}-J\sum_{j}\vec{S_{j}}.\vec{\sigma}\delta(x-x_{j})
	+\Delta_{p}k\tau_{y}\ ,
	\label{lowHamiltonian}
	\end{eqnarray}
	where, $\xi_{k}=\frac{k^{2}}{2m}-\mu$, $k$ is the momentum along the 1D chain and $\Delta_{p}$ is the superconducting gap. 
	Particle-hole and spin degrees of freedom can be expressed in terms of Pauli matrices 
	$\tau$ and $\sigma$ respectively. The BdG Hamiltonian can be expressed in terms of Nambu basis
	$\Phi=(\phi_{\uparrow},\phi_{\downarrow},\phi_{\downarrow}^{\dagger},
	-\phi_{\uparrow}^{\dagger})$.
	
	At first, we calculate the eigenvalues and eigenvectors for a single
	magnetic impurity placed on a $p$-wave superconductor. The motivation for this part is to understand 
	the spinors that are needed for our final calculation of the multiple impurities. For arbitrary
	orientation of the single impurity spin, our Hamiltonian reads,
	\begin{eqnarray}
	H=\xi_{k}\tau_{z}-J\vec{S}.\vec{\sigma}+\Delta_{p}k\tau_{y}\ ,
	\label{lowHamiltoniansingle}
	\end{eqnarray}
	The eigenvalues of Eq.~(\ref{lowHamiltoniansingle}) are
	\begin{align} \lambda_{1}
	&=-(JS+\sqrt{\xi_{k}^{2}+\Delta_{p}^{2}k^{2}})\ ,\\
	\lambda_{2}&=
	JS-\sqrt{\xi_{k}^{2}+\Delta_{p}^{2}k^{2}}\ ,\\
	\lambda_{3} &=-(JS-\sqrt{\xi_{k}^{2}+\Delta_{p}^{2}k^{2}})\ ,\\ \lambda_{4}&=
	(JS+\sqrt{\xi_{k}^{2}+\Delta_{p}^{2}k^{2}})\ ,
	\end{align}
	The corresponding BdG spinors of the eigenvectors can be written as,
	\begin{align}
	&&\alpha_{1}=
	\begin{pmatrix}
	\frac{i(-\xi_{k}+\sqrt{\xi_{k}^{2}+\Delta_{p}^{2}k^{2}})}{k\Delta_{p}}
	\cos(\frac{\theta}{2})
	\\
	-\frac{i(\xi_{k}-\sqrt{\xi_{k}^{2}+\Delta_{p}^{2}k^{2}})}{k\Delta_{p}}e^{i\phi}
	\sin(\frac{\theta}{2})
	\\ \cos\frac{\theta}{2}\\
	e^{i\phi}\sin(\frac{\theta}{2})
	\end{pmatrix} \ ,
	\end{align}			
	\begin{align}
	&&\alpha_{2}=
	\begin{pmatrix}
	\frac{i(\xi_{k}-\sqrt{\xi_{k}^{2}+\Delta_{p}^{2}k^{2}})}{k\Delta_{p}}e^{-i\phi}
	\sin(\frac{\theta}{2})
	\\
	-\frac{i(\xi_{k}-\sqrt{\xi_{k}^{2}+\Delta_{p}^{2}k^{2}})}{k\Delta_{p}}
	\cos(\frac{\theta}{2})
	\\ -e^{-i\phi}\sin\frac{\theta}{2}\\
	\cos(\frac{\theta}{2})
	\end{pmatrix}\ ,
	\end{align}	
	\begin{align}
	&&\alpha_{3}=
	\begin{pmatrix}
	-\frac{i(\xi_{k}+\sqrt{\xi_{k}^{2}+\Delta_{p}^{2}k^{2}})}{k\Delta_{p}}
	\cos(\frac{\theta}{2})
	\\
	-\frac{i(\xi_{k}+\sqrt{\xi_{k}^{2}+\Delta_{p}^{2}k^{2}})}{k\Delta_{p}}e^{i\phi}
	\sin(\frac{\theta}{2})
	\\ \cos\frac{\theta}{2}\\
	e^{i\phi}\sin(\frac{\theta}{2}) 
	\end{pmatrix}\ ,
	\end{align}	
	\begin{align}
	&&\alpha_{4}=
	\begin{pmatrix}
	\frac{i(\xi_{k}+\sqrt{\xi_{k}^{2}+\Delta_{p}^{2}k^{2}})}{k\Delta_{p}}e^{-i\phi}
	\sin(\frac{\theta}{2})
	\\
	-\frac{i(\xi_{k}+\sqrt{\xi_{k}^{2}+\Delta_{p}^{2}k^{2}})}{k\Delta_{p}}
	\cos(\frac{\theta}{2})
	\\ -e^{-i\phi}\sin\frac{\theta}{2}\\
	\cos(\frac{\theta}{2})
	\end{pmatrix}\ .
	\end{align}
	
	Under the approximation $\xi_{k}\ll \Delta_{p}$, these spinors take the form, 
	\begin{align} \begin{matrix} \psi_{1}= \begin{pmatrix}
	i\cos(\frac{\theta}{2})          \\
	ie^{i\phi}\sin(\frac{\theta}{2}) \\
	\cos\frac{\theta}{2}             \\
	e^{i\phi}\sin(\frac{\theta}{2})\end{pmatrix}\ , &&
	\psi_{3}=\begin{pmatrix} -ie^{-i\phi}\sin(\frac{\theta}{2}) \\
	i\cos(\frac{\theta}{2})            \\
	-e^{-i\phi}\sin\frac{\theta}{2}    \\
	\cos(\frac{\theta}{2})\end{pmatrix}\end{matrix}\ ,
	\end{align}
	\begin{align} \begin{matrix} \psi_{2}= \begin{pmatrix} -i\cos(\frac{\theta}{2})
	\\
	-ie^{i\phi}\sin(\frac{\theta}{2}) \\
	\cos\frac{\theta}{2}              \\
	e^{i\phi}\sin(\frac{\theta}{2})\end{pmatrix}\ , &&   
	\psi_{4}=\begin{pmatrix}
	ie^{-i\phi}\sin(\frac{\theta}{2}) \\
	-i\cos(\frac{\theta}{2})          \\
	-e^{-i\phi}\sin\frac{\theta}{2}   \\
	\cos(\frac{\theta}{2})\end{pmatrix}
	\end{matrix}\ . \end{align}
	
	The eigenvectors of $\boldsymbol{\sigma}.\boldsymbol{S}$ are, 
	\begin{align}
	\begin{matrix} \ket{\uparrow} & = \begin{pmatrix}
	\cos(\frac{\theta}{2})          \\
	e^{i\phi}\sin(\frac{\theta}{2})\end{pmatrix}\ , &&&
	\ket{\downarrow} & =
	\begin{pmatrix} e^{-i\phi}\sin(\frac{\theta}{2}) \\
	-\cos(\frac{\theta}{2})
	\end{pmatrix}\end{matrix} \ ,
	\end{align}
	Hence, we can write the BdG spinors in terms of the eigenstates of $\boldsymbol{\sigma}.\boldsymbol{S}$ as
	\begin{align} &&\begin{matrix} \psi_{1} =\begin{pmatrix} i\ket{\uparrow} \\
	\ket{\uparrow}\end{pmatrix}\ , &&&&& \psi_{3} =\begin{pmatrix}
	-i\ket{\downarrow} \\
	-\ket{\downarrow}\end{pmatrix}\ , \end{matrix} 
	\end{align}
	\begin{align}
	&&\begin{matrix} \psi_{2} =\begin{pmatrix} -i\ket{\uparrow} \\ \ket{\uparrow}
	\end{pmatrix}\ , &&& \psi_{4} =\begin{pmatrix} i\ket{\downarrow} \\
	-\ket{\downarrow}\end{pmatrix} \end{matrix} \ .
	\end{align}
	
	As for $p$-wave superconductor the constituents quasi-particles exhibit symmetric spin \ie
	triplet pairing, therefore one has to choose the spatial wave-functions as, 
	\begin{align}
	\begin{matrix} \psi_{1}(r_{j})= \begin{pmatrix} i\ket{\uparrow,j} \\
	\ket{\uparrow,j}\end{pmatrix}\ , &   &   &
	\psi_{2}(r_{j})=
	\begin{pmatrix} -i\ket{\uparrow,j} \\ \ket{\uparrow,j}
	\end{pmatrix}\end{matrix}\label{wavefun}\ .	
	\end{align} 
	
	Hence, for multiple impurities we start from the Schr\"odinger equation $H\psi(x)=E\psi(x)$. 
	Now, following Ref.~\cite{Felix} we can write from Eq.~(\ref{lowHamiltonian}),
	\begin{align}
	[E-\xi_{k}\tau_{z}-\Delta_{p}k\tau_{y}]\psi(x)=-JS\sum_{j}\hat{S_{j}}\cdot
	\vec{\sigma}\delta(x-x_{j})\psi(x)\ ,
	\end{align}
	Thus, one can further obtain
	\begin{eqnarray} 
	\psi(x_{i})
	&=&-JS\sum_{j}\int
	\frac{dk}{2\pi}\frac{e^{ik(x_{i}-x_{j})}}{[E-\xi_{k}\tau_{z}-\Delta_{p}k\tau_{y
		}]}\hat{S_{j}}.\vec{\sigma}\psi(x_{j}) \nonumber \\
	&=&-JS\sum_{j}\int\frac{dk}{2\pi} 
	\begin{pmatrix}
	\frac{E+\xi_{k}}{E^{2}-\xi_{k}^{2}-\Delta_{k}^{2}k^{2}}
	&
	\frac{-i\Delta_{k}k}{E^{2}-\xi_{k}^{2}-\Delta_{k}^{2}k^{2}} \\
	\frac{i\Delta_{k}k}{E^{2}-\xi_{k}^{2}-\Delta_{k}^{2}k^{2}}
	&
	\frac{E-\xi_{k}}{E^{2}-\xi_{k}^{2}-\Delta_{k}^{2}k^{2}}
	\end{pmatrix} \nonumber \\
	&\times& e^{ik(x_{i}-x_{j})}\hat{S_{j}}.\vec{\sigma}\psi(x_{j}) \ ,
	\label{B18}
	\end{eqnarray}
	From Eq.~(\ref{B18}), we can further find
	\begin{eqnarray} 
	&&    \psi(x_{i})
	=-\sum_{j}J_{E}(\lvert
	x_{i}-x_{j}\rvert)\hat{S_{j}}.\vec{\sigma}\psi(x_{j})\ ,
	\label{psieq.}
	\end{eqnarray}
	where,
	\begin{eqnarray} J_{E}(\lvert
	x_{i}-x_{j}\rvert)&=&J_{E}(x_{ij})=f_{E}^{(1)}(x_{i,j})\mathbb{I}+f_{E}^{(2)}(x_{
		i,j})\tau_{z}\nonumber\\
	&+& f_{E}^{(3)}(x_{i,j})\tau_{y}\ ,
	\end{eqnarray}
	and 
	\begin{eqnarray} 
	&&		f_{E}^{(1)}(x_{ij})=JS\int\frac{dk}{2\pi}\frac{Ee^{ikx_{ij}}}{E^{2}-\xi_{k}^{2}
		-\Delta_{k}^{2}k^{2}}\ ,
	\nonumber \\
	&&		f_{E}^{(2)}(x_{ij})=JS\int\frac{dk}{2\pi}\frac{\xi_{k}e^{ikx_{ij}}}{E^{2}-\xi_{
			k}^{2}-\Delta_{k}^{2}k^{2}}\ ,
	\nonumber \\
	&&		f_{E}^{(3)}(x_{ij})=JS\int\frac{dk}{2\pi}\frac{\Delta_{k}ke^{ikx_{ij}}}{E^{2}-\
		xi_{k}^{2}-\Delta_{k}^{2}k^{2}}\ .
	\end{eqnarray}
	
	Afterwards, we linearize the spectrum around the Fermi momentum and thus obtain $\xi_{k}=V_{F}(k-k_{F})$. 
	Here $V_{F}$ and $k_{F}$ are the Fermi velocity and Fermi momentum respectively. Under this approximation, 
	the results of the integrals are given by (see Ref.~\cite{Kaladzhyan_2016})
	\begin{eqnarray}
	f_{E}^{(1)}(x)
	&=&-\frac{E}{V_{F}}\frac{JS}{1+\tilde{\Delta}_{p}^{2}}\frac{\cos(k_{F}^{\prime}x
		)}{\omega(E)}e^{-\frac{\omega(E)x}{V_{F}}}\ , \nonumber \\
	f_{E}^{(2)}(x)
	&=&\frac{1}{V_{F}}\frac{JS}{1+\tilde{\Delta}_{p}^{2}}\left[\frac{\gamma\Delta_{t
	}}{\omega(E)}\cos(k_{F}^{\prime}x)+\sin(k_{F}^{\prime}x)\right] e^{-\frac{\omega(E)x}{V_{F}}}\ , \nonumber \\
	f_{E}^{(3)}(x)
	&=&-\frac{i}{V_{F}}\frac{\tilde{\Delta}_{p}JS}{1+\tilde{\Delta}_{p}^{2}}\left[
	\frac{k_{F}^{\prime}V_{F}}{\omega(E)}\sin(k_{F}^{\prime}x)
	+{\rm{sgn}}(x)\cos(k_{F}^{\prime}x)\right]\nonumber \\
	&\times& e^{-\frac{\omega(E)x}{V_{F}}}. 
	\end{eqnarray}
	where, $k_{F}^{\prime}=\frac{k_{F}}{\sqrt{1+\tilde{\Delta}_{p}^{2}}}$, $\tilde{\Delta}_{p}=\frac{\Delta_{p}}{V_{F}}$,
	$\omega(E)=\frac{\sqrt{\Delta_{t}^{2}-E^{2}}}{\sqrt{1+\tilde{\Delta}_{p}^{2}}}$,
	$\Delta_{t}=\frac{\Delta_{p}k_{F}}{\sqrt{1+\tilde{\Delta}_{p}^{2}}}$,
	$\gamma=\frac{\tilde{\Delta}_{p}}{\sqrt{1+\tilde{\Delta}_{p}^{2}}}$ as introduced in the main text also.
	
	From Eq.~(\ref{psieq.}) we can write,
	\begin{equation}
	\psi(x_{i})+J_{E}(0)\hat{S_{i}}.\vec{\sigma}\psi(x_{i})=-\sum_{j\ne
		i}J_{E}(\lvert
	x_{i}-x_{j}\rvert)\hat{S_{j}}.\vec{\sigma}\psi\left(x_{j}\right)\ .
	\label{centraleq}
	\end{equation}
	Therefore, at $\lvert x_{i}-x_{j}\rvert=0$, these functions become,
	\begin{eqnarray}
	f_{E}^{(1)}(0) &=&
	-\frac{JS}{V_{F}}\frac{E}{\sqrt{1+\tilde{\Delta}_{p}^{2}}}\frac{1}{\sqrt{\Delta
			_{t}^{2}-E^{2}}}\ , \nonumber \\
	f_{E}^{(2)}(0) &=&
	\frac{JS}{V_{F}}\frac{\gamma\Delta_{t}}{\sqrt{1+\tilde{\Delta}_{p}^{2}}}\frac{1
	}{\sqrt{\Delta_{t}^{2}-E^{2}}}\ , \nonumber \\
	f_{E}^{(3)}(x) &=& 0 \ .
	\label{B24}
	\end{eqnarray}
	We can define the impurity strength in terms of a new parameter $\alpha=JS/(V_{F}\sqrt{1+\tilde{\Delta}_{p}^{2}})$. 
	We further assume that impurities are sufficiently dilute such that the resulting impurity band remains well 
	within the superconducting gap \ie $\Delta_{t}\gg E$. Using Eq.~(\ref{B24}) and substituting $E=0$, $\alpha=1/\gamma$ 
	in the RHS of the Eq.~(\ref{centraleq}), we obtain~(following Ref.~\cite{Kaladzhyan_2016})
	\begin{eqnarray} 
	\left[\hat{S_{i}}.\vec{\sigma}-\left(\frac{\alpha
		E}{\Delta_{t}}-\alpha\gamma\tau_{z}\right)\right]\psi(x_{i})
	&=&-\sum_{j\ne
		i}[f_{0}^{(2)}(x_{ij})\tau_{z}\nonumber \\
	+f_{0}^{(3)}(x_{ij})\tau_{y}](\hat{S_{i}}.\vec{\sigma})
	(\hat{S_{j}}.\vec{\sigma })\psi(x_{j})\ ,
	\label{central} 
	\end{eqnarray} 
	Hence, Eq.~(\ref{central}) can be written as
	\begin{eqnarray}
	\tilde{H}_{\rm {eff}}\phi=E\phi
	\end{eqnarray}
	Now projecting Eq.~(\ref{wavefun}) into Eq.~(\ref{central}), we can find the diagonal and off-diagonal elements of $H_{\rm{eff}}$ as
	\begin{eqnarray}
	{\tilde{h}^{\rm{eff}}}_{ij} &=&
	(\epsilon_{0}/2)\delta_{ij}
	+(1-\delta_{ij})\nonumber \\
	&&    \left[k_{F}^{\prime}\Delta_{p}\sin(k_{F}^{\prime}\lvert
	x_{ij}\rvert)  + 	\gamma\Delta_{t}{\rm{sgn}}(x_{ij})\cos(k_{F}^{\prime}x_{ij})\right]
	\nonumber\\
	&& \times e^{-\frac{\lvert x_{ij}\rvert}{\xi_{0}}}\bra{\uparrow,i}\ket{\uparrow,j}\ ,
	\label{h}
	\end{eqnarray}
	and
	\begin{eqnarray}
	(\tilde{\Delta}^{\rm{eff}})_{ij}&=&-(\epsilon_{0}/2)\delta_{ij}-(1-\delta_{ij})
	\nonumber\\
	&&\left[\frac{\Delta_{t}}{\sqrt{1+\tilde{\Delta}_{p}^{2}}}\sin(k_{F}^{\prime}
	\lvert
	x_{ij}\rvert)+\gamma\Delta_{t}\cos(k_{F}^{\prime}x_{ij})\right]\nonumber\\
	&& \times~e^{-\frac{\lvert
			x_{ij}\rvert}{\xi_{0}}}\bra{\uparrow,i}\ket{\uparrow,j}\ ,
	\label{delt}
	\end{eqnarray}
	where, $\epsilon_{0}=2\Delta_{t}/\alpha$ denotes the Shiba bound state energy for a single magnetic impurity 
	in the deep Shiba limit and $\xi_{0}=V_{F}/\omega(0)$ is the phase coherence length for a $p$-wave superconductor.
	Here, $\tilde{\Delta}^{\rm{eff}}_{ij}$ denotes the effective pairing gap in real space which we have mentioned in the main text 
	(see Eq.~(\ref{effdel})).
	
	Therefore, considering N\'eel-type rotation of the magnetic impurities \ie $\phi_{i}=0$ and $\theta_{i}=G_{s}ja$ 
	($j$ is the lattice site index and $a$ is the lattice spacing), one can obtain
	\[\bra{\uparrow,i}\ket{\uparrow,j}=\cos(G_{s}x_{ij}/2)\]
	and \[\bra{\uparrow,i}\ket{\downarrow,j}=-\sin(G_{s}x_{ij}/2) \ .\]
	Hence, the effective $k$-space $2\times2$ BdG Hamiltonian reads,
	\begin{eqnarray} H_{k}= \begin{pmatrix} h_{k} &
	\Delta_{k}\\ \Delta_{k}^{*} & -h_{-k}^{*}\end{pmatrix}\ ,
	\end{eqnarray} 
	where, \begin{eqnarray}
	h_{k}=\sum_{j}{(\tilde{h}^{{\rm{eff}}})}_{ij}e^{ikx_{ij}}\ ,
	\end{eqnarray} 
	\begin{eqnarray}
	\Delta_{k}=\sum_{j}{(\tilde{\Delta}^{{\rm{eff}}})}_{ij}e^{ikx_{ij}}\ ,
	\end{eqnarray} 
	Inserting Eqs.~(\ref{h}-\ref{delt}) into the above equations, we finally obtain Eqs.~(\ref{Feffonsite}-\ref{FeffDelta})
	which we have introduced in the main text to explain our analytical results.
	
 \section{Analytical solution of topological phase boundary for our 1D lattice model Hamiltonian}\label{gap_structure}
              As the tight-binding model [Eq.~(\ref{NuModel}) in our main text] doesn't respect translational symmetry, hence we cannot directly perform 
	      Fourier transform to obtain a spectrum in $k$-space. Thus following Ref.~\cite{Ali}, we transform our Hamiltonian to the 
	      Majorana basis in order to diagonalize it in momentum space. Under the new basis $m_k=(m_{k,1,\uparrow},m_{k,1,\downarrow},m_{k,2,\uparrow},m_{k,2,\downarrow})$, our $k$-space 
	      Hamiltonian can be written for the N\'eel-type rotation as

	      \begin{eqnarray}
		      H(k)=\frac{i}{4}\sum_{k}m_k^{\dagger}h(k)m_k\ ,
		      \label{hamiltonian}
	      \end{eqnarray}
	      where, non-zero elements of Eq.~(\ref{hamiltonian}) are,
	      \begin{eqnarray}
		      &&h_{13}=(2t+\Delta_p)\alpha\cos(k)+B_0-\mu=h_{31}^*\ , \\
		      &&h_{14}=2it\beta\sin(k)-\Delta_p\beta\cos(k)=h_{41}^*\ , \\
		      &&h_{23}=-2it\beta\sin(k)+\Delta_p\beta\cos(k)=h_{32}^*\ , \\
		      &&h_{24}=(2t+\Delta_p)\alpha\cos(k)-B_0-\mu=h_{42}^*\ .
		      \label{elements}
	      \end{eqnarray}
	      For $k=\pi$, the eigenvalues of the above Hamiltonian [Eq.~(\ref{hamiltonian})] can be written as,
	      \begin{eqnarray}
		      &&E=\pm\frac{1}{4}\left[B_0 \pm \sqrt{(\mu+2\alpha t)^{2}+\Delta_p^{2}(\alpha^2+\beta^2)+2\alpha\Delta_p(\mu+2\alpha t]}\right]\ ,
		      \label{eigenvalue1}
	      \end{eqnarray}
	      where, $\alpha=\cos(G_s/2)$ and $\beta=\sin(G_s/2)$. In presence of a $p$-wave superconductor, the system is initially topological and after a critical value of the exchange field $B_{0}=JS$, 
	      it goes to a non-topological phase. We obtain this critical exchange field from the gap-closing condition of the two minimum bands. From Eq.~(\ref{eigenvalue1}) we obtain,
	      \begin{eqnarray}
		      \lvert B_{c1}\rvert=\sqrt{(\mu+2\alpha t)^{2}+\Delta_p^{2}(\alpha^2+\beta^2)+2\alpha\Delta_p(\mu+2\alpha t)}\ ,
		      \label{critfield1}
	      \end{eqnarray}
	      On the other hand, for $k=0$ the eigenvalues of the Hamiltonian become,
	      \begin{eqnarray}
		      E=\pm\frac{1}{4}\left[B_0\pm\sqrt{(\mu-2\alpha t)^{2}+\Delta_p^{2}(\alpha^2+\beta^2)-2\alpha\Delta_p(\mu-2\alpha t)}\right]\ ,
		      \label{eigenvalue2}
	      \end{eqnarray}
	      Similarly, from the gap closing condition (considering the lowest two bands) the critical exchange field becomes,
	      \begin{eqnarray}
		      \lvert B_{c2}\rvert=\sqrt{(\mu-2\alpha t)^{2}+\Delta_p^{2}(\alpha^2+\beta^2)-2\alpha\Delta_p(\mu-2\alpha t)}\ .
		      \label{critfield2}
	      \end{eqnarray}
	      
                Hence, considering the parameters values $\mu=4t$, $B_0=5t$, $\Delta_p=t$ and employing Eq.~(\ref{critfield1}), Eq.~(\ref{critfield2}) we obtain $G_{s1}= 2.52639$ 
                and $G_{s2}=3.7568$. This matches well with the numerically obtained gap closing points as well as sharp jump of the topological invariant (see Fig.~\ref{t_c}(b)).

               Out of four cases, here we present the calculation for N\'eel-like rotation ($\theta_{l}=G_sl$ and $\phi_{l}=0$) only. For the other two cases, Bloch-type ($\theta_{l}=G_sl, \phi_{l}=\pi/2$) 
	      and conical-spin spiral ($\theta_{l}=\pi/2, \phi_{l}=G_hl$), our results remain the same. 

\bibliography{bibfile}{}

\begin{thebibliography}{93}%
\makeatletter
\providecommand \@ifxundefined [1]{%
 \@ifx{#1\undefined}
}%
\providecommand \@ifnum [1]{%
 \ifnum #1\expandafter \@firstoftwo
 \else \expandafter \@secondoftwo
 \fi
}%
\providecommand \@ifx [1]{%
 \ifx #1\expandafter \@firstoftwo
 \else \expandafter \@secondoftwo
 \fi
}%
\providecommand \natexlab [1]{#1}%
\providecommand \enquote  [1]{``#1''}%
\providecommand \bibnamefont  [1]{#1}%
\providecommand \bibfnamefont [1]{#1}%
\providecommand \citenamefont [1]{#1}%
\providecommand \href@noop [0]{\@secondoftwo}%
\providecommand \href [0]{\begingroup \@sanitize@url \@href}%
\providecommand \@href[1]{\@@startlink{#1}\@@href}%
\providecommand \@@href[1]{\endgroup#1\@@endlink}%
\providecommand \@sanitize@url [0]{\catcode `\\12\catcode `\$12\catcode
  `\&12\catcode `\#12\catcode `\^12\catcode `\_12\catcode `\%12\relax}%
\providecommand \@@startlink[1]{}%
\providecommand \@@endlink[0]{}%
\providecommand \url  [0]{\begingroup\@sanitize@url \@url }%
\providecommand \@url [1]{\endgroup\@href {#1}{\urlprefix }}%
\providecommand \urlprefix  [0]{URL }%
\providecommand \Eprint [0]{\href }%
\providecommand \doibase [0]{http://dx.doi.org/}%
\providecommand \selectlanguage [0]{\@gobble}%
\providecommand \bibinfo  [0]{\@secondoftwo}%
\providecommand \bibfield  [0]{\@secondoftwo}%
\providecommand \translation [1]{[#1]}%
\providecommand \BibitemOpen [0]{}%
\providecommand \bibitemStop [0]{}%
\providecommand \bibitemNoStop [0]{.\EOS\space}%
\providecommand \EOS [0]{\spacefactor3000\relax}%
\providecommand \BibitemShut  [1]{\csname bibitem#1\endcsname}%
\let\auto@bib@innerbib\@empty
\bibitem [{\citenamefont {Thouless}\ \emph {et~al.}(1982)\citenamefont
  {Thouless}, \citenamefont {Kohmoto}, \citenamefont {Nightingale},\ and\
  \citenamefont {den Nijs}}]{Thouless}%
  \BibitemOpen
  \bibfield  {author} {\bibinfo {author} {\bibfnamefont {D.~J.}\ \bibnamefont
  {Thouless}}, \bibinfo {author} {\bibfnamefont {M.}~\bibnamefont {Kohmoto}},
  \bibinfo {author} {\bibfnamefont {M.~P.}\ \bibnamefont {Nightingale}}, \ and\
  \bibinfo {author} {\bibfnamefont {M.}~\bibnamefont {den Nijs}},\ }\bibfield
  {title} {\enquote {\bibinfo {title} {Quantized hall conductance in a
  two-dimensional periodic potential},}\ }\href {\doibase
  10.1103/PhysRevLett.49.405} {\bibfield  {journal} {\bibinfo  {journal} {Phys.
  Rev. Lett.}\ }\textbf {\bibinfo {volume} {49}},\ \bibinfo {pages} {405--408}
  (\bibinfo {year} {1982})}\BibitemShut {NoStop}%
\bibitem [{\citenamefont {Kane}\ and\ \citenamefont
  {Mele}(2005{\natexlab{a}})}]{kane2005quantum}%
  \BibitemOpen
  \bibfield  {author} {\bibinfo {author} {\bibfnamefont {Charles~L}\
  \bibnamefont {Kane}}\ and\ \bibinfo {author} {\bibfnamefont {Eugene~J}\
  \bibnamefont {Mele}},\ }\bibfield  {title} {\enquote {\bibinfo {title}
  {Quantum spin hall effect in graphene},}\ }\href {\doibase
  10.1103/PhysRevLett.95.226801} {\bibfield  {journal} {\bibinfo  {journal}
  {Phys. Rev. Lett.}\ }\textbf {\bibinfo {volume} {95}},\ \bibinfo {pages}
  {226801} (\bibinfo {year} {2005}{\natexlab{a}})}\BibitemShut {NoStop}%
\bibitem [{\citenamefont {Kane}\ and\ \citenamefont
  {Mele}(2005{\natexlab{b}})}]{kane2005quantumZ2}%
  \BibitemOpen
  \bibfield  {author} {\bibinfo {author} {\bibfnamefont {Charles~L}\
  \bibnamefont {Kane}}\ and\ \bibinfo {author} {\bibfnamefont {Eugene~J}\
  \bibnamefont {Mele}},\ }\bibfield  {title} {\enquote {\bibinfo {title} {Z2
  topological order and the quantum spin hall effect},}\ }\href {\doibase
  10.1103/PhysRevLett.95.146802} {\bibfield  {journal} {\bibinfo  {journal}
  {Phys. Rev. Lett.}\ }\textbf {\bibinfo {volume} {95}},\ \bibinfo {pages}
  {146802} (\bibinfo {year} {2005}{\natexlab{b}})}\BibitemShut {NoStop}%
\bibitem [{\citenamefont {Bernevig}\ \emph {et~al.}(2006)\citenamefont
  {Bernevig}, \citenamefont {Hughes},\ and\ \citenamefont
  {Zhang}}]{bernevig2006quantum}%
  \BibitemOpen
  \bibfield  {author} {\bibinfo {author} {\bibfnamefont {B~Andrei}\
  \bibnamefont {Bernevig}}, \bibinfo {author} {\bibfnamefont {Taylor~L}\
  \bibnamefont {Hughes}}, \ and\ \bibinfo {author} {\bibfnamefont {Shou-Cheng}\
  \bibnamefont {Zhang}},\ }\bibfield  {title} {\enquote {\bibinfo {title}
  {Quantum spin hall effect and topological phase transition in
  {$\mathrm{HgTe}$} quantum wells},}\ }\href
  {https://doi.org/10.1126/science.1133734} {\bibfield  {journal} {\bibinfo
  {journal} {Science}\ }\textbf {\bibinfo {volume} {314}},\ \bibinfo {pages}
  {1757--1761} (\bibinfo {year} {2006})}\BibitemShut {NoStop}%
\bibitem [{\citenamefont {Qi}\ and\ \citenamefont
  {Zhang}(2011)}]{qi2011topological}%
  \BibitemOpen
  \bibfield  {author} {\bibinfo {author} {\bibfnamefont {Xiao-Liang}\
  \bibnamefont {Qi}}\ and\ \bibinfo {author} {\bibfnamefont {Shou-Cheng}\
  \bibnamefont {Zhang}},\ }\bibfield  {title} {\enquote {\bibinfo {title}
  {Topological insulators and superconductors},}\ }\href
  {https://doi.org/10.1103/RevModPhys.83.1057} {\bibfield  {journal} {\bibinfo
  {journal} {Rev. Mod. Phys.}\ }\textbf {\bibinfo {volume} {83}},\ \bibinfo
  {pages} {1057} (\bibinfo {year} {2011})}\BibitemShut {NoStop}%
\bibitem [{\citenamefont {König}\ \emph {et~al.}(2007)\citenamefont {König},
  \citenamefont {Wiedmann}, \citenamefont {Brüne}, \citenamefont {Roth},
  \citenamefont {Buhmann}, \citenamefont {Molenkamp}, \citenamefont {Qi},\ and\
  \citenamefont {Zhang}}]{doi:10.1126/science.1148047}%
  \BibitemOpen
  \bibfield  {author} {\bibinfo {author} {\bibfnamefont {Markus}\ \bibnamefont
  {König}}, \bibinfo {author} {\bibfnamefont {Steffen}\ \bibnamefont
  {Wiedmann}}, \bibinfo {author} {\bibfnamefont {Christoph}\ \bibnamefont
  {Brüne}}, \bibinfo {author} {\bibfnamefont {Andreas}\ \bibnamefont {Roth}},
  \bibinfo {author} {\bibfnamefont {Hartmut}\ \bibnamefont {Buhmann}}, \bibinfo
  {author} {\bibfnamefont {Laurens~W.}\ \bibnamefont {Molenkamp}}, \bibinfo
  {author} {\bibfnamefont {Xiao-Liang}\ \bibnamefont {Qi}}, \ and\ \bibinfo
  {author} {\bibfnamefont {Shou-Cheng}\ \bibnamefont {Zhang}},\ }\bibfield
  {title} {\enquote {\bibinfo {title} {Quantum spin hall insulator state in
  hgte quantum wells},}\ }\href {\doibase 10.1126/science.1148047} {\bibfield
  {journal} {\bibinfo  {journal} {Science}\ }\textbf {\bibinfo {volume}
  {318}},\ \bibinfo {pages} {766--770} (\bibinfo {year} {2007})}\BibitemShut
  {NoStop}%
\bibitem [{\citenamefont {Kohmoto}(1985)}]{KOHMOTO1985343}%
  \BibitemOpen
  \bibfield  {author} {\bibinfo {author} {\bibfnamefont {Mahito}\ \bibnamefont
  {Kohmoto}},\ }\bibfield  {title} {\enquote {\bibinfo {title} {Topological
  invariant and the quantization of the hall conductance},}\ }\href {\doibase
  10.1016/0003-4916(85)90148-4} {\bibfield  {journal} {\bibinfo  {journal}
  {Annals of Physics}\ }\textbf {\bibinfo {volume} {160}},\ \bibinfo {pages}
  {343–354} (\bibinfo {year} {1985})}\BibitemShut {NoStop}%
\bibitem [{KAN(2013)}]{KANE20133}%
  \BibitemOpen
  \bibfield  {title} {\enquote {\bibinfo {title} {Chapter 1 - topological band
  theory and the {Z2} invariant},}\ }in\ \href {\doibase
  https://doi.org/10.1016/B978-0-444-63314-9.00001-9} {\emph {\bibinfo
  {booktitle} {Topological Insulators}}},\ \bibinfo {series} {Contemporary
  Concepts of Condensed Matter Science}, Vol.~\bibinfo {volume} {6},\ \bibinfo
  {editor} {edited by\ \bibinfo {editor} {\bibfnamefont {Marcel}\ \bibnamefont
  {Franz}}\ and\ \bibinfo {editor} {\bibfnamefont {Laurens}\ \bibnamefont
  {Molenkamp}}}\ (\bibinfo  {publisher} {Elsevier},\ \bibinfo {year} {2013})\
  pp.\ \bibinfo {pages} {3--34}\BibitemShut {NoStop}%
\bibitem [{\citenamefont {Moore}\ and\ \citenamefont
  {Balents}(2007)}]{moore2007topological}%
  \BibitemOpen
  \bibfield  {author} {\bibinfo {author} {\bibfnamefont {Joel~E}\ \bibnamefont
  {Moore}}\ and\ \bibinfo {author} {\bibfnamefont {Leon}\ \bibnamefont
  {Balents}},\ }\bibfield  {title} {\enquote {\bibinfo {title} {Topological
  invariants of time-reversal-invariant band structures},}\ }\href {\doibase
  10.1103/PhysRevB.75.121306} {\bibfield  {journal} {\bibinfo  {journal} {Phys.
  Rev. B}\ }\textbf {\bibinfo {volume} {75}},\ \bibinfo {pages} {121306}
  (\bibinfo {year} {2007})}\BibitemShut {NoStop}%
\bibitem [{\citenamefont {Fu}\ \emph {et~al.}(2007)\citenamefont {Fu},
  \citenamefont {Kane},\ and\ \citenamefont {Mele}}]{fu2007topological}%
  \BibitemOpen
  \bibfield  {author} {\bibinfo {author} {\bibfnamefont {Liang}\ \bibnamefont
  {Fu}}, \bibinfo {author} {\bibfnamefont {Charles~L}\ \bibnamefont {Kane}}, \
  and\ \bibinfo {author} {\bibfnamefont {Eugene~J}\ \bibnamefont {Mele}},\
  }\bibfield  {title} {\enquote {\bibinfo {title} {Topological insulators in
  three dimensions},}\ }\href {\doibase 10.1103/PhysRevLett.98.106803}
  {\bibfield  {journal} {\bibinfo  {journal} {Phys. Rev. Lett.}\ }\textbf
  {\bibinfo {volume} {98}},\ \bibinfo {pages} {106803} (\bibinfo {year}
  {2007})}\BibitemShut {NoStop}%
\bibitem [{\citenamefont {Roy}(2009)}]{roy2009topological}%
  \BibitemOpen
  \bibfield  {author} {\bibinfo {author} {\bibfnamefont {Rahul}\ \bibnamefont
  {Roy}},\ }\bibfield  {title} {\enquote {\bibinfo {title} {Topological phases
  and the quantum spin hall effect in three dimensions},}\ }\href {\doibase
  10.1103/PhysRevB.79.195322} {\bibfield  {journal} {\bibinfo  {journal} {Phys.
  Rev. B}\ }\textbf {\bibinfo {volume} {79}},\ \bibinfo {pages} {195322}
  (\bibinfo {year} {2009})}\BibitemShut {NoStop}%
\bibitem [{\citenamefont {Ando}(2013)}]{ando2013topological}%
  \BibitemOpen
  \bibfield  {author} {\bibinfo {author} {\bibfnamefont {Yoichi}\ \bibnamefont
  {Ando}},\ }\bibfield  {title} {\enquote {\bibinfo {title} {Topological
  insulator materials},}\ }\href {https://doi.org/10.7566/JPSJ.82.102001}
  {\bibfield  {journal} {\bibinfo  {journal} {Journal of the Physical Society
  of Japan}\ }\textbf {\bibinfo {volume} {82}},\ \bibinfo {pages} {102001}
  (\bibinfo {year} {2013})}\BibitemShut {NoStop}%
\bibitem [{\citenamefont {Fu}\ and\ \citenamefont {Kane}(2007)}]{fukaneTI2007}%
  \BibitemOpen
  \bibfield  {author} {\bibinfo {author} {\bibfnamefont {L.}~\bibnamefont
  {Fu}}\ and\ \bibinfo {author} {\bibfnamefont {C.~L.}\ \bibnamefont {Kane}},\
  }\bibfield  {title} {\enquote {\bibinfo {title} {Topological insulators with
  inversion symmetry},}\ }\href {\doibase 10.1103/PhysRevB.76.045302}
  {\bibfield  {journal} {\bibinfo  {journal} {Phys. Rev. B}\ }\textbf {\bibinfo
  {volume} {76}},\ \bibinfo {pages} {045302} (\bibinfo {year}
  {2007})}\BibitemShut {NoStop}%
\bibitem [{\citenamefont {Teo}\ \emph {et~al.}(2008)\citenamefont {Teo},
  \citenamefont {Fu},\ and\ \citenamefont {Kane}}]{teofukaneTI2008}%
  \BibitemOpen
  \bibfield  {author} {\bibinfo {author} {\bibfnamefont {Jeffrey C.~Y.}\
  \bibnamefont {Teo}}, \bibinfo {author} {\bibfnamefont {Liang}\ \bibnamefont
  {Fu}}, \ and\ \bibinfo {author} {\bibfnamefont {C.~L.}\ \bibnamefont
  {Kane}},\ }\bibfield  {title} {\enquote {\bibinfo {title} {Surface states and
  topological invariants in three-dimensional topological insulators:
  Application to {${\mathrm{Bi}}_{1\ensuremath{-}x}{\mathrm{Sb}}_{x}$}},}\
  }\href {\doibase 10.1103/PhysRevB.78.045426} {\bibfield  {journal} {\bibinfo
  {journal} {Phys. Rev. B}\ }\textbf {\bibinfo {volume} {78}},\ \bibinfo
  {pages} {045426} (\bibinfo {year} {2008})}\BibitemShut {NoStop}%
\bibitem [{\citenamefont {Qi}\ \emph {et~al.}(2008)\citenamefont {Qi},
  \citenamefont {Hughes},\ and\ \citenamefont {Zhang}}]{qi2008topological}%
  \BibitemOpen
  \bibfield  {author} {\bibinfo {author} {\bibfnamefont {Xiao-Liang}\
  \bibnamefont {Qi}}, \bibinfo {author} {\bibfnamefont {Taylor~L}\ \bibnamefont
  {Hughes}}, \ and\ \bibinfo {author} {\bibfnamefont {Shou-Cheng}\ \bibnamefont
  {Zhang}},\ }\bibfield  {title} {\enquote {\bibinfo {title} {Topological field
  theory of time-reversal invariant insulators},}\ }\href {\doibase
  10.1103/PhysRevB.78.195424} {\bibfield  {journal} {\bibinfo  {journal} {Phys.
  Rev. B}\ }\textbf {\bibinfo {volume} {78}},\ \bibinfo {pages} {195424}
  (\bibinfo {year} {2008})}\BibitemShut {NoStop}%
\bibitem [{\citenamefont {Hosen}\ \emph {et~al.}(2018)\citenamefont {Hosen},
  \citenamefont {Dimitri}, \citenamefont {Nandy}, \citenamefont {Aperis},
  \citenamefont {Sankar}, \citenamefont {Dhakal}, \citenamefont {Maldonado},
  \citenamefont {Kabir}, \citenamefont {Sims}, \citenamefont {Chou},
  \citenamefont {Kaczorowski}, \citenamefont {Durakiewicz}, \citenamefont
  {Oppeneer},\ and\ \citenamefont {Neupane}}]{Hosen2018}%
  \BibitemOpen
  \bibfield  {author} {\bibinfo {author} {\bibfnamefont {M.~Mofazzel}\
  \bibnamefont {Hosen}}, \bibinfo {author} {\bibfnamefont {Klauss}\
  \bibnamefont {Dimitri}}, \bibinfo {author} {\bibfnamefont {Ashis~K.}\
  \bibnamefont {Nandy}}, \bibinfo {author} {\bibfnamefont {Alex}\ \bibnamefont
  {Aperis}}, \bibinfo {author} {\bibfnamefont {Raman}\ \bibnamefont {Sankar}},
  \bibinfo {author} {\bibfnamefont {Gyanendra}\ \bibnamefont {Dhakal}},
  \bibinfo {author} {\bibfnamefont {Pablo}\ \bibnamefont {Maldonado}}, \bibinfo
  {author} {\bibfnamefont {Firoza}\ \bibnamefont {Kabir}}, \bibinfo {author}
  {\bibfnamefont {Christopher}\ \bibnamefont {Sims}}, \bibinfo {author}
  {\bibfnamefont {Fangcheng}\ \bibnamefont {Chou}}, \bibinfo {author}
  {\bibfnamefont {Dariusz}\ \bibnamefont {Kaczorowski}}, \bibinfo {author}
  {\bibfnamefont {Tomasz}\ \bibnamefont {Durakiewicz}}, \bibinfo {author}
  {\bibfnamefont {Peter~M.}\ \bibnamefont {Oppeneer}}, \ and\ \bibinfo {author}
  {\bibfnamefont {Madhab}\ \bibnamefont {Neupane}},\ }\bibfield  {title}
  {\enquote {\bibinfo {title} {Distinct multiple fermionic states in a single
  topological metal},}\ }\href {\doibase 10.1038/s41467-018-05233-1} {\bibfield
   {journal} {\bibinfo  {journal} {Nature Communications}\ }\textbf {\bibinfo
  {volume} {9}},\ \bibinfo {pages} {3002} (\bibinfo {year} {2018})}\BibitemShut
  {NoStop}%
\bibitem [{\citenamefont {Dhakal}\ \emph {et~al.}(2022)\citenamefont {Dhakal},
  \citenamefont {Kabir}, \citenamefont {Nandy}, \citenamefont {Aperis},
  \citenamefont {Sakhya}, \citenamefont {Pradhan}, \citenamefont {Dimitri},
  \citenamefont {Sims}, \citenamefont {Regmi}, \citenamefont {Hosen},
  \citenamefont {Liu}, \citenamefont {Persaud}, \citenamefont {Kaczorowski},
  \citenamefont {Oppeneer},\ and\ \citenamefont
  {Neupane}}]{PhysRevB.106.125124}%
  \BibitemOpen
  \bibfield  {author} {\bibinfo {author} {\bibfnamefont {Gyanendra}\
  \bibnamefont {Dhakal}}, \bibinfo {author} {\bibfnamefont {Firoza}\
  \bibnamefont {Kabir}}, \bibinfo {author} {\bibfnamefont {Ashis~K.}\
  \bibnamefont {Nandy}}, \bibinfo {author} {\bibfnamefont {Alex}\ \bibnamefont
  {Aperis}}, \bibinfo {author} {\bibfnamefont {Anup~Pradhan}\ \bibnamefont
  {Sakhya}}, \bibinfo {author} {\bibfnamefont {Subhadip}\ \bibnamefont
  {Pradhan}}, \bibinfo {author} {\bibfnamefont {Klauss}\ \bibnamefont
  {Dimitri}}, \bibinfo {author} {\bibfnamefont {Christopher}\ \bibnamefont
  {Sims}}, \bibinfo {author} {\bibfnamefont {Sabin}\ \bibnamefont {Regmi}},
  \bibinfo {author} {\bibfnamefont {M.~Mofazzel}\ \bibnamefont {Hosen}},
  \bibinfo {author} {\bibfnamefont {Yangyang}\ \bibnamefont {Liu}}, \bibinfo
  {author} {\bibfnamefont {Luis}\ \bibnamefont {Persaud}}, \bibinfo {author}
  {\bibfnamefont {Dariusz}\ \bibnamefont {Kaczorowski}}, \bibinfo {author}
  {\bibfnamefont {Peter~M.}\ \bibnamefont {Oppeneer}}, \ and\ \bibinfo {author}
  {\bibfnamefont {Madhab}\ \bibnamefont {Neupane}},\ }\bibfield  {title}
  {\enquote {\bibinfo {title} {Observation of anisotropic dirac cones in the
  topological material {${\mathrm{Ti}}_{2}{\mathrm{Te}}_{2}\mathrm{P}$}},}\
  }\href {\doibase 10.1103/PhysRevB.106.125124} {\bibfield  {journal} {\bibinfo
   {journal} {Phys. Rev. B}\ }\textbf {\bibinfo {volume} {106}},\ \bibinfo
  {pages} {125124} (\bibinfo {year} {2022})}\BibitemShut {NoStop}%
\bibitem [{\citenamefont {Altland}\ and\ \citenamefont
  {Zirnbauer}(1997)}]{PhysRevB.55.1142}%
  \BibitemOpen
  \bibfield  {author} {\bibinfo {author} {\bibfnamefont {Alexander}\
  \bibnamefont {Altland}}\ and\ \bibinfo {author} {\bibfnamefont {Martin~R.}\
  \bibnamefont {Zirnbauer}},\ }\bibfield  {title} {\enquote {\bibinfo {title}
  {Nonstandard symmetry classes in mesoscopic normal-superconducting hybrid
  structures},}\ }\href {\doibase 10.1103/PhysRevB.55.1142} {\bibfield
  {journal} {\bibinfo  {journal} {Phys. Rev. B}\ }\textbf {\bibinfo {volume}
  {55}},\ \bibinfo {pages} {1142--1161} (\bibinfo {year} {1997})}\BibitemShut
  {NoStop}%
\bibitem [{\citenamefont {Kitaev}(2001)}]{kitaev2001unpaired}%
  \BibitemOpen
  \bibfield  {author} {\bibinfo {author} {\bibfnamefont {A~Yu}\ \bibnamefont
  {Kitaev}},\ }\bibfield  {title} {\enquote {\bibinfo {title} {Unpaired
  {M}ajorana fermions in quantum wires},}\ }\href {\doibase
  10.1070/1063-7869/44/10s/s29} {\bibfield  {journal} {\bibinfo  {journal}
  {Physics-Uspekhi}\ }\textbf {\bibinfo {volume} {44}},\ \bibinfo {pages} {131}
  (\bibinfo {year} {2001})}\BibitemShut {NoStop}%
\bibitem [{\citenamefont {Kitaev}(2009)}]{kitaev2009periodic}%
  \BibitemOpen
  \bibfield  {author} {\bibinfo {author} {\bibfnamefont {Alexei}\ \bibnamefont
  {Kitaev}},\ }\bibfield  {title} {\enquote {\bibinfo {title} {Periodic table
  for topological insulators and superconductors},}\ }in\ \href
  {https://doi.org/10.1063/1.3149495} {\emph {\bibinfo {booktitle} {AIP
  conference proceedings}}},\ Vol.\ \bibinfo {volume} {1134}\ (\bibinfo
  {publisher} {American Institute of Physics},\ \bibinfo {year} {2009})\ pp.\
  \bibinfo {pages} {22--30}\BibitemShut {NoStop}%
\bibitem [{\citenamefont {Read}\ and\ \citenamefont
  {Green}(2000)}]{read2000paired}%
  \BibitemOpen
  \bibfield  {author} {\bibinfo {author} {\bibfnamefont {Nicholas}\
  \bibnamefont {Read}}\ and\ \bibinfo {author} {\bibfnamefont {Dmitry}\
  \bibnamefont {Green}},\ }\bibfield  {title} {\enquote {\bibinfo {title}
  {Paired states of fermions in two dimensions with breaking of parity and
  time-reversal symmetries and the fractional quantum hall effect},}\ }\href
  {\doibase 10.1103/PhysRevB.61.10267} {\bibfield  {journal} {\bibinfo
  {journal} {Phys. Rev. B}\ }\textbf {\bibinfo {volume} {61}},\ \bibinfo
  {pages} {10267} (\bibinfo {year} {2000})}\BibitemShut {NoStop}%
\bibitem [{\citenamefont {Fu}\ and\ \citenamefont
  {Kane}(2008)}]{fu2008superconducting}%
  \BibitemOpen
  \bibfield  {author} {\bibinfo {author} {\bibfnamefont {Liang}\ \bibnamefont
  {Fu}}\ and\ \bibinfo {author} {\bibfnamefont {Charles~L}\ \bibnamefont
  {Kane}},\ }\bibfield  {title} {\enquote {\bibinfo {title} {Superconducting
  proximity effect and {M}ajorana fermions at the surface of a topological
  insulator},}\ }\href {\doibase 10.1103/PhysRevLett.100.096407} {\bibfield
  {journal} {\bibinfo  {journal} {Phys. Rev. Lett.}\ }\textbf {\bibinfo
  {volume} {100}},\ \bibinfo {pages} {096407} (\bibinfo {year}
  {2008})}\BibitemShut {NoStop}%
\bibitem [{\citenamefont {Sato}(2003)}]{sato2003non}%
  \BibitemOpen
  \bibfield  {author} {\bibinfo {author} {\bibfnamefont {Masatoshi}\
  \bibnamefont {Sato}},\ }\bibfield  {title} {\enquote {\bibinfo {title}
  {Non-abelian statistics of axion strings},}\ }\href
  {https://doi.org/10.1016/j.physletb.2003.09.047} {\bibfield  {journal}
  {\bibinfo  {journal} {Phys. Lett. B}\ }\textbf {\bibinfo {volume} {575}},\
  \bibinfo {pages} {126--130} (\bibinfo {year} {2003})}\BibitemShut {NoStop}%
\bibitem [{\citenamefont {Simon}(1983)}]{simon1983holonomy}%
  \BibitemOpen
  \bibfield  {author} {\bibinfo {author} {\bibfnamefont {Barry}\ \bibnamefont
  {Simon}},\ }\bibfield  {title} {\enquote {\bibinfo {title} {Holonomy, the
  quantum adiabatic theorem, and {B}erry's phase},}\ }\href {\doibase
  10.1103/PhysRevLett.51.2167} {\bibfield  {journal} {\bibinfo  {journal}
  {Phys. Rev. Lett.}\ }\textbf {\bibinfo {volume} {51}},\ \bibinfo {pages}
  {2167} (\bibinfo {year} {1983})}\BibitemShut {NoStop}%
\bibitem [{\citenamefont {Avron}\ \emph {et~al.}(1983)\citenamefont {Avron},
  \citenamefont {Seiler},\ and\ \citenamefont {Simon}}]{avron1983homotopy}%
  \BibitemOpen
  \bibfield  {author} {\bibinfo {author} {\bibfnamefont {Joseph~E}\
  \bibnamefont {Avron}}, \bibinfo {author} {\bibfnamefont {Ruedi}\ \bibnamefont
  {Seiler}}, \ and\ \bibinfo {author} {\bibfnamefont {Barry}\ \bibnamefont
  {Simon}},\ }\bibfield  {title} {\enquote {\bibinfo {title} {Homotopy and
  quantization in condensed matter physics},}\ }\href {\doibase
  10.1103/PhysRevLett.51.51} {\bibfield  {journal} {\bibinfo  {journal} {Phys.
  Rev. Lett.}\ }\textbf {\bibinfo {volume} {51}},\ \bibinfo {pages} {51}
  (\bibinfo {year} {1983})}\BibitemShut {NoStop}%
\bibitem [{\citenamefont {Nayak}\ \emph {et~al.}(2008)\citenamefont {Nayak},
  \citenamefont {Simon}, \citenamefont {Stern}, \citenamefont {Freedman},\ and\
  \citenamefont {Sarma}}]{nayak2008non}%
  \BibitemOpen
  \bibfield  {author} {\bibinfo {author} {\bibfnamefont {Chetan}\ \bibnamefont
  {Nayak}}, \bibinfo {author} {\bibfnamefont {Steven~H}\ \bibnamefont {Simon}},
  \bibinfo {author} {\bibfnamefont {Ady}\ \bibnamefont {Stern}}, \bibinfo
  {author} {\bibfnamefont {Michael}\ \bibnamefont {Freedman}}, \ and\ \bibinfo
  {author} {\bibfnamefont {Sankar~Das}\ \bibnamefont {Sarma}},\ }\bibfield
  {title} {\enquote {\bibinfo {title} {Non-abelian anyons and topological
  quantum computation},}\ }\href {\doibase 10.1103/RevModPhys.80.1083}
  {\bibfield  {journal} {\bibinfo  {journal} {Rev. Mod. Phys.}\ }\textbf
  {\bibinfo {volume} {80}},\ \bibinfo {pages} {1083} (\bibinfo {year}
  {2008})}\BibitemShut {NoStop}%
\bibitem [{\citenamefont {Oreg}\ \emph {et~al.}(2010)\citenamefont {Oreg},
  \citenamefont {Refael},\ and\ \citenamefont {von
  Oppen}}]{PhysRevLett.105.177002}%
  \BibitemOpen
  \bibfield  {author} {\bibinfo {author} {\bibfnamefont {Yuval}\ \bibnamefont
  {Oreg}}, \bibinfo {author} {\bibfnamefont {Gil}\ \bibnamefont {Refael}}, \
  and\ \bibinfo {author} {\bibfnamefont {Felix}\ \bibnamefont {von Oppen}},\
  }\bibfield  {title} {\enquote {\bibinfo {title} {Helical liquids and
  {M}ajorana bound states in quantum wires},}\ }\href {\doibase
  10.1103/PhysRevLett.105.177002} {\bibfield  {journal} {\bibinfo  {journal}
  {Phys. Rev. Lett.}\ }\textbf {\bibinfo {volume} {105}},\ \bibinfo {pages}
  {177002} (\bibinfo {year} {2010})}\BibitemShut {NoStop}%
\bibitem [{\citenamefont {Lutchyn}\ \emph {et~al.}(2010)\citenamefont
  {Lutchyn}, \citenamefont {Sau},\ and\ \citenamefont
  {Das~Sarma}}]{PhysRevLett.105.077001}%
  \BibitemOpen
  \bibfield  {author} {\bibinfo {author} {\bibfnamefont {Roman~M.}\
  \bibnamefont {Lutchyn}}, \bibinfo {author} {\bibfnamefont {Jay~D.}\
  \bibnamefont {Sau}}, \ and\ \bibinfo {author} {\bibfnamefont
  {S.}~\bibnamefont {Das~Sarma}},\ }\bibfield  {title} {\enquote {\bibinfo
  {title} {Majorana fermions and a topological phase transition in
  semiconductor-superconductor heterostructures},}\ }\href {\doibase
  10.1103/PhysRevLett.105.077001} {\bibfield  {journal} {\bibinfo  {journal}
  {Phys. Rev. Lett.}\ }\textbf {\bibinfo {volume} {105}},\ \bibinfo {pages}
  {077001} (\bibinfo {year} {2010})}\BibitemShut {NoStop}%
\bibitem [{\citenamefont {Mourik}\ \emph {et~al.}(2012)\citenamefont {Mourik},
  \citenamefont {Zuo}, \citenamefont {Frolov}, \citenamefont {Plissard},
  \citenamefont {Bakkers},\ and\ \citenamefont
  {Kouwenhoven}}]{mourik2012signatures}%
  \BibitemOpen
  \bibfield  {author} {\bibinfo {author} {\bibfnamefont {Vincent}\ \bibnamefont
  {Mourik}}, \bibinfo {author} {\bibfnamefont {Kun}\ \bibnamefont {Zuo}},
  \bibinfo {author} {\bibfnamefont {Sergey~M}\ \bibnamefont {Frolov}}, \bibinfo
  {author} {\bibfnamefont {SR}~\bibnamefont {Plissard}}, \bibinfo {author}
  {\bibfnamefont {EPAM}\ \bibnamefont {Bakkers}}, \ and\ \bibinfo {author}
  {\bibfnamefont {Leo~P}\ \bibnamefont {Kouwenhoven}},\ }\bibfield  {title}
  {\enquote {\bibinfo {title} {Signatures of {M}ajorana fermions in hybrid
  superconductor-semiconductor nanowire devices},}\ }\href {\doibase
  10.1126/science.1222360} {\bibfield  {journal} {\bibinfo  {journal}
  {Science}\ }\textbf {\bibinfo {volume} {336}},\ \bibinfo {pages} {1003--1007}
  (\bibinfo {year} {2012})}\BibitemShut {NoStop}%
\bibitem [{\citenamefont {Das}\ \emph {et~al.}(2012)\citenamefont {Das},
  \citenamefont {Ronen}, \citenamefont {Most}, \citenamefont {Oreg},
  \citenamefont {Heiblum},\ and\ \citenamefont {Shtrikman}}]{das2012zero}%
  \BibitemOpen
  \bibfield  {author} {\bibinfo {author} {\bibfnamefont {Anindya}\ \bibnamefont
  {Das}}, \bibinfo {author} {\bibfnamefont {Yuval}\ \bibnamefont {Ronen}},
  \bibinfo {author} {\bibfnamefont {Yonatan}\ \bibnamefont {Most}}, \bibinfo
  {author} {\bibfnamefont {Yuval}\ \bibnamefont {Oreg}}, \bibinfo {author}
  {\bibfnamefont {Moty}\ \bibnamefont {Heiblum}}, \ and\ \bibinfo {author}
  {\bibfnamefont {Hadas}\ \bibnamefont {Shtrikman}},\ }\bibfield  {title}
  {\enquote {\bibinfo {title} {Zero-bias peaks and splitting in an
  {$\mathrm{Al-InAs}$} nanowire topological superconductor as a signature of
  {M}ajorana fermions},}\ }\href {\doibase 10.1038/nphys2479} {\bibfield
  {journal} {\bibinfo  {journal} {Nat. Phys.}\ }\textbf {\bibinfo {volume}
  {8}},\ \bibinfo {pages} {887--895} (\bibinfo {year} {2012})}\BibitemShut
  {NoStop}%
\bibitem [{\citenamefont {Rokhinson}\ \emph {et~al.}(2012)\citenamefont
  {Rokhinson}, \citenamefont {Liu},\ and\ \citenamefont
  {Furdyna}}]{rokhinson2012fractional}%
  \BibitemOpen
  \bibfield  {author} {\bibinfo {author} {\bibfnamefont {Leonid~P}\
  \bibnamefont {Rokhinson}}, \bibinfo {author} {\bibfnamefont {Xinyu}\
  \bibnamefont {Liu}}, \ and\ \bibinfo {author} {\bibfnamefont {Jacek~K}\
  \bibnamefont {Furdyna}},\ }\bibfield  {title} {\enquote {\bibinfo {title}
  {The fractional ac josephson effect in a semiconductor-superconductor
  nanowire as a signature of {M}ajorana particles},}\ }\href {\doibase
  10.1038/nphys2429} {\bibfield  {journal} {\bibinfo  {journal} {Nat. Phys.}\
  }\textbf {\bibinfo {volume} {8}},\ \bibinfo {pages} {795--799} (\bibinfo
  {year} {2012})}\BibitemShut {NoStop}%
\bibitem [{\citenamefont {Finck}\ \emph {et~al.}(2013)\citenamefont {Finck},
  \citenamefont {Van~Harlingen}, \citenamefont {Mohseni}, \citenamefont
  {Jung},\ and\ \citenamefont {Li}}]{finck2013anomalous}%
  \BibitemOpen
  \bibfield  {author} {\bibinfo {author} {\bibfnamefont {ADK}\ \bibnamefont
  {Finck}}, \bibinfo {author} {\bibfnamefont {DJ}~\bibnamefont
  {Van~Harlingen}}, \bibinfo {author} {\bibfnamefont {PK}~\bibnamefont
  {Mohseni}}, \bibinfo {author} {\bibfnamefont {K}~\bibnamefont {Jung}}, \ and\
  \bibinfo {author} {\bibfnamefont {X}~\bibnamefont {Li}},\ }\bibfield  {title}
  {\enquote {\bibinfo {title} {Anomalous modulation of a zero-bias peak in a
  hybrid nanowire-superconductor device},}\ }\href {\doibase
  10.1103/PhysRevLett.110.126406} {\bibfield  {journal} {\bibinfo  {journal}
  {Phys. Rev. Lett.}\ }\textbf {\bibinfo {volume} {110}},\ \bibinfo {pages}
  {126406} (\bibinfo {year} {2013})}\BibitemShut {NoStop}%
\bibitem [{\citenamefont {Albrecht}\ \emph {et~al.}(2016)\citenamefont
  {Albrecht}, \citenamefont {Higginbotham}, \citenamefont {Madsen},
  \citenamefont {Kuemmeth}, \citenamefont {Jespersen}, \citenamefont
  {Nyg{\aa}rd}, \citenamefont {Krogstrup},\ and\ \citenamefont
  {Marcus}}]{Albrecht2016}%
  \BibitemOpen
  \bibfield  {author} {\bibinfo {author} {\bibfnamefont {S.~M.}\ \bibnamefont
  {Albrecht}}, \bibinfo {author} {\bibfnamefont {A.~P.}\ \bibnamefont
  {Higginbotham}}, \bibinfo {author} {\bibfnamefont {M.}~\bibnamefont
  {Madsen}}, \bibinfo {author} {\bibfnamefont {F.}~\bibnamefont {Kuemmeth}},
  \bibinfo {author} {\bibfnamefont {T.~S.}\ \bibnamefont {Jespersen}}, \bibinfo
  {author} {\bibfnamefont {J.}~\bibnamefont {Nyg{\aa}rd}}, \bibinfo {author}
  {\bibfnamefont {P.}~\bibnamefont {Krogstrup}}, \ and\ \bibinfo {author}
  {\bibfnamefont {C.~M.}\ \bibnamefont {Marcus}},\ }\bibfield  {title}
  {\enquote {\bibinfo {title} {Exponential protection of zero modes in majorana
  islands},}\ }\href {\doibase 10.1038/nature17162} {\bibfield  {journal}
  {\bibinfo  {journal} {Nature}\ }\textbf {\bibinfo {volume} {531}},\ \bibinfo
  {pages} {206--209} (\bibinfo {year} {2016})}\BibitemShut {NoStop}%
\bibitem [{\citenamefont {Deng}\ \emph {et~al.}(2016)\citenamefont {Deng},
  \citenamefont {Vaitiek{\"A}nas}, \citenamefont {Hansen}, \citenamefont
  {Danon}, \citenamefont {Leijnse}, \citenamefont {Flensberg}, \citenamefont
  {Nyg{\aa}rd}, \citenamefont {Krogstrup},\ and\ \citenamefont
  {Marcus}}]{doi:10.1126/science.aaf3961}%
  \BibitemOpen
  \bibfield  {author} {\bibinfo {author} {\bibfnamefont {M.~T.}\ \bibnamefont
  {Deng}}, \bibinfo {author} {\bibfnamefont {S.}~\bibnamefont
  {Vaitiek{\"A}nas}}, \bibinfo {author} {\bibfnamefont {E.~B.}\ \bibnamefont
  {Hansen}}, \bibinfo {author} {\bibfnamefont {J.}~\bibnamefont {Danon}},
  \bibinfo {author} {\bibfnamefont {M.}~\bibnamefont {Leijnse}}, \bibinfo
  {author} {\bibfnamefont {K.}~\bibnamefont {Flensberg}}, \bibinfo {author}
  {\bibfnamefont {J.}~\bibnamefont {Nyg{\aa}rd}}, \bibinfo {author}
  {\bibfnamefont {P.}~\bibnamefont {Krogstrup}}, \ and\ \bibinfo {author}
  {\bibfnamefont {C.~M.}\ \bibnamefont {Marcus}},\ }\bibfield  {title}
  {\enquote {\bibinfo {title} {Majorana bound state in a coupled quantum-dot
  hybrid-nanowire system},}\ }\href {\doibase 10.1126/science.aaf3961}
  {\bibfield  {journal} {\bibinfo  {journal} {Science}\ }\textbf {\bibinfo
  {volume} {354}},\ \bibinfo {pages} {1557--1562} (\bibinfo {year}
  {2016})}\BibitemShut {NoStop}%
\bibitem [{\citenamefont {Nadj-Perge}\ \emph {et~al.}(2013)\citenamefont
  {Nadj-Perge}, \citenamefont {Drozdov}, \citenamefont {Bernevig},\ and\
  \citenamefont {Yazdani}}]{Ali}%
  \BibitemOpen
  \bibfield  {author} {\bibinfo {author} {\bibfnamefont {S.}~\bibnamefont
  {Nadj-Perge}}, \bibinfo {author} {\bibfnamefont {I.~K.}\ \bibnamefont
  {Drozdov}}, \bibinfo {author} {\bibfnamefont {B.~A.}\ \bibnamefont
  {Bernevig}}, \ and\ \bibinfo {author} {\bibfnamefont {Ali}\ \bibnamefont
  {Yazdani}},\ }\bibfield  {title} {\enquote {\bibinfo {title} {Proposal for
  realizing majorana fermions in chains of magnetic atoms on a
  superconductor},}\ }\href {\doibase 10.1103/PhysRevB.88.020407} {\bibfield
  {journal} {\bibinfo  {journal} {Phys. Rev. B}\ }\textbf {\bibinfo {volume}
  {88}},\ \bibinfo {pages} {020407} (\bibinfo {year} {2013})}\BibitemShut
  {NoStop}%
\bibitem [{\citenamefont {Choy}\ \emph {et~al.}(2011)\citenamefont {Choy},
  \citenamefont {Edge}, \citenamefont {Akhmerov},\ and\ \citenamefont
  {Beenakker}}]{nanoplates}%
  \BibitemOpen
  \bibfield  {author} {\bibinfo {author} {\bibfnamefont {T.-P.}\ \bibnamefont
  {Choy}}, \bibinfo {author} {\bibfnamefont {J.~M.}\ \bibnamefont {Edge}},
  \bibinfo {author} {\bibfnamefont {A.~R.}\ \bibnamefont {Akhmerov}}, \ and\
  \bibinfo {author} {\bibfnamefont {C.~W.~J.}\ \bibnamefont {Beenakker}},\
  }\bibfield  {title} {\enquote {\bibinfo {title} {Majorana fermions emerging
  from magnetic nanoparticles on a superconductor without spin-orbit
  coupling},}\ }\href {\doibase 10.1103/PhysRevB.84.195442} {\bibfield
  {journal} {\bibinfo  {journal} {Phys. Rev. B}\ }\textbf {\bibinfo {volume}
  {84}},\ \bibinfo {pages} {195442} (\bibinfo {year} {2011})}\BibitemShut
  {NoStop}%
\bibitem [{\citenamefont {Rex}\ \emph {et~al.}(2020)\citenamefont {Rex},
  \citenamefont {Gornyi},\ and\ \citenamefont {Mirlin}}]{Majoranacontrol}%
  \BibitemOpen
  \bibfield  {author} {\bibinfo {author} {\bibfnamefont {Stefan}\ \bibnamefont
  {Rex}}, \bibinfo {author} {\bibfnamefont {Igor~V.}\ \bibnamefont {Gornyi}}, \
  and\ \bibinfo {author} {\bibfnamefont {Alexander~D.}\ \bibnamefont
  {Mirlin}},\ }\bibfield  {title} {\enquote {\bibinfo {title} {Majorana modes
  in emergent-wire phases of helical and cycloidal magnet-superconductor
  hybrids},}\ }\href {\doibase 10.1103/PhysRevB.102.224501} {\bibfield
  {journal} {\bibinfo  {journal} {Phys. Rev. B}\ }\textbf {\bibinfo {volume}
  {102}},\ \bibinfo {pages} {224501} (\bibinfo {year} {2020})}\BibitemShut
  {NoStop}%
\bibitem [{\citenamefont {Pientka}\ \emph {et~al.}(2013)\citenamefont
  {Pientka}, \citenamefont {Glazman},\ and\ \citenamefont {von Oppen}}]{Felix}%
  \BibitemOpen
  \bibfield  {author} {\bibinfo {author} {\bibfnamefont {Falko}\ \bibnamefont
  {Pientka}}, \bibinfo {author} {\bibfnamefont {Leonid~I.}\ \bibnamefont
  {Glazman}}, \ and\ \bibinfo {author} {\bibfnamefont {Felix}\ \bibnamefont
  {von Oppen}},\ }\bibfield  {title} {\enquote {\bibinfo {title} {Topological
  superconducting phase in helical {S}hiba chains},}\ }\href {\doibase
  10.1103/PhysRevB.88.155420} {\bibfield  {journal} {\bibinfo  {journal} {Phys.
  Rev. B}\ }\textbf {\bibinfo {volume} {88}},\ \bibinfo {pages} {155420}
  (\bibinfo {year} {2013})}\BibitemShut {NoStop}%
\bibitem [{\citenamefont {Pientka}\ \emph {et~al.}(2014)\citenamefont
  {Pientka}, \citenamefont {Glazman},\ and\ \citenamefont {von
  Oppen}}]{Felix2}%
  \BibitemOpen
  \bibfield  {author} {\bibinfo {author} {\bibfnamefont {Falko}\ \bibnamefont
  {Pientka}}, \bibinfo {author} {\bibfnamefont {Leonid~I.}\ \bibnamefont
  {Glazman}}, \ and\ \bibinfo {author} {\bibfnamefont {Felix}\ \bibnamefont
  {von Oppen}},\ }\bibfield  {title} {\enquote {\bibinfo {title}
  {Unconventional topological phase transitions in helical {S}hiba chains},}\
  }\href {\doibase 10.1103/PhysRevB.89.180505} {\bibfield  {journal} {\bibinfo
  {journal} {Phys. Rev. B}\ }\textbf {\bibinfo {volume} {89}},\ \bibinfo
  {pages} {180505} (\bibinfo {year} {2014})}\BibitemShut {NoStop}%
\bibitem [{\citenamefont {Mohanta}\ \emph {et~al.}(2021)\citenamefont
  {Mohanta}, \citenamefont {Okamoto},\ and\ \citenamefont
  {Dagotto}}]{natcomms}%
  \BibitemOpen
  \bibfield  {author} {\bibinfo {author} {\bibfnamefont {Narayan}\ \bibnamefont
  {Mohanta}}, \bibinfo {author} {\bibfnamefont {Satoshi}\ \bibnamefont
  {Okamoto}}, \ and\ \bibinfo {author} {\bibfnamefont {Elbio}\ \bibnamefont
  {Dagotto}},\ }\bibfield  {title} {\enquote {\bibinfo {title} {Skyrmion
  control of majorana states in planar josephson junctions},}\ }\href {\doibase
  10.1038/s42005-021-00666-5} {\bibfield  {journal} {\bibinfo  {journal}
  {Communications Physics}\ }\textbf {\bibinfo {volume} {4}},\ \bibinfo {pages}
  {163} (\bibinfo {year} {2021})}\BibitemShut {NoStop}%
\bibitem [{\citenamefont {Klinovaja}\ \emph {et~al.}(2013)\citenamefont
  {Klinovaja}, \citenamefont {Stano}, \citenamefont {Yazdani},\ and\
  \citenamefont {Loss}}]{RKKY}%
  \BibitemOpen
  \bibfield  {author} {\bibinfo {author} {\bibfnamefont {Jelena}\ \bibnamefont
  {Klinovaja}}, \bibinfo {author} {\bibfnamefont {Peter}\ \bibnamefont
  {Stano}}, \bibinfo {author} {\bibfnamefont {Ali}\ \bibnamefont {Yazdani}}, \
  and\ \bibinfo {author} {\bibfnamefont {Daniel}\ \bibnamefont {Loss}},\
  }\bibfield  {title} {\enquote {\bibinfo {title} {Topological
  superconductivity and majorana fermions in {RKKY} systems},}\ }\href
  {\doibase 10.1103/PhysRevLett.111.186805} {\bibfield  {journal} {\bibinfo
  {journal} {Phys. Rev. Lett.}\ }\textbf {\bibinfo {volume} {111}},\ \bibinfo
  {pages} {186805} (\bibinfo {year} {2013})}\BibitemShut {NoStop}%
\bibitem [{\citenamefont {Andolina}\ and\ \citenamefont
  {Simon}(2017)}]{pascal}%
  \BibitemOpen
  \bibfield  {author} {\bibinfo {author} {\bibfnamefont {Gian~Marcello}\
  \bibnamefont {Andolina}}\ and\ \bibinfo {author} {\bibfnamefont {Pascal}\
  \bibnamefont {Simon}},\ }\bibfield  {title} {\enquote {\bibinfo {title}
  {Topological properties of chains of magnetic impurities on a superconducting
  substrate: Interplay between the {S}hiba band and ferromagnetic wire
  limits},}\ }\href {\doibase 10.1103/PhysRevB.96.235411} {\bibfield  {journal}
  {\bibinfo  {journal} {Phys. Rev. B}\ }\textbf {\bibinfo {volume} {96}},\
  \bibinfo {pages} {235411} (\bibinfo {year} {2017})}\BibitemShut {NoStop}%
\bibitem [{\citenamefont {Hoffman}\ \emph {et~al.}(2016)\citenamefont
  {Hoffman}, \citenamefont {Klinovaja},\ and\ \citenamefont
  {Loss}}]{Daniel_Loss}%
  \BibitemOpen
  \bibfield  {author} {\bibinfo {author} {\bibfnamefont {Silas}\ \bibnamefont
  {Hoffman}}, \bibinfo {author} {\bibfnamefont {Jelena}\ \bibnamefont
  {Klinovaja}}, \ and\ \bibinfo {author} {\bibfnamefont {Daniel}\ \bibnamefont
  {Loss}},\ }\bibfield  {title} {\enquote {\bibinfo {title} {Topological phases
  of inhomogeneous superconductivity},}\ }\href {\doibase
  10.1103/PhysRevB.93.165418} {\bibfield  {journal} {\bibinfo  {journal} {Phys.
  Rev. B}\ }\textbf {\bibinfo {volume} {93}},\ \bibinfo {pages} {165418}
  (\bibinfo {year} {2016})}\BibitemShut {NoStop}%
\bibitem [{\citenamefont {Sticlet}\ and\ \citenamefont
  {Morari}(2019)}]{PhysRevB.100.075420}%
  \BibitemOpen
  \bibfield  {author} {\bibinfo {author} {\bibfnamefont {Doru}\ \bibnamefont
  {Sticlet}}\ and\ \bibinfo {author} {\bibfnamefont {Cristian}\ \bibnamefont
  {Morari}},\ }\bibfield  {title} {\enquote {\bibinfo {title} {Topological
  superconductivity from magnetic impurities on monolayer
  {${\mathrm{NbSe}}_{2}$}},}\ }\href {\doibase 10.1103/PhysRevB.100.075420}
  {\bibfield  {journal} {\bibinfo  {journal} {Phys. Rev. B}\ }\textbf {\bibinfo
  {volume} {100}},\ \bibinfo {pages} {075420} (\bibinfo {year}
  {2019})}\BibitemShut {NoStop}%
\bibitem [{\citenamefont {Sau}\ and\ \citenamefont
  {Demler}(2013)}]{PhysRevB.88.205402}%
  \BibitemOpen
  \bibfield  {author} {\bibinfo {author} {\bibfnamefont {Jay~D.}\ \bibnamefont
  {Sau}}\ and\ \bibinfo {author} {\bibfnamefont {Eugene}\ \bibnamefont
  {Demler}},\ }\bibfield  {title} {\enquote {\bibinfo {title} {Bound states at
  impurities as a probe of topological superconductivity in nanowires},}\
  }\href {\doibase 10.1103/PhysRevB.88.205402} {\bibfield  {journal} {\bibinfo
  {journal} {Phys. Rev. B}\ }\textbf {\bibinfo {volume} {88}},\ \bibinfo
  {pages} {205402} (\bibinfo {year} {2013})}\BibitemShut {NoStop}%
\bibitem [{\citenamefont {Mashkoori}\ and\ \citenamefont
  {Black-Schaffer}(2019)}]{PhysRevB.99.024505}%
  \BibitemOpen
  \bibfield  {author} {\bibinfo {author} {\bibfnamefont {Mahdi}\ \bibnamefont
  {Mashkoori}}\ and\ \bibinfo {author} {\bibfnamefont {Annica}\ \bibnamefont
  {Black-Schaffer}},\ }\bibfield  {title} {\enquote {\bibinfo {title} {Majorana
  bound states in magnetic impurity chains: Effects of $d$-wave pairing},}\
  }\href {\doibase 10.1103/PhysRevB.99.024505} {\bibfield  {journal} {\bibinfo
  {journal} {Phys. Rev. B}\ }\textbf {\bibinfo {volume} {99}},\ \bibinfo
  {pages} {024505} (\bibinfo {year} {2019})}\BibitemShut {NoStop}%
\bibitem [{\citenamefont {Mashkoori}\ \emph {et~al.}(2020)\citenamefont
  {Mashkoori}, \citenamefont {Pradhan}, \citenamefont {Bj\"ornson},
  \citenamefont {Fransson},\ and\ \citenamefont
  {Black-Schaffer}}]{PhysRevB.102.104501}%
  \BibitemOpen
  \bibfield  {author} {\bibinfo {author} {\bibfnamefont {Mahdi}\ \bibnamefont
  {Mashkoori}}, \bibinfo {author} {\bibfnamefont {Saurabh}\ \bibnamefont
  {Pradhan}}, \bibinfo {author} {\bibfnamefont {Kristofer}\ \bibnamefont
  {Bj\"ornson}}, \bibinfo {author} {\bibfnamefont {Jonas}\ \bibnamefont
  {Fransson}}, \ and\ \bibinfo {author} {\bibfnamefont {Annica~M.}\
  \bibnamefont {Black-Schaffer}},\ }\bibfield  {title} {\enquote {\bibinfo
  {title} {Identification of topological superconductivity in magnetic impurity
  systems using bulk spin polarization},}\ }\href {\doibase
  10.1103/PhysRevB.102.104501} {\bibfield  {journal} {\bibinfo  {journal}
  {Phys. Rev. B}\ }\textbf {\bibinfo {volume} {102}},\ \bibinfo {pages}
  {104501} (\bibinfo {year} {2020})}\BibitemShut {NoStop}%
\bibitem [{\citenamefont {Teixeira}\ \emph {et~al.}(2020)\citenamefont
  {Teixeira}, \citenamefont {Kuzmanovski}, \citenamefont {Black-Schaffer},\
  and\ \citenamefont {da~Silva}}]{PhysRevB.102.165312}%
  \BibitemOpen
  \bibfield  {author} {\bibinfo {author} {\bibfnamefont {Raphael L. R.~C.}\
  \bibnamefont {Teixeira}}, \bibinfo {author} {\bibfnamefont {Dushko}\
  \bibnamefont {Kuzmanovski}}, \bibinfo {author} {\bibfnamefont {Annica~M.}\
  \bibnamefont {Black-Schaffer}}, \ and\ \bibinfo {author} {\bibfnamefont {Luis
  G. G. V.~Dias}\ \bibnamefont {da~Silva}},\ }\bibfield  {title} {\enquote
  {\bibinfo {title} {Enhanced majorana bound states in magnetic chains on
  superconducting topological insulator edges},}\ }\href {\doibase
  10.1103/PhysRevB.102.165312} {\bibfield  {journal} {\bibinfo  {journal}
  {Phys. Rev. B}\ }\textbf {\bibinfo {volume} {102}},\ \bibinfo {pages}
  {165312} (\bibinfo {year} {2020})}\BibitemShut {NoStop}%
\bibitem [{\citenamefont {Theiler}\ \emph {et~al.}(2019)\citenamefont
  {Theiler}, \citenamefont {Bj\"ornson},\ and\ \citenamefont
  {Black-Schaffer}}]{PhysRevB.100.214504}%
  \BibitemOpen
  \bibfield  {author} {\bibinfo {author} {\bibfnamefont {Andreas}\ \bibnamefont
  {Theiler}}, \bibinfo {author} {\bibfnamefont {Kristofer}\ \bibnamefont
  {Bj\"ornson}}, \ and\ \bibinfo {author} {\bibfnamefont {Annica~M.}\
  \bibnamefont {Black-Schaffer}},\ }\bibfield  {title} {\enquote {\bibinfo
  {title} {Majorana bound state localization and energy oscillations for
  magnetic impurity chains on conventional superconductors},}\ }\href {\doibase
  10.1103/PhysRevB.100.214504} {\bibfield  {journal} {\bibinfo  {journal}
  {Phys. Rev. B}\ }\textbf {\bibinfo {volume} {100}},\ \bibinfo {pages}
  {214504} (\bibinfo {year} {2019})}\BibitemShut {NoStop}%
\bibitem [{\citenamefont {Hui}\ \emph {et~al.}(2015)\citenamefont {Hui},
  \citenamefont {Brydon}, \citenamefont {Sau}, \citenamefont {Tewari},\ and\
  \citenamefont {Sarma}}]{Hui2015}%
  \BibitemOpen
  \bibfield  {author} {\bibinfo {author} {\bibfnamefont {Hoi-Yin}\ \bibnamefont
  {Hui}}, \bibinfo {author} {\bibfnamefont {P.~M.~R.}\ \bibnamefont {Brydon}},
  \bibinfo {author} {\bibfnamefont {Jay~D.}\ \bibnamefont {Sau}}, \bibinfo
  {author} {\bibfnamefont {S.}~\bibnamefont {Tewari}}, \ and\ \bibinfo {author}
  {\bibfnamefont {S.~Das}\ \bibnamefont {Sarma}},\ }\bibfield  {title}
  {\enquote {\bibinfo {title} {Majorana fermions in ferromagnetic chains on the
  surface of bulk spin-orbit coupled s-wave superconductors},}\ }\href
  {\doibase 10.1038/srep08880} {\bibfield  {journal} {\bibinfo  {journal}
  {Scientific Reports}\ }\textbf {\bibinfo {volume} {5}},\ \bibinfo {pages}
  {8880} (\bibinfo {year} {2015})}\BibitemShut {NoStop}%
\bibitem [{\citenamefont {Perrin}\ \emph {et~al.}(2021)\citenamefont {Perrin},
  \citenamefont {Civelli},\ and\ \citenamefont {Simon}}]{PhysRevB.104.L121406}%
  \BibitemOpen
  \bibfield  {author} {\bibinfo {author} {\bibfnamefont {Vivien}\ \bibnamefont
  {Perrin}}, \bibinfo {author} {\bibfnamefont {Marcello}\ \bibnamefont
  {Civelli}}, \ and\ \bibinfo {author} {\bibfnamefont {Pascal}\ \bibnamefont
  {Simon}},\ }\bibfield  {title} {\enquote {\bibinfo {title} {Identifying
  majorana bound states by tunneling shot-noise tomography},}\ }\href {\doibase
  10.1103/PhysRevB.104.L121406} {\bibfield  {journal} {\bibinfo  {journal}
  {Phys. Rev. B}\ }\textbf {\bibinfo {volume} {104}},\ \bibinfo {pages}
  {L121406} (\bibinfo {year} {2021})}\BibitemShut {NoStop}%
\bibitem [{\citenamefont {Christensen}\ \emph {et~al.}(2016)\citenamefont
  {Christensen}, \citenamefont {Schecter}, \citenamefont {Flensberg},
  \citenamefont {Andersen},\ and\ \citenamefont {Paaske}}]{PhysRevB.94.144509}%
  \BibitemOpen
  \bibfield  {author} {\bibinfo {author} {\bibfnamefont {Morten~H.}\
  \bibnamefont {Christensen}}, \bibinfo {author} {\bibfnamefont {Michael}\
  \bibnamefont {Schecter}}, \bibinfo {author} {\bibfnamefont {Karsten}\
  \bibnamefont {Flensberg}}, \bibinfo {author} {\bibfnamefont {Brian~M.}\
  \bibnamefont {Andersen}}, \ and\ \bibinfo {author} {\bibfnamefont {Jens}\
  \bibnamefont {Paaske}},\ }\bibfield  {title} {\enquote {\bibinfo {title}
  {Spiral magnetic order and topological superconductivity in a chain of
  magnetic adatoms on a two-dimensional superconductor},}\ }\href {\doibase
  10.1103/PhysRevB.94.144509} {\bibfield  {journal} {\bibinfo  {journal} {Phys.
  Rev. B}\ }\textbf {\bibinfo {volume} {94}},\ \bibinfo {pages} {144509}
  (\bibinfo {year} {2016})}\BibitemShut {NoStop}%
\bibitem [{\citenamefont {Sharma}\ and\ \citenamefont
  {Tewari}(2016)}]{PhysRevB.94.094515}%
  \BibitemOpen
  \bibfield  {author} {\bibinfo {author} {\bibfnamefont {Girish}\ \bibnamefont
  {Sharma}}\ and\ \bibinfo {author} {\bibfnamefont {Sumanta}\ \bibnamefont
  {Tewari}},\ }\bibfield  {title} {\enquote {\bibinfo {title}
  {{Y}u-{S}hiba-{R}usinov states and topological superconductivity in ising
  paired superconductors},}\ }\href {\doibase 10.1103/PhysRevB.94.094515}
  {\bibfield  {journal} {\bibinfo  {journal} {Phys. Rev. B}\ }\textbf {\bibinfo
  {volume} {94}},\ \bibinfo {pages} {094515} (\bibinfo {year}
  {2016})}\BibitemShut {NoStop}%
\bibitem [{\citenamefont {{M\'enard, Gerbold C.}}\ \emph
  {et~al.}(2019)\citenamefont {{M\'enard, Gerbold C.}}, \citenamefont {{Brun,
  Christophe}}, \citenamefont {{Leriche, Rapha\"el}}, \citenamefont {{Trif,
  Mircea}}, \citenamefont {{Debontridder, Fran\c{c}ois}}, \citenamefont
  {{Demaille, Dominique}}, \citenamefont {{Roditchev, Dimitri}}, \citenamefont
  {{Simon, Pascal}},\ and\ \citenamefont {{Cren, Tristan}}}]{refId0}%
  \BibitemOpen
  \bibfield  {author} {\bibinfo {author} {\bibnamefont {{M\'enard, Gerbold
  C.}}}, \bibinfo {author} {\bibnamefont {{Brun, Christophe}}}, \bibinfo
  {author} {\bibnamefont {{Leriche, Rapha\"el}}}, \bibinfo {author}
  {\bibnamefont {{Trif, Mircea}}}, \bibinfo {author} {\bibnamefont
  {{Debontridder, Fran\c{c}ois}}}, \bibinfo {author} {\bibnamefont {{Demaille,
  Dominique}}}, \bibinfo {author} {\bibnamefont {{Roditchev, Dimitri}}},
  \bibinfo {author} {\bibnamefont {{Simon, Pascal}}}, \ and\ \bibinfo {author}
  {\bibnamefont {{Cren, Tristan}}},\ }\bibfield  {title} {\enquote {\bibinfo
  {title} {{Y}u-{S}hiba-{R}usinov bound states versus topological edge states
  in {Pb/Si(111)}},}\ }\href {\doibase 10.1140/epjst/e2018-800056-3} {\bibfield
   {journal} {\bibinfo  {journal} {Eur. Phys. J. Special Topics}\ }\textbf
  {\bibinfo {volume} {227}} (\bibinfo {year} {2019}),\
  10.1140/epjst/e2018-800056-3}\BibitemShut {NoStop}%
\bibitem [{\citenamefont {Shiba}(1968)}]{10.1143/PTP.40.435}%
  \BibitemOpen
  \bibfield  {author} {\bibinfo {author} {\bibfnamefont {Hiroyuki}\
  \bibnamefont {Shiba}},\ }\bibfield  {title} {\enquote {\bibinfo {title}
  {{Classical Spins in Superconductors}},}\ }\href {\doibase
  10.1143/PTP.40.435} {\bibfield  {journal} {\bibinfo  {journal} {Progress of
  Theoretical Physics}\ }\textbf {\bibinfo {volume} {40}},\ \bibinfo {pages}
  {435--451} (\bibinfo {year} {1968})}\BibitemShut {NoStop}%
\bibitem [{\citenamefont {Deb}\ \emph {et~al.}(2021)\citenamefont {Deb},
  \citenamefont {Hoffman}, \citenamefont {Loss},\ and\ \citenamefont
  {Klinovaja}}]{PhysRevB.103.165403}%
  \BibitemOpen
  \bibfield  {author} {\bibinfo {author} {\bibfnamefont {Oindrila}\
  \bibnamefont {Deb}}, \bibinfo {author} {\bibfnamefont {Silas}\ \bibnamefont
  {Hoffman}}, \bibinfo {author} {\bibfnamefont {Daniel}\ \bibnamefont {Loss}},
  \ and\ \bibinfo {author} {\bibfnamefont {Jelena}\ \bibnamefont {Klinovaja}},\
  }\bibfield  {title} {\enquote {\bibinfo {title} {{Y}u-{S}hiba-{R}usinov
  states and ordering of magnetic impurities near the boundary of a
  superconducting nanowire},}\ }\href {\doibase 10.1103/PhysRevB.103.165403}
  {\bibfield  {journal} {\bibinfo  {journal} {Phys. Rev. B}\ }\textbf {\bibinfo
  {volume} {103}},\ \bibinfo {pages} {165403} (\bibinfo {year}
  {2021})}\BibitemShut {NoStop}%
\bibitem [{\citenamefont {von Oppen}\ and\ \citenamefont
  {Franke}(2021)}]{PhysRevB.103.205424}%
  \BibitemOpen
  \bibfield  {author} {\bibinfo {author} {\bibfnamefont {Felix}\ \bibnamefont
  {von Oppen}}\ and\ \bibinfo {author} {\bibfnamefont {Katharina~J.}\
  \bibnamefont {Franke}},\ }\bibfield  {title} {\enquote {\bibinfo {title}
  {{Y}u-{S}hiba-{R}usinov states in real metals},}\ }\href {\doibase
  10.1103/PhysRevB.103.205424} {\bibfield  {journal} {\bibinfo  {journal}
  {Phys. Rev. B}\ }\textbf {\bibinfo {volume} {103}},\ \bibinfo {pages}
  {205424} (\bibinfo {year} {2021})}\BibitemShut {NoStop}%
\bibitem [{\citenamefont {Ortuzar}\ \emph {et~al.}(2022)\citenamefont
  {Ortuzar}, \citenamefont {Trivini}, \citenamefont {Alvarado}, \citenamefont
  {Rouco}, \citenamefont {Zaldivar}, \citenamefont {Yeyati}, \citenamefont
  {Pascual},\ and\ \citenamefont {Bergeret}}]{PhysRevB.105.245403}%
  \BibitemOpen
  \bibfield  {author} {\bibinfo {author} {\bibfnamefont {Jon}\ \bibnamefont
  {Ortuzar}}, \bibinfo {author} {\bibfnamefont {Stefano}\ \bibnamefont
  {Trivini}}, \bibinfo {author} {\bibfnamefont {Miguel}\ \bibnamefont
  {Alvarado}}, \bibinfo {author} {\bibfnamefont {Mikel}\ \bibnamefont {Rouco}},
  \bibinfo {author} {\bibfnamefont {Javier}\ \bibnamefont {Zaldivar}}, \bibinfo
  {author} {\bibfnamefont {Alfredo~Levy}\ \bibnamefont {Yeyati}}, \bibinfo
  {author} {\bibfnamefont {Jose~Ignacio}\ \bibnamefont {Pascual}}, \ and\
  \bibinfo {author} {\bibfnamefont {F.~Sebastian}\ \bibnamefont {Bergeret}},\
  }\bibfield  {title} {\enquote {\bibinfo {title} {{Y}u-{S}hiba-{R}usinov
  states in two-dimensional superconductors with arbitrary fermi contours},}\
  }\href {\doibase 10.1103/PhysRevB.105.245403} {\bibfield  {journal} {\bibinfo
   {journal} {Phys. Rev. B}\ }\textbf {\bibinfo {volume} {105}},\ \bibinfo
  {pages} {245403} (\bibinfo {year} {2022})}\BibitemShut {NoStop}%
\bibitem [{\citenamefont {Beenakker}(2013)}]{Beenakker}%
  \BibitemOpen
  \bibfield  {author} {\bibinfo {author} {\bibfnamefont {C.W.J.}\ \bibnamefont
  {Beenakker}},\ }\bibfield  {title} {\enquote {\bibinfo {title} {Search for
  majorana fermions in superconductors},}\ }\href {\doibase
  10.1146/annurev-conmatphys-030212-184337} {\bibfield  {journal} {\bibinfo
  {journal} {Annual Review of Condensed Matter Physics}\ }\textbf {\bibinfo
  {volume} {4}},\ \bibinfo {pages} {113–136} (\bibinfo {year}
  {2013})}\BibitemShut {NoStop}%
\bibitem [{\citenamefont {Alicea}(2012)}]{Alicea_2012}%
  \BibitemOpen
  \bibfield  {author} {\bibinfo {author} {\bibfnamefont {Jason}\ \bibnamefont
  {Alicea}},\ }\bibfield  {title} {\enquote {\bibinfo {title} {New directions
  in the pursuit of majorana fermions in solid state systems},}\ }\href
  {\doibase 10.1088/0034-4885/75/7/076501} {\bibfield  {journal} {\bibinfo
  {journal} {Reports on Progress in Physics}\ }\textbf {\bibinfo {volume}
  {75}},\ \bibinfo {pages} {076501} (\bibinfo {year} {2012})}\BibitemShut
  {NoStop}%
\bibitem [{\citenamefont {Leijnse}\ and\ \citenamefont
  {Flensberg}(2012)}]{Leijnse_2012}%
  \BibitemOpen
  \bibfield  {author} {\bibinfo {author} {\bibfnamefont {Martin}\ \bibnamefont
  {Leijnse}}\ and\ \bibinfo {author} {\bibfnamefont {Karsten}\ \bibnamefont
  {Flensberg}},\ }\bibfield  {title} {\enquote {\bibinfo {title} {Introduction
  to topological superconductivity and majorana fermions},}\ }\href {\doibase
  10.1088/0268-1242/27/12/124003} {\bibfield  {journal} {\bibinfo  {journal}
  {Semiconductor Science and Technology}\ }\textbf {\bibinfo {volume} {27}},\
  \bibinfo {pages} {124003} (\bibinfo {year} {2012})}\BibitemShut {NoStop}%
\bibitem [{\citenamefont {Wang}\ \emph {et~al.}(2021)\citenamefont {Wang},
  \citenamefont {Wiebe}, \citenamefont {Zhong}, \citenamefont {Gu},\ and\
  \citenamefont {Wiesendanger}}]{PhysRevLett.126.076802}%
  \BibitemOpen
  \bibfield  {author} {\bibinfo {author} {\bibfnamefont {Dongfei}\ \bibnamefont
  {Wang}}, \bibinfo {author} {\bibfnamefont {Jens}\ \bibnamefont {Wiebe}},
  \bibinfo {author} {\bibfnamefont {Ruidan}\ \bibnamefont {Zhong}}, \bibinfo
  {author} {\bibfnamefont {Genda}\ \bibnamefont {Gu}}, \ and\ \bibinfo {author}
  {\bibfnamefont {Roland}\ \bibnamefont {Wiesendanger}},\ }\bibfield  {title}
  {\enquote {\bibinfo {title} {Spin-polarized {Y}u-{S}hiba-{R}usinov states in
  an iron-based superconductor},}\ }\href {\doibase
  10.1103/PhysRevLett.126.076802} {\bibfield  {journal} {\bibinfo  {journal}
  {Phys. Rev. Lett.}\ }\textbf {\bibinfo {volume} {126}},\ \bibinfo {pages}
  {076802} (\bibinfo {year} {2021})}\BibitemShut {NoStop}%
\bibitem [{\citenamefont {Yazdani}\ \emph {et~al.}(1999)\citenamefont
  {Yazdani}, \citenamefont {Howald}, \citenamefont {Lutz}, \citenamefont
  {Kapitulnik},\ and\ \citenamefont {Eigler}}]{PhysRevLett.83.176}%
  \BibitemOpen
  \bibfield  {author} {\bibinfo {author} {\bibfnamefont {Ali}\ \bibnamefont
  {Yazdani}}, \bibinfo {author} {\bibfnamefont {C.~M.}\ \bibnamefont {Howald}},
  \bibinfo {author} {\bibfnamefont {C.~P.}\ \bibnamefont {Lutz}}, \bibinfo
  {author} {\bibfnamefont {A.}~\bibnamefont {Kapitulnik}}, \ and\ \bibinfo
  {author} {\bibfnamefont {D.~M.}\ \bibnamefont {Eigler}},\ }\bibfield  {title}
  {\enquote {\bibinfo {title} {Impurity-induced bound excitations on the
  surface of
  {${\mathrm{Bi}}_{2}{\mathrm{Sr}}_{2}{\mathrm{CaCu}}_{2}{\mathrm{O}}_{8}$}},}\
  }\href {\doibase 10.1103/PhysRevLett.83.176} {\bibfield  {journal} {\bibinfo
  {journal} {Phys. Rev. Lett.}\ }\textbf {\bibinfo {volume} {83}},\ \bibinfo
  {pages} {176--179} (\bibinfo {year} {1999})}\BibitemShut {NoStop}%
\bibitem [{\citenamefont {Yazdani}\ \emph {et~al.}(1997)\citenamefont
  {Yazdani}, \citenamefont {Jones}, \citenamefont {Lutz}, \citenamefont
  {Crommie},\ and\ \citenamefont {Eigler}}]{doi:10.1126/science.275.5307.1767}%
  \BibitemOpen
  \bibfield  {author} {\bibinfo {author} {\bibfnamefont {Ali}\ \bibnamefont
  {Yazdani}}, \bibinfo {author} {\bibfnamefont {B.~A.}\ \bibnamefont {Jones}},
  \bibinfo {author} {\bibfnamefont {C.~P.}\ \bibnamefont {Lutz}}, \bibinfo
  {author} {\bibfnamefont {M.~F.}\ \bibnamefont {Crommie}}, \ and\ \bibinfo
  {author} {\bibfnamefont {D.~M.}\ \bibnamefont {Eigler}},\ }\bibfield  {title}
  {\enquote {\bibinfo {title} {Probing the local effects of magnetic impurities
  on superconductivity},}\ }\href {\doibase 10.1126/science.275.5307.1767}
  {\bibfield  {journal} {\bibinfo  {journal} {Science}\ }\textbf {\bibinfo
  {volume} {275}},\ \bibinfo {pages} {1767--1770} (\bibinfo {year}
  {1997})}\BibitemShut {NoStop}%
\bibitem [{\citenamefont {Yazdani}(2015)}]{Yazdani_2015}%
  \BibitemOpen
  \bibfield  {author} {\bibinfo {author} {\bibfnamefont {Ali}\ \bibnamefont
  {Yazdani}},\ }\bibfield  {title} {\enquote {\bibinfo {title} {Visualizing
  {M}ajorana fermions in a chain of magnetic atoms on a superconductor},}\
  }\href {\doibase 10.1088/0031-8949/2015/t164/014012} {\bibfield  {journal}
  {\bibinfo  {journal} {Physica Scripta}\ }\textbf {\bibinfo {volume} {T164}},\
  \bibinfo {pages} {014012} (\bibinfo {year} {2015})}\BibitemShut {NoStop}%
\bibitem [{\citenamefont {Nadj-Perge}\ \emph {et~al.}(2014)\citenamefont
  {Nadj-Perge}, \citenamefont {Drozdov}, \citenamefont {Li}, \citenamefont
  {Chen}, \citenamefont {Jeon}, \citenamefont {Seo}, \citenamefont {MacDonald},
  \citenamefont {Bernevig},\ and\ \citenamefont {Yazdani}}]{science}%
  \BibitemOpen
  \bibfield  {author} {\bibinfo {author} {\bibfnamefont {S.}~\bibnamefont
  {Nadj-Perge}}, \bibinfo {author} {\bibfnamefont {I.~K.}\ \bibnamefont
  {Drozdov}}, \bibinfo {author} {\bibfnamefont {J.}~\bibnamefont {Li}},
  \bibinfo {author} {\bibfnamefont {H.}~\bibnamefont {Chen}}, \bibinfo {author}
  {\bibfnamefont {S.}~\bibnamefont {Jeon}}, \bibinfo {author} {\bibfnamefont
  {J.}~\bibnamefont {Seo}}, \bibinfo {author} {\bibfnamefont {A.~H.}\
  \bibnamefont {MacDonald}}, \bibinfo {author} {\bibfnamefont {B.~A.}\
  \bibnamefont {Bernevig}}, \ and\ \bibinfo {author} {\bibfnamefont
  {A.}~\bibnamefont {Yazdani}},\ }\bibfield  {title} {\enquote {\bibinfo
  {title} {Observation of majorana fermions in ferromagnetic atomic chains on a
  superconductor},}\ }\href {\doibase 10.1126/science.1259327} {\bibfield
  {journal} {\bibinfo  {journal} {Science}\ }\textbf {\bibinfo {volume}
  {346}},\ \bibinfo {pages} {602–607} (\bibinfo {year} {2014})}\BibitemShut
  {NoStop}%
\bibitem [{\citenamefont {M\'enard}\ \emph {et~al.}(2017)\citenamefont
  {M\'enard}, \citenamefont {Guissart}, \citenamefont {Brun}, \citenamefont
  {Leriche}, \citenamefont {Trif}, \citenamefont {Debontridder}, \citenamefont
  {Demaille}, \citenamefont {Roditchev}, \citenamefont {Simon},\ and\
  \citenamefont {Cren}}]{Menard2017}%
  \BibitemOpen
  \bibfield  {author} {\bibinfo {author} {\bibfnamefont {Gerbold~C.}\
  \bibnamefont {M\'enard}}, \bibinfo {author} {\bibfnamefont {S\'ebastien}\
  \bibnamefont {Guissart}}, \bibinfo {author} {\bibfnamefont {Christophe}\
  \bibnamefont {Brun}}, \bibinfo {author} {\bibfnamefont {Rapha\"el~T.}\
  \bibnamefont {Leriche}}, \bibinfo {author} {\bibfnamefont {Mircea}\
  \bibnamefont {Trif}}, \bibinfo {author} {\bibfnamefont {Francois}\
  \bibnamefont {Debontridder}}, \bibinfo {author} {\bibfnamefont {Dominique}\
  \bibnamefont {Demaille}}, \bibinfo {author} {\bibfnamefont {Dimitri}\
  \bibnamefont {Roditchev}}, \bibinfo {author} {\bibfnamefont {Pascal}\
  \bibnamefont {Simon}}, \ and\ \bibinfo {author} {\bibfnamefont {Tristan}\
  \bibnamefont {Cren}},\ }\bibfield  {title} {\enquote {\bibinfo {title}
  {Two-dimensional topological superconductivity in {Pb/Co/Si(111)}},}\ }\href
  {\doibase 10.1038/s41467-017-02192-x} {\bibfield  {journal} {\bibinfo
  {journal} {Nature Communications}\ }\textbf {\bibinfo {volume} {8}},\
  \bibinfo {pages} {2040} (\bibinfo {year} {2017})}\BibitemShut {NoStop}%
\bibitem [{\citenamefont {Hor}\ \emph {et~al.}(2010)\citenamefont {Hor},
  \citenamefont {Williams}, \citenamefont {Checkelsky}, \citenamefont
  {Roushan}, \citenamefont {Seo}, \citenamefont {Xu}, \citenamefont
  {Zandbergen}, \citenamefont {Yazdani}, \citenamefont {Ong},\ and\
  \citenamefont {Cava}}]{PhysRevLett.104.057001}%
  \BibitemOpen
  \bibfield  {author} {\bibinfo {author} {\bibfnamefont {Y.~S.}\ \bibnamefont
  {Hor}}, \bibinfo {author} {\bibfnamefont {A.~J.}\ \bibnamefont {Williams}},
  \bibinfo {author} {\bibfnamefont {J.~G.}\ \bibnamefont {Checkelsky}},
  \bibinfo {author} {\bibfnamefont {P.}~\bibnamefont {Roushan}}, \bibinfo
  {author} {\bibfnamefont {J.}~\bibnamefont {Seo}}, \bibinfo {author}
  {\bibfnamefont {Q.}~\bibnamefont {Xu}}, \bibinfo {author} {\bibfnamefont
  {H.~W.}\ \bibnamefont {Zandbergen}}, \bibinfo {author} {\bibfnamefont
  {A.}~\bibnamefont {Yazdani}}, \bibinfo {author} {\bibfnamefont {N.~P.}\
  \bibnamefont {Ong}}, \ and\ \bibinfo {author} {\bibfnamefont {R.~J.}\
  \bibnamefont {Cava}},\ }\bibfield  {title} {\enquote {\bibinfo {title}
  {Superconductivity in {${\mathrm{Cu}}_{x}{\mathrm{Bi}}_{2}{\mathrm{Se}}_{3}$}
  and its implications for pairing in the undoped topological insulator},}\
  }\href {\doibase 10.1103/PhysRevLett.104.057001} {\bibfield  {journal}
  {\bibinfo  {journal} {Phys. Rev. Lett.}\ }\textbf {\bibinfo {volume} {104}},\
  \bibinfo {pages} {057001} (\bibinfo {year} {2010})}\BibitemShut {NoStop}%
\bibitem [{\citenamefont {Fu}\ and\ \citenamefont
  {Berg}(2010)}]{PhysRevLett.105.097001}%
  \BibitemOpen
  \bibfield  {author} {\bibinfo {author} {\bibfnamefont {Liang}\ \bibnamefont
  {Fu}}\ and\ \bibinfo {author} {\bibfnamefont {Erez}\ \bibnamefont {Berg}},\
  }\bibfield  {title} {\enquote {\bibinfo {title} {Odd-parity topological
  superconductors: Theory and application to
  {${\mathrm{Cu}}_{x}{\mathrm{Bi}}_{2}{\mathrm{Se}}_{3}$}},}\ }\href {\doibase
  10.1103/PhysRevLett.105.097001} {\bibfield  {journal} {\bibinfo  {journal}
  {Phys. Rev. Lett.}\ }\textbf {\bibinfo {volume} {105}},\ \bibinfo {pages}
  {097001} (\bibinfo {year} {2010})}\BibitemShut {NoStop}%
\bibitem [{\citenamefont {Wray}\ \emph {et~al.}(2011)\citenamefont {Wray},
  \citenamefont {Xu}, \citenamefont {Xia}, \citenamefont {Qian}, \citenamefont
  {Fedorov}, \citenamefont {Lin}, \citenamefont {Bansil}, \citenamefont {Fu},
  \citenamefont {Hor}, \citenamefont {Cava},\ and\ \citenamefont
  {Hasan}}]{PhysRevB.83.224516}%
  \BibitemOpen
  \bibfield  {author} {\bibinfo {author} {\bibfnamefont {L.~Andrew}\
  \bibnamefont {Wray}}, \bibinfo {author} {\bibfnamefont {Suyang}\ \bibnamefont
  {Xu}}, \bibinfo {author} {\bibfnamefont {Yuqi}\ \bibnamefont {Xia}}, \bibinfo
  {author} {\bibfnamefont {Dong}\ \bibnamefont {Qian}}, \bibinfo {author}
  {\bibfnamefont {Alexei~V.}\ \bibnamefont {Fedorov}}, \bibinfo {author}
  {\bibfnamefont {Hsin}\ \bibnamefont {Lin}}, \bibinfo {author} {\bibfnamefont
  {Arun}\ \bibnamefont {Bansil}}, \bibinfo {author} {\bibfnamefont {Liang}\
  \bibnamefont {Fu}}, \bibinfo {author} {\bibfnamefont {Yew~San}\ \bibnamefont
  {Hor}}, \bibinfo {author} {\bibfnamefont {Robert~J.}\ \bibnamefont {Cava}}, \
  and\ \bibinfo {author} {\bibfnamefont {M.~Zahid}\ \bibnamefont {Hasan}},\
  }\bibfield  {title} {\enquote {\bibinfo {title} {Spin-orbital ground states
  of superconducting doped topological insulators: A majorana platform},}\
  }\href {\doibase 10.1103/PhysRevB.83.224516} {\bibfield  {journal} {\bibinfo
  {journal} {Phys. Rev. B}\ }\textbf {\bibinfo {volume} {83}},\ \bibinfo
  {pages} {224516} (\bibinfo {year} {2011})}\BibitemShut {NoStop}%
\bibitem [{\citenamefont {Kaladzhyan}\ \emph {et~al.}(2018)\citenamefont
  {Kaladzhyan}, \citenamefont {Bena},\ and\ \citenamefont
  {Simon}}]{nontriviality}%
  \BibitemOpen
  \bibfield  {author} {\bibinfo {author} {\bibfnamefont {V.}~\bibnamefont
  {Kaladzhyan}}, \bibinfo {author} {\bibfnamefont {Cristina}\ \bibnamefont
  {Bena}}, \ and\ \bibinfo {author} {\bibfnamefont {Pascal}\ \bibnamefont
  {Simon}},\ }\bibfield  {title} {\enquote {\bibinfo {title} {Topology from
  triviality},}\ }\href {\doibase 10.1103/PhysRevB.97.104512} {\bibfield
  {journal} {\bibinfo  {journal} {Phys. Rev. B}\ }\textbf {\bibinfo {volume}
  {97}},\ \bibinfo {pages} {104512} (\bibinfo {year} {2018})}\BibitemShut
  {NoStop}%
\bibitem [{\citenamefont {Sedlmayr}\ and\ \citenamefont {Bena}(2015)}]{theta}%
  \BibitemOpen
  \bibfield  {author} {\bibinfo {author} {\bibfnamefont {N.}~\bibnamefont
  {Sedlmayr}}\ and\ \bibinfo {author} {\bibfnamefont {C.}~\bibnamefont
  {Bena}},\ }\bibfield  {title} {\enquote {\bibinfo {title} {Visualizing
  majorana bound states in one and two dimensions using the generalized
  majorana polarization},}\ }\href {\doibase 10.1103/PhysRevB.92.115115}
  {\bibfield  {journal} {\bibinfo  {journal} {Phys. Rev. B}\ }\textbf {\bibinfo
  {volume} {92}},\ \bibinfo {pages} {115115} (\bibinfo {year}
  {2015})}\BibitemShut {NoStop}%
\bibitem [{\citenamefont {Vazifeh}\ and\ \citenamefont
  {Franz}(2013)}]{phi_Franz}%
  \BibitemOpen
  \bibfield  {author} {\bibinfo {author} {\bibfnamefont {M.~M.}\ \bibnamefont
  {Vazifeh}}\ and\ \bibinfo {author} {\bibfnamefont {M.}~\bibnamefont
  {Franz}},\ }\bibfield  {title} {\enquote {\bibinfo {title} {Self-organized
  topological state with majorana fermions},}\ }\href {\doibase
  10.1103/PhysRevLett.111.206802} {\bibfield  {journal} {\bibinfo  {journal}
  {Phys. Rev. Lett.}\ }\textbf {\bibinfo {volume} {111}},\ \bibinfo {pages}
  {206802} (\bibinfo {year} {2013})}\BibitemShut {NoStop}%
\bibitem [{\citenamefont {Nandy}\ \emph {et~al.}(2016)\citenamefont {Nandy},
  \citenamefont {Kiselev},\ and\ \citenamefont
  {Bl\"ugel}}]{PhysRevLett.116.177202}%
  \BibitemOpen
  \bibfield  {author} {\bibinfo {author} {\bibfnamefont {Ashis~Kumar}\
  \bibnamefont {Nandy}}, \bibinfo {author} {\bibfnamefont {Nikolai~S.}\
  \bibnamefont {Kiselev}}, \ and\ \bibinfo {author} {\bibfnamefont {Stefan}\
  \bibnamefont {Bl\"ugel}},\ }\bibfield  {title} {\enquote {\bibinfo {title}
  {Interlayer exchange coupling: A general scheme turning chiral magnets into
  magnetic multilayers carrying atomic-scale {S}kyrmions},}\ }\href {\doibase
  10.1103/PhysRevLett.116.177202} {\bibfield  {journal} {\bibinfo  {journal}
  {Phys. Rev. Lett.}\ }\textbf {\bibinfo {volume} {116}},\ \bibinfo {pages}
  {177202} (\bibinfo {year} {2016})}\BibitemShut {NoStop}%
\bibitem [{\citenamefont {Singh}\ \emph {et~al.}(2020)\citenamefont {Singh},
  \citenamefont {Singh}, \citenamefont {Pradhan}, \citenamefont {Srihari},
  \citenamefont {Poswal}, \citenamefont {Nath}, \citenamefont {Nandy},\ and\
  \citenamefont {Nayak}}]{PhysRevResearch.2.043366}%
  \BibitemOpen
  \bibfield  {author} {\bibinfo {author} {\bibfnamefont {Charanpreet}\
  \bibnamefont {Singh}}, \bibinfo {author} {\bibfnamefont {Vikram}\
  \bibnamefont {Singh}}, \bibinfo {author} {\bibfnamefont {Gyandeep}\
  \bibnamefont {Pradhan}}, \bibinfo {author} {\bibfnamefont {Velaga}\
  \bibnamefont {Srihari}}, \bibinfo {author} {\bibfnamefont {Himanshu~Kumar}\
  \bibnamefont {Poswal}}, \bibinfo {author} {\bibfnamefont {Ramesh}\
  \bibnamefont {Nath}}, \bibinfo {author} {\bibfnamefont {Ashis~K.}\
  \bibnamefont {Nandy}}, \ and\ \bibinfo {author} {\bibfnamefont {Ajaya~K.}\
  \bibnamefont {Nayak}},\ }\bibfield  {title} {\enquote {\bibinfo {title}
  {Pressure controlled trimerization for switching of anomalous hall effect in
  triangular antiferromagnet {${\mathrm{Mn}}_{3}\mathrm{Sn}$}},}\ }\href
  {\doibase 10.1103/PhysRevResearch.2.043366} {\bibfield  {journal} {\bibinfo
  {journal} {Phys. Rev. Research}\ }\textbf {\bibinfo {volume} {2}},\ \bibinfo
  {pages} {043366} (\bibinfo {year} {2020})}\BibitemShut {NoStop}%
\bibitem [{\citenamefont {Kitagawa}\ \emph {et~al.}(2010)\citenamefont
  {Kitagawa}, \citenamefont {Berg}, \citenamefont {Rudner},\ and\ \citenamefont
  {Demler}}]{Demler2010}%
  \BibitemOpen
  \bibfield  {author} {\bibinfo {author} {\bibfnamefont {Takuya}\ \bibnamefont
  {Kitagawa}}, \bibinfo {author} {\bibfnamefont {Erez}\ \bibnamefont {Berg}},
  \bibinfo {author} {\bibfnamefont {Mark}\ \bibnamefont {Rudner}}, \ and\
  \bibinfo {author} {\bibfnamefont {Eugene}\ \bibnamefont {Demler}},\
  }\bibfield  {title} {\enquote {\bibinfo {title} {Topological characterization
  of periodically driven quantum systems},}\ }\href {\doibase
  10.1103/PhysRevB.82.235114} {\bibfield  {journal} {\bibinfo  {journal} {Phys.
  Rev. B}\ }\textbf {\bibinfo {volume} {82}},\ \bibinfo {pages} {235114}
  (\bibinfo {year} {2010})}\BibitemShut {NoStop}%
\bibitem [{\citenamefont {Tewari}\ and\ \citenamefont
  {Sau}(2012)}]{PhysRevLett.109.150408}%
  \BibitemOpen
  \bibfield  {author} {\bibinfo {author} {\bibfnamefont {Sumanta}\ \bibnamefont
  {Tewari}}\ and\ \bibinfo {author} {\bibfnamefont {Jay~D.}\ \bibnamefont
  {Sau}},\ }\bibfield  {title} {\enquote {\bibinfo {title} {Topological
  invariants for spin-orbit coupled superconductor nanowires},}\ }\href
  {\doibase 10.1103/PhysRevLett.109.150408} {\bibfield  {journal} {\bibinfo
  {journal} {Phys. Rev. Lett.}\ }\textbf {\bibinfo {volume} {109}},\ \bibinfo
  {pages} {150408} (\bibinfo {year} {2012})}\BibitemShut {NoStop}%
\bibitem [{\citenamefont {Singh}\ \emph {et~al.}()\citenamefont {Singh},
  \citenamefont {Jamaluddin}, \citenamefont {Nandy}, \citenamefont {Tokunaga},
  \citenamefont {Avdeev},\ and\ \citenamefont {Nayak}}]{Charanpreet}%
  \BibitemOpen
  \bibfield  {author} {\bibinfo {author} {\bibfnamefont {Charanpreet}\
  \bibnamefont {Singh}}, \bibinfo {author} {\bibfnamefont {Sk}~\bibnamefont
  {Jamaluddin}}, \bibinfo {author} {\bibfnamefont {Ashis~K.}\ \bibnamefont
  {Nandy}}, \bibinfo {author} {\bibfnamefont {Masashi}\ \bibnamefont
  {Tokunaga}}, \bibinfo {author} {\bibfnamefont {Maxim}\ \bibnamefont
  {Avdeev}}, \ and\ \bibinfo {author} {\bibfnamefont {Ajaya~K.}\ \bibnamefont
  {Nayak}},\ }\href@noop {} {\enquote {\bibinfo {title} {Higher order exchange
  driven noncoplanar magnetic state and large anomalous hall effects in
  electron doped kagome magnet {Mn$_3$Sn}},}\ }\Eprint
  {http://arxiv.org/abs/2211.12722} {arXiv:2211.12722} \BibitemShut {NoStop}%
\bibitem [{\citenamefont {Zheng}\ \emph {et~al.}(2021)\citenamefont {Zheng},
  \citenamefont {Wang}, \citenamefont {Zhu}, \citenamefont {Tan}, \citenamefont
  {Wang}, \citenamefont {Albarakati}, \citenamefont {Aloufi}, \citenamefont
  {Algarni}, \citenamefont {Farrar}, \citenamefont {Wu}, \citenamefont {Yao},
  \citenamefont {Tian}, \citenamefont {Zhou},\ and\ \citenamefont
  {Wang}}]{Zheng2021}%
  \BibitemOpen
  \bibfield  {author} {\bibinfo {author} {\bibfnamefont {Guolin}\ \bibnamefont
  {Zheng}}, \bibinfo {author} {\bibfnamefont {Maoyuan}\ \bibnamefont {Wang}},
  \bibinfo {author} {\bibfnamefont {Xiangde}\ \bibnamefont {Zhu}}, \bibinfo
  {author} {\bibfnamefont {Cheng}\ \bibnamefont {Tan}}, \bibinfo {author}
  {\bibfnamefont {Jie}\ \bibnamefont {Wang}}, \bibinfo {author} {\bibfnamefont
  {Sultan}\ \bibnamefont {Albarakati}}, \bibinfo {author} {\bibfnamefont
  {Nuriyah}\ \bibnamefont {Aloufi}}, \bibinfo {author} {\bibfnamefont {Meri}\
  \bibnamefont {Algarni}}, \bibinfo {author} {\bibfnamefont {Lawrence}\
  \bibnamefont {Farrar}}, \bibinfo {author} {\bibfnamefont {Min}\ \bibnamefont
  {Wu}}, \bibinfo {author} {\bibfnamefont {Yugui}\ \bibnamefont {Yao}},
  \bibinfo {author} {\bibfnamefont {Mingliang}\ \bibnamefont {Tian}}, \bibinfo
  {author} {\bibfnamefont {Jianhui}\ \bibnamefont {Zhou}}, \ and\ \bibinfo
  {author} {\bibfnamefont {Lan}\ \bibnamefont {Wang}},\ }\bibfield  {title}
  {\enquote {\bibinfo {title} {Tailoring dzyaloshinskii--moriya interaction in
  a transition metal dichalcogenide by dual-intercalation},}\ }\href {\doibase
  10.1038/s41467-021-23658-z} {\bibfield  {journal} {\bibinfo  {journal}
  {Nature Communications}\ }\textbf {\bibinfo {volume} {12}},\ \bibinfo {pages}
  {3639} (\bibinfo {year} {2021})}\BibitemShut {NoStop}%
\bibitem [{\citenamefont {Ma}\ \emph {et~al.}()\citenamefont {Ma},
  \citenamefont {Sharma}, \citenamefont {Saha}, \citenamefont {Srivastava},
  \citenamefont {Werner}, \citenamefont {Vir}, \citenamefont {Kumar},
  \citenamefont {Felser},\ and\ \citenamefont
  {Parkin}}]{https://doi.org/10.1002/adma.202002043}%
  \BibitemOpen
  \bibfield  {author} {\bibinfo {author} {\bibfnamefont {Tianping}\
  \bibnamefont {Ma}}, \bibinfo {author} {\bibfnamefont {Ankit~K.}\ \bibnamefont
  {Sharma}}, \bibinfo {author} {\bibfnamefont {Rana}\ \bibnamefont {Saha}},
  \bibinfo {author} {\bibfnamefont {Abhay~K.}\ \bibnamefont {Srivastava}},
  \bibinfo {author} {\bibfnamefont {Peter}\ \bibnamefont {Werner}}, \bibinfo
  {author} {\bibfnamefont {Praveen}\ \bibnamefont {Vir}}, \bibinfo {author}
  {\bibfnamefont {Vivek}\ \bibnamefont {Kumar}}, \bibinfo {author}
  {\bibfnamefont {Claudia}\ \bibnamefont {Felser}}, \ and\ \bibinfo {author}
  {\bibfnamefont {Stuart S.~P.}\ \bibnamefont {Parkin}},\ }\bibfield  {title}
  {\enquote {\bibinfo {title} {Tunable magnetic antiskyrmion size and helical
  period from nanometers to micrometers in a {$\rm D$$\rm 2d$} heusler
  compound},}\ }\href {\doibase https://doi.org/10.1002/adma.202002043}
  {\bibfield  {journal} {\bibinfo  {journal} {Advanced Materials}\ }\textbf
  {\bibinfo {volume} {32}},\ \bibinfo {pages} {2002043}}\BibitemShut {NoStop}%
\bibitem [{\citenamefont {Hsu}\ \emph {et~al.}(2016)\citenamefont {Hsu},
  \citenamefont {Finco}, \citenamefont {Schmidt}, \citenamefont {Kubetzka},
  \citenamefont {von Bergmann},\ and\ \citenamefont
  {Wiesendanger}}]{PhysRevLett.116.017201}%
  \BibitemOpen
  \bibfield  {author} {\bibinfo {author} {\bibfnamefont {Pin-Jui}\ \bibnamefont
  {Hsu}}, \bibinfo {author} {\bibfnamefont {Aurore}\ \bibnamefont {Finco}},
  \bibinfo {author} {\bibfnamefont {Lorenz}\ \bibnamefont {Schmidt}}, \bibinfo
  {author} {\bibfnamefont {Andr\'e}\ \bibnamefont {Kubetzka}}, \bibinfo
  {author} {\bibfnamefont {Kirsten}\ \bibnamefont {von Bergmann}}, \ and\
  \bibinfo {author} {\bibfnamefont {Roland}\ \bibnamefont {Wiesendanger}},\
  }\bibfield  {title} {\enquote {\bibinfo {title} {Guiding spin spirals by
  local uniaxial strain relief},}\ }\href {\doibase
  10.1103/PhysRevLett.116.017201} {\bibfield  {journal} {\bibinfo  {journal}
  {Phys. Rev. Lett.}\ }\textbf {\bibinfo {volume} {116}},\ \bibinfo {pages}
  {017201} (\bibinfo {year} {2016})}\BibitemShut {NoStop}%
\bibitem [{\citenamefont {Yang}\ \emph {et~al.}(2018)\citenamefont {Yang},
  \citenamefont {Boulle}, \citenamefont {Cros}, \citenamefont {Fert},\ and\
  \citenamefont {Chshiev}}]{Yang2018}%
  \BibitemOpen
  \bibfield  {author} {\bibinfo {author} {\bibfnamefont {Hongxin}\ \bibnamefont
  {Yang}}, \bibinfo {author} {\bibfnamefont {Olivier}\ \bibnamefont {Boulle}},
  \bibinfo {author} {\bibfnamefont {Vincent}\ \bibnamefont {Cros}}, \bibinfo
  {author} {\bibfnamefont {Albert}\ \bibnamefont {Fert}}, \ and\ \bibinfo
  {author} {\bibfnamefont {Mairbek}\ \bibnamefont {Chshiev}},\ }\bibfield
  {title} {\enquote {\bibinfo {title} {Controlling dzyaloshinskii-moriya
  interaction via chirality dependent atomic-layer stacking, insulator capping
  and electric field},}\ }\href {\doibase 10.1038/s41598-018-30063-y}
  {\bibfield  {journal} {\bibinfo  {journal} {Scientific Reports}\ }\textbf
  {\bibinfo {volume} {8}},\ \bibinfo {pages} {12356} (\bibinfo {year}
  {2018})}\BibitemShut {NoStop}%
\bibitem [{\citenamefont {Khajetoorians}\ \emph {et~al.}(2016)\citenamefont
  {Khajetoorians}, \citenamefont {Steinbrecher}, \citenamefont {Ternes},
  \citenamefont {Bouhassoune}, \citenamefont {dos Santos~Dias}, \citenamefont
  {Lounis}, \citenamefont {Wiebe},\ and\ \citenamefont
  {Wiesendanger}}]{Khajetoorians2016}%
  \BibitemOpen
  \bibfield  {author} {\bibinfo {author} {\bibfnamefont {A.~A.}\ \bibnamefont
  {Khajetoorians}}, \bibinfo {author} {\bibfnamefont {M.}~\bibnamefont
  {Steinbrecher}}, \bibinfo {author} {\bibfnamefont {M.}~\bibnamefont
  {Ternes}}, \bibinfo {author} {\bibfnamefont {M.}~\bibnamefont {Bouhassoune}},
  \bibinfo {author} {\bibfnamefont {M.}~\bibnamefont {dos Santos~Dias}},
  \bibinfo {author} {\bibfnamefont {S.}~\bibnamefont {Lounis}}, \bibinfo
  {author} {\bibfnamefont {J.}~\bibnamefont {Wiebe}}, \ and\ \bibinfo {author}
  {\bibfnamefont {R.}~\bibnamefont {Wiesendanger}},\ }\bibfield  {title}
  {\enquote {\bibinfo {title} {Tailoring the chiral magnetic interaction
  between two individual atoms},}\ }\href {\doibase 10.1038/ncomms10620}
  {\bibfield  {journal} {\bibinfo  {journal} {Nature Communications}\ }\textbf
  {\bibinfo {volume} {7}},\ \bibinfo {pages} {10620} (\bibinfo {year}
  {2016})}\BibitemShut {NoStop}%
\bibitem [{\citenamefont {Zyuzin}\ and\ \citenamefont
  {Loss}(2014)}]{PhysRevB.90.125443}%
  \BibitemOpen
  \bibfield  {author} {\bibinfo {author} {\bibfnamefont {Alexander~A.}\
  \bibnamefont {Zyuzin}}\ and\ \bibinfo {author} {\bibfnamefont {Daniel}\
  \bibnamefont {Loss}},\ }\bibfield  {title} {\enquote {\bibinfo {title}
  {{RKKY} interaction on surfaces of topological insulators with
  superconducting proximity effect},}\ }\href {\doibase
  10.1103/PhysRevB.90.125443} {\bibfield  {journal} {\bibinfo  {journal} {Phys.
  Rev. B}\ }\textbf {\bibinfo {volume} {90}},\ \bibinfo {pages} {125443}
  (\bibinfo {year} {2014})}\BibitemShut {NoStop}%
\bibitem [{\citenamefont {Kaladzhyan}\ \emph {et~al.}(2016)\citenamefont
  {Kaladzhyan}, \citenamefont {Bena},\ and\ \citenamefont
  {Simon}}]{Kaladzhyan_2016}%
  \BibitemOpen
  \bibfield  {author} {\bibinfo {author} {\bibfnamefont {V}~\bibnamefont
  {Kaladzhyan}}, \bibinfo {author} {\bibfnamefont {C}~\bibnamefont {Bena}}, \
  and\ \bibinfo {author} {\bibfnamefont {P}~\bibnamefont {Simon}},\ }\bibfield
  {title} {\enquote {\bibinfo {title} {Asymptotic behavior of impurity-induced
  bound states in low-dimensional topological superconductors},}\ }\href
  {\doibase 10.1088/0953-8984/28/48/485701} {\bibfield  {journal} {\bibinfo
  {journal} {Journal of Physics: Condensed Matter}\ }\textbf {\bibinfo {volume}
  {28}},\ \bibinfo {pages} {485701} (\bibinfo {year} {2016})}\BibitemShut
  {NoStop}%
\bibitem [{\citenamefont {Uchoa}\ and\ \citenamefont
  {Castro~Neto}(2007)}]{CastroNeto}%
  \BibitemOpen
  \bibfield  {author} {\bibinfo {author} {\bibfnamefont {Bruno}\ \bibnamefont
  {Uchoa}}\ and\ \bibinfo {author} {\bibfnamefont {A.~H.}\ \bibnamefont
  {Castro~Neto}},\ }\bibfield  {title} {\enquote {\bibinfo {title}
  {Superconducting states of pure and doped graphene},}\ }\href {\doibase
  10.1103/PhysRevLett.98.146801} {\bibfield  {journal} {\bibinfo  {journal}
  {Phys. Rev. Lett.}\ }\textbf {\bibinfo {volume} {98}},\ \bibinfo {pages}
  {146801} (\bibinfo {year} {2007})}\BibitemShut {NoStop}%
\bibitem [{\citenamefont {Zhang}\ \emph {et~al.}(2019)\citenamefont {Zhang},
  \citenamefont {Wang},\ and\ \citenamefont {Song}}]{Zhang2019}%
  \BibitemOpen
  \bibfield  {author} {\bibinfo {author} {\bibfnamefont {K.~L.}\ \bibnamefont
  {Zhang}}, \bibinfo {author} {\bibfnamefont {P.}~\bibnamefont {Wang}}, \ and\
  \bibinfo {author} {\bibfnamefont {Z.}~\bibnamefont {Song}},\ }\bibfield
  {title} {\enquote {\bibinfo {title} {Majorana flat band edge modes of
  topological gapless phase in 2d kitaev square lattice},}\ }\href {\doibase
  10.1038/s41598-019-41529-y} {\bibfield  {journal} {\bibinfo  {journal}
  {Scientific Reports}\ }\textbf {\bibinfo {volume} {9}},\ \bibinfo {pages}
  {4978} (\bibinfo {year} {2019})}\BibitemShut {NoStop}%
\bibitem [{\citenamefont {Wang}\ \emph {et~al.}(2017)\citenamefont {Wang},
  \citenamefont {Lin}, \citenamefont {Zhang},\ and\ \citenamefont
  {Song}}]{Wang2017}%
  \BibitemOpen
  \bibfield  {author} {\bibinfo {author} {\bibfnamefont {P.}~\bibnamefont
  {Wang}}, \bibinfo {author} {\bibfnamefont {S.}~\bibnamefont {Lin}}, \bibinfo
  {author} {\bibfnamefont {G.}~\bibnamefont {Zhang}}, \ and\ \bibinfo {author}
  {\bibfnamefont {Z.}~\bibnamefont {Song}},\ }\bibfield  {title} {\enquote
  {\bibinfo {title} {Topological gapless phase in kitaev model on square
  lattice},}\ }\href {\doibase 10.1038/s41598-017-17334-w} {\bibfield
  {journal} {\bibinfo  {journal} {Scientific Reports}\ }\textbf {\bibinfo
  {volume} {7}},\ \bibinfo {pages} {17179} (\bibinfo {year}
  {2017})}\BibitemShut {NoStop}%
\bibitem [{\citenamefont {K\"uster}\ \emph {et~al.}(2021)\citenamefont
  {K\"uster}, \citenamefont {Brinker}, \citenamefont {Lounis}, \citenamefont
  {Parkin},\ and\ \citenamefont {Sessi}}]{Kaesster2021}%
  \BibitemOpen
  \bibfield  {author} {\bibinfo {author} {\bibfnamefont {Felix}\ \bibnamefont
  {K\"uster}}, \bibinfo {author} {\bibfnamefont {Sascha}\ \bibnamefont
  {Brinker}}, \bibinfo {author} {\bibfnamefont {Samir}\ \bibnamefont {Lounis}},
  \bibinfo {author} {\bibfnamefont {Stuart S.~P.}\ \bibnamefont {Parkin}}, \
  and\ \bibinfo {author} {\bibfnamefont {Paolo}\ \bibnamefont {Sessi}},\
  }\bibfield  {title} {\enquote {\bibinfo {title} {Long range and highly
  tunable interaction between local spins coupled to a superconducting
  condensate},}\ }\href {\doibase 10.1038/s41467-021-26802-x} {\bibfield
  {journal} {\bibinfo  {journal} {Nature Communications}\ }\textbf {\bibinfo
  {volume} {12}},\ \bibinfo {pages} {6722} (\bibinfo {year}
  {2021})}\BibitemShut {NoStop}%
\bibitem [{\citenamefont {K\"uster}\ \emph {et~al.}(2022)\citenamefont
  {K\"uster}, \citenamefont {Brinker}, \citenamefont {Hess}, \citenamefont
  {Loss}, \citenamefont {Parkin}, \citenamefont {Klinovaja}, \citenamefont
  {Lounis},\ and\ \citenamefont {Sessi}}]{doi:10.1073/pnas.2210589119}%
  \BibitemOpen
  \bibfield  {author} {\bibinfo {author} {\bibfnamefont {Felix}\ \bibnamefont
  {K\"uster}}, \bibinfo {author} {\bibfnamefont {Sascha}\ \bibnamefont
  {Brinker}}, \bibinfo {author} {\bibfnamefont {Richard}\ \bibnamefont {Hess}},
  \bibinfo {author} {\bibfnamefont {Daniel}\ \bibnamefont {Loss}}, \bibinfo
  {author} {\bibfnamefont {Stuart S.~P.}\ \bibnamefont {Parkin}}, \bibinfo
  {author} {\bibfnamefont {Jelena}\ \bibnamefont {Klinovaja}}, \bibinfo
  {author} {\bibfnamefont {Samir}\ \bibnamefont {Lounis}}, \ and\ \bibinfo
  {author} {\bibfnamefont {Paolo}\ \bibnamefont {Sessi}},\ }\bibfield  {title}
  {\enquote {\bibinfo {title} {Non-majorana modes in diluted spin chains
  proximitized to a superconductor},}\ }\href {\doibase
  10.1073/pnas.2210589119} {\bibfield  {journal} {\bibinfo  {journal}
  {Proceedings of the National Academy of Sciences}\ }\textbf {\bibinfo
  {volume} {119}},\ \bibinfo {pages} {e2210589119} (\bibinfo {year}
  {2022})}\BibitemShut {NoStop}%
\bibitem [{\citenamefont {Schneider}\ \emph {et~al.}(2021)\citenamefont
  {Schneider}, \citenamefont {Beck}, \citenamefont {Posske}, \citenamefont
  {Crawford}, \citenamefont {Mascot}, \citenamefont {Rachel}, \citenamefont
  {Wiesendanger},\ and\ \citenamefont {Wiebe}}]{Schneider2021}%
  \BibitemOpen
  \bibfield  {author} {\bibinfo {author} {\bibfnamefont {Lucas}\ \bibnamefont
  {Schneider}}, \bibinfo {author} {\bibfnamefont {Philip}\ \bibnamefont
  {Beck}}, \bibinfo {author} {\bibfnamefont {Thore}\ \bibnamefont {Posske}},
  \bibinfo {author} {\bibfnamefont {Daniel}\ \bibnamefont {Crawford}}, \bibinfo
  {author} {\bibfnamefont {Eric}\ \bibnamefont {Mascot}}, \bibinfo {author}
  {\bibfnamefont {Stephan}\ \bibnamefont {Rachel}}, \bibinfo {author}
  {\bibfnamefont {Roland}\ \bibnamefont {Wiesendanger}}, \ and\ \bibinfo
  {author} {\bibfnamefont {Jens}\ \bibnamefont {Wiebe}},\ }\bibfield  {title}
  {\enquote {\bibinfo {title} {Topological {S}hiba bands in artificial spin
  chains on superconductors},}\ }\href {\doibase 10.1038/s41567-021-01234-y}
  {\bibfield  {journal} {\bibinfo  {journal} {Nature Physics}\ }\textbf
  {\bibinfo {volume} {17}},\ \bibinfo {pages} {943--948} (\bibinfo {year}
  {2021})}\BibitemShut {NoStop}%
\bibitem [{\citenamefont {Beck}\ \emph {et~al.}(2021)\citenamefont {Beck},
  \citenamefont {Schneider}, \citenamefont {R\'{o}zsa}, \citenamefont
  {Palot\'{a}s}, \citenamefont {L\'{a}szl\'{o}ffy}, \citenamefont {Szunyogh},
  \citenamefont {Wiebe},\ and\ \citenamefont {Wiesendanger}}]{Beck2021}%
  \BibitemOpen
  \bibfield  {author} {\bibinfo {author} {\bibfnamefont {Philip}\ \bibnamefont
  {Beck}}, \bibinfo {author} {\bibfnamefont {Lucas}\ \bibnamefont {Schneider}},
  \bibinfo {author} {\bibfnamefont {Levente}\ \bibnamefont {R\'{o}zsa}},
  \bibinfo {author} {\bibfnamefont {Kriszti\'{a}n}\ \bibnamefont
  {Palot\'{a}s}}, \bibinfo {author} {\bibfnamefont {Andr\'{a}s}\ \bibnamefont
  {L\'{a}szl\'{o}ffy}}, \bibinfo {author} {\bibfnamefont {L\'{a}szl\'{o}}\
  \bibnamefont {Szunyogh}}, \bibinfo {author} {\bibfnamefont {Jens}\
  \bibnamefont {Wiebe}}, \ and\ \bibinfo {author} {\bibfnamefont {Roland}\
  \bibnamefont {Wiesendanger}},\ }\bibfield  {title} {\enquote {\bibinfo
  {title} {Spin-orbit coupling induced splitting of {Y}u-{S}hiba-{R}usinov
  states in antiferromagnetic dimers},}\ }\href {\doibase
  10.1038/s41467-021-22261-6} {\bibfield  {journal} {\bibinfo  {journal}
  {Nature Communications}\ }\textbf {\bibinfo {volume} {12}},\ \bibinfo {pages}
  {2040} (\bibinfo {year} {2021})}\BibitemShut {NoStop}%
\bibitem [{\citenamefont {Schneider}\ \emph {et~al.}(2022)\citenamefont
  {Schneider}, \citenamefont {Beck}, \citenamefont {Neuhaus-Steinmetz},
  \citenamefont {R\'{o}zsa}, \citenamefont {Posske}, \citenamefont {Wiebe},\
  and\ \citenamefont {Wiesendanger}}]{Schneider2022}%
  \BibitemOpen
  \bibfield  {author} {\bibinfo {author} {\bibfnamefont {Lucas}\ \bibnamefont
  {Schneider}}, \bibinfo {author} {\bibfnamefont {Philip}\ \bibnamefont
  {Beck}}, \bibinfo {author} {\bibfnamefont {Jannis}\ \bibnamefont
  {Neuhaus-Steinmetz}}, \bibinfo {author} {\bibfnamefont {Levente}\
  \bibnamefont {R\'{o}zsa}}, \bibinfo {author} {\bibfnamefont {Thore}\
  \bibnamefont {Posske}}, \bibinfo {author} {\bibfnamefont {Jens}\ \bibnamefont
  {Wiebe}}, \ and\ \bibinfo {author} {\bibfnamefont {Roland}\ \bibnamefont
  {Wiesendanger}},\ }\bibfield  {title} {\enquote {\bibinfo {title} {Precursors
  of majorana modes and their length-dependent energy oscillations probed at
  both ends of atomic shiba chains},}\ }\href {\doibase
  10.1038/s41565-022-01078-4} {\bibfield  {journal} {\bibinfo  {journal}
  {Nature Nanotechnology}\ }\textbf {\bibinfo {volume} {17}},\ \bibinfo {pages}
  {384--389} (\bibinfo {year} {2022})}\BibitemShut {NoStop}%
\end{thebibliography}%
\end{document}